\documentclass[10]{iopart}

\usepackage{epsf}  
\usepackage{natbib}
\usepackage{graphicx}
\graphicspath{{./fig/}}
\include{natbib}

\begin{document}
\def\newblock{\hskip .11em plus .33em minus .07em} 

\topmargin 25pt


\newcommand{\Index}[1]{#1\index{#1}} 
\newcommand{\J}[6]{#1\ (#2). #3, \textit{#4}\ \textbf{#5}, {#6}.}

\newcommand\ars{Adv.\ Radio Sci.}
\newcommand\asr{Adv.\ Space.\ Res.}
\newcommand\Aa{Astron.\ Astrophys.}             
\newcommand\aap{Astron.\ Astrophys.}             
\newcommand\Aar{Astron.\ Astrophys.\ Rev.}    
\newcommand\aapr{Astron.\ Astrophys.\ Review}   
\newcommand\aas{Astron.\ Astrophys.\ Suppl.}
\newcommand\aaps{Astron.\ Astrophys.\ Suppl.}
\newcommand\anl{Astron.\ Lett.}
\newcommand\al{Astrophys.\ Lett.}
\newcommand\aplett{Astrophys.\ Lett.}
\newcommand\araa{Ann.\ Rev.\ Astron.\ Astrophys.}
\newcommand\arfm{Ann.\ Rev.\ Fluid Mech.}
\newcommand\aj{Astron.\ J.}
\newcommand\an{Astron.\ Nachr.}
\newcommand\apj{Astrophys.\ J.}
\newcommand\apjl{Astrophys.\ J.\ Lett.}
\newcommand\apjs{Astrophys.\ J.\ Suppl.}
\newcommand\apss{Astophys.\ Space Sci.}
\newcommand\ajp{Aust.\ J.\ Phys.}
\newcommand\azh{Astron.\ Zh.}

\newcommand\basi{Bull.\ Astron.\ Soc.\ India}

\newcommand\casp{Comm.\ Astrophys.\ and Space Phys.}
\newcommand\cppcf{Comm.\ Plasma Phys.\ Controlled Fusion}
\newcommand\cpc{Computer\ Phys.\ Comm.}
\newcommand\cs{Current Science}

\newcommand\dan{Dokl.\ Akad.\ Nauk SSSR}

\newcommand\fcp{Fund.\ Cosmic Phys.}

\newcommand\geae{Geomagn.\ Aeron.}
\newcommand\gafd{Geophys.\ Astrophys.\ Fluid Dyn.}
\newcommand\grl{Geophys. \ Res. \ Lett.}

\newcommand\ijmpd{Int.\ J.\ Mod.\ Phys.\ D}

\newcommand\jas{J.\ Aero.\ Sci.}
\newcommand\jaa{J.\ Astrophys.\ Astron.}
\newcommand\zhetp{JETP}
\newcommand\jcap{J.\ Cosmol.\ Astropart.\  Phys.}
\newcommand\jfm{J.\ Fluid Mech.}
\newcommand\jqsrt{J.\ Quantit.\ Spectroscopy Radiative Transfer}
\newcommand\jt{J.\ Turbulence}

\newcommand\mhd{Magnetohydrodynamics}
\newcommand\mnras{MNRAS}

\newcommand\nat{Nature}
\newcommand\na{New Astron.}
\newcommand\nar{New.\ Astron.\ Rev.}
\newcommand\njp{New.\ J.\ Phys.}

\newcommand\obs{Observatory}

\newcommand\prev{Phys.\ Rev.}
\newcommand\prd{Phys.\ Rev.\ D}
\newcommand\pre{Phys.\ Rev.\ E}
\newcommand\prl{Phys.\ Rev.\ Lett.}
\newcommand\pha{Physica A}
\newcommand\phd{Physica D}
\newcommand\phla{Phys.\ Lett.\ A}
\newcommand\phlb{Phys.\ Lett.\ B}
\newcommand\physrep{Phys.\ Rep.}
\newcommand\pf{Phys.\ Fluids}
\newcommand\ppl{Phys.\ Plasmas}
\newcommand\prs{Proc.\ Roy.\ Soc.\ Lond.\ A}
\newcommand\ptrsl{Phil.\ Trans.\ Roy.\ Soc.\ Lond.\ A}
\newcommand\pasa{Publ.\ Astron.\ Soc.\ Australia}
\newcommand\pasj{Publ.\ Astron.\ Soc.\ Japan}
\newcommand\pasp{Publ.\ Astron.\ Soc.\ Pacific}

\newcommand\qjras{Quart.\ J.\ Roy.\ Astron.\ Soc.}

\newcommand\rfz{Radiofizika}
\newcommand\rfqe{Radiophys.\ Quantum Electronics}
\newcommand\rjmp{Russ.\ J.\ Math.\ Phys.}
\newcommand\rmp{Rev.\ Mod.\ Phys.}
\newcommand\rpp{Rev.\ Plasma Phys.}

\newcommand\sphys{Solar\ Phys.}
\newcommand\sa{Sov.\ Astron.}
\newcommand\sovast{Sov.\ Astron.}
\newcommand\sal{Sov.\ Astron.\ Lett.}
\newcommand\jetp{Sov.\ Phys.\ JETP}
\newcommand\spd{Sov.\ Phys.\ Dokl.}
\newcommand\spu{Sov.\ Phys.\ Usp.}
\newcommand\ssr{Space Sci.\ Rev.}
\newcommand\sgg{Studia Geophysica Geodaetica}

\newcommand\zn{Z.\ Naturforsch.\ A}

\newcommand{\cf}{see} 
\newcommand{\ie}{{i.e.}}            
\newcommand{\eg}{{e.g.}}            
\newcommand{\etc}{{etc}} 

\newcommand{\Al}{_\mathrm{A}}   
\newcommand{\cor}{_0} 	    
\newcommand{\crit}{_\mathrm{c}} 
\newcommand{\cra}{_\mathrm{cr}} 
\newcommand{\dif}{_\mathrm{d}}  
\newcommand{\eff}{_\mathrm{e}}  
\newcommand{\el}{_\mathrm{e}}   
\newcommand{\eq}{_\mathrm{eq}}  
\newcommand{\h}{_\mathrm{h}}    
\newcommand{\I}{_\mathrm{I}}    
\newcommand{\ion}{_\mathrm{i}}  
\newcommand{\kin}{_\mathrm{K}}  
\newcommand{\loc}{_\mathrm{L}}  
\newcommand{\M}{_\mathrm{m}}    
\newcommand{\maxi}{_\mathrm{max}}    
\newcommand{\sound}{_\mathrm{s}}
\newcommand{\pro}{_\mathrm{p}}  
\newcommand{\pr}{_\mathrm{p}}  
\newcommand{\rms}{_0}  					
\newcommand{\thermal}{_\mathrm{th}}  
\newcommand{\turb}{_\mathrm{t}} 
\newcommand{\w}{_\mathrm{w}}    

\newcommand{\EQ}{\begin{equation}}
\newcommand{\EN}{\end{equation}}
\newcommand{\EQA}{\begin{eqnarray}}
\newcommand{\ENA}{\end{eqnarray}}
%
\newcommand{\App}[1]{Appendix~\ref{#1}}
\newcommand{\Sec}[1]{Section~\ref{#1}}
\newcommand{\Chap}[1]{Chapter~\ref{#1}}
\newcommand{\Eq}[1]{Eq.~(\ref{#1})}
\newcommand{\Fig}[1]{Fig.~\ref{#1}}
\newcommand{\Figs}[2]{Figures~\ref{#1} and \ref{#2}}
\newcommand{\Eqs}[2]{Eq.~(\ref{#1}) and (\ref{#2})}
\newcommand{\bra}[1]{\langle #1\rangle}
\newcommand{\bbra}[1]{\left\langle #1\right\rangle}
\newcommand{\mean}[1]{\overline{#1}}

\newcommand{\const}{{\rm const}  {}} 
\newcommand{\emf}{{\cal E}}          
\newcommand{\tildemeanEMF}{\tilde{\mbox{\boldmath ${\cal E}$}}}
\newcommand{\totalE}{\tilde{\mbox{$E$}}}
\newcommand{\totalH}{\tilde{\mbox{$H$}}}
\newcommand\Fo[1]{{\widehat{#1}}}    
\newcommand\col[2]{\left(\begin{array}{c}#1\\#2\end{array}\right)}
\newcommand{\conj}{^\dagger} 					
\newcommand{\myvcenter}[1]{\ensuremath{\vcenter{\hbox{#1}}}} 

\newcommand\deriv[2]{\displaystyle\frac{\partial #1}{\partial #2} }
\newcommand{\dd}{\mathrm{d}} 
\newcommand{\ii}{\mathrm{i}} 
\newcommand{\DD}{\mathrm{D}} 

\newcommand\sfrac[2]{{\textstyle{\frac{#1}{#2}}}}
%
\renewcommand{\vec}[1]{{{\mbox{\boldmath $#1$}}}}
\newcommand\uvec[1]{{\widehat{\vec{#1}}}}        
\newcommand{\aaa}{{\vec{a}}} 
\newcommand{\AAA}{{\vec{A}}} 
\newcommand{\bb}{{\vec{b}}}
\newcommand{\BBB}{{\vec{B}}}
\newcommand{\cc}{{\vec{c}}}
\newcommand{\CC}{{\vec{C}}}
\newcommand{\EE}{{\vec{E}}}
\newcommand{\ee}{{\vec{e}}}
\newcommand{\ff}{{\vec{f}}}
\newcommand{\FF}{{\vec{F}}}
\newcommand{\HH}{{\vec{H}}}
\newcommand{\TT}{{\vec{T}}}
\newcommand{\GG}{{\vec{G}}}
\newcommand{\GGGG}{\mbox{\boldmath ${\sf G}$} {}}
\newcommand{\jj}{{\vec{j}}}
\newcommand{\JJ}{{\vec{J}}}
\newcommand{\kk}{{\vec{k}}}
\newcommand{\KK}{{\vec{K}}}
\newcommand{\lv}{{\vec{l}}}
\newcommand{\nn}{{\vec{n}}}
\newcommand{\nnn}{{\uvec{n}}}
\newcommand{\kkk}{{\uvec{k}}}
\newcommand{\NN}{{\vec{N}}}
\newcommand{\pp}{{\vec{p}}}
\newcommand{\qq}{{\vec{q}}}
\newcommand{\QQ}{{\vec{Q}}}
\newcommand{\rr}{{\vec{r}}}
\newcommand{\RR}{{\vec{R}}}
\newcommand{\uu}{{\vec{u}}}
\newcommand{\UU}{{\vec{U}}}
\newcommand{\vv}{{\vec{v}}}
\newcommand{\VV}{{\vec{V}}}
\newcommand{\ww}{{\vec{w}}}
\newcommand{\WW}{{\vec{W}}}
\newcommand{\xx}{{\vec{x}}}
\newcommand{\XX}{{\vec{X}}}
\newcommand{\yy}{{\vec{y}}}
\newcommand{\YY}{{\vec{Y}}}
\newcommand{\zz}{{\vec{z}}}
\newcommand{\ZZ}{{\vec{Z}}}
\newcommand{\SSS}{\vec{S}}
\newcommand{\OO}{\vec{\Omega}}
\newcommand{\OOO}{\uvec{\Omega}}
\newcommand{\oo}{\vec{\omega}}
\newcommand{\ggrav}{\uvec{g}}
\newcommand{\grav}{\vec{g}}
\newcommand{\nab}{{{\nabla}}}
\newcommand{\fjk}{F_{jk}}
%
\newcommand{\tensor}[1]{{{\mbox{\boldmath ${\sf #1}$}}}}

\def\la{\mathrel{\mathchoice {\vcenter{\offinterlineskip\halign{\hfil
$\displaystyle##$\hfil\cr<\cr\sim\cr}}}
{\vcenter{\offinterlineskip\halign{\hfil$\textstyle##$\hfil\cr<\cr\sim\cr}}}
{\vcenter{\offinterlineskip\halign{\hfil$\scriptstyle##$\hfil\cr<\cr\sim\cr}}}
{\vcenter{\offinterlineskip\halign{\hfil$\scriptscriptstyle##$\hfil\cr<\cr\sim\cr}}}}}
\def\ga{\mathrel{\mathchoice {\vcenter{\offinterlineskip\halign{\hfil
$\displaystyle##$\hfil\cr>\cr\sim\cr}}}
{\vcenter{\offinterlineskip\halign{\hfil$\textstyle##$\hfil\cr>\cr\sim\cr}}}
{\vcenter{\offinterlineskip\halign{\hfil$\scriptstyle##$\hfil\cr>\cr\sim\cr}}}
{\vcenter{\offinterlineskip\halign{\hfil$\scriptscriptstyle##$\hfil\cr>\cr\sim\cr}}}}}
%
\newcommand{\Pra}{\mathrm{Pr}}          
\newcommand{\Prm}{\mathrm{Pr}_\mathrm{m}}
\newcommand{\Pm}{\mathrm{Pr}_\mathrm{m}}
\newcommand{\Rey}{\mbox{\rm Re}}        
\newcommand{\Rm}{R_\mathrm{m}}          
\newcommand{\Rmc}{R_\mathrm{m, c}}      
\newcommand{\Ro}{\mathrm{Ro}}           
%
\newcommand{\cp}{{\cal P}}            
\newcommand{\DM}{{\rm DM}}            
\newcommand{\DP}{{\rm DP}}            
\newcommand{\EM}{{\rm EM}}            
\newcommand{\fd}{F}                   
\newcommand{\kB}{k_\mathrm{B}}        
\newcommand{\po}{p_\mathrm{i}}        
\newcommand{\PI}{P}                   
\newcommand{\pRM}{{\cal R}}           
\newcommand{\RM}{{\rm RM}}            

\newcommand{\HI}{\rm H\,{\sc i}}      
\newcommand{\HII}{\rm H\,{\sc ii}}    

%
\newcommand{\A}{\,{\rm \AA}}    
\newcommand{\AU}{\,{\rm AU}}    
\newcommand{\mkm}{\,\mu{\rm m}} 
\newcommand{\mm}{\,{\rm mm}}    
\newcommand{\cm}{\,{\rm cm}}    
\newcommand{\km}{\,{\rm km}}    
\newcommand{\m}{\,{\rm m}}      
\newcommand{\Mm}{\,{\rm Mm}}    
\newcommand{\p}{\,{\rm pc}}     
\newcommand{\kpc}{\,{\rm kpc}}  
\newcommand{\Mpc}{\,{\rm Mpc}}  

\newcommand{\g}{\,{\rm g}}      
\newcommand{\kg}{\,{\rm kg}}    
\newcommand{\mol}{\,{\rm mol}}  

\newcommand{\days}{\,{\rm d}}   
\newcommand{\GHz}{\,{\rm GHz}}  
\newcommand{\Hz}{\,{\rm Hz}}    
\newcommand{\kHz}{\,{\rm kHz}}  
\newcommand{\MHz}{\, {\rm MHz}} 
\newcommand{\nHz}{\,{\rm nHz}}  
\newcommand{\s}{\,{\rm s}}      
\newcommand{\yr}{\,{\rm yr}}    
\newcommand{\Myr}{\,{\rm Myr}}  
\newcommand{\Gyr}{\,{\rm Gyr}}  

\newcommand{\cms}{\cm\s^{-1}}    
\newcommand{\kms}{\km\s^{-1}}    

\newcommand{\G}{\,{\rm G}}      
\newcommand{\kG}{\,{\rm kG}}    
\newcommand{\mG}{\,{\rm mG}}    
\newcommand{\mkG}{\,\mu{\rm G}} 
\newcommand{\nG}{\,{\rm nG}}    
\newcommand{\Mx}{\,{\rm Mx}}    
\newcommand{\V}{\,{\rm V}}      
\newcommand{\kV}{\,{\rm kV}}    
\newcommand{\T}{\,{\rm T}}      

\newcommand{\C}{\,{\rm C}}      
\newcommand{\dyn}{\,{\rm dyn}}  
\newcommand{\erg}{\,{\rm erg}}  
\newcommand{\eV}{\,{\rm eV}}    
\newcommand{\keV}{\,{\rm keV}}  
\newcommand{\MeV}{\,{\rm MeV}}  
\newcommand{\GeV}{\,{\rm GeV}}  
\newcommand{\TeV}{\,{\rm TeV}}  
\newcommand{\K}{\,{\rm K}}      
\newcommand{\Jo}{\,{\rm J}}      
\newcommand{\Jy}{\,{\rm Jy}}    
\newcommand{\Jyb}{\,{\rm Jy/beam}}      
\newcommand{\mJy}{\,{\rm mJy}}          
\newcommand{\mJyb}{\,{\rm mJy/beam}}    
\newcommand{\mkJy}{\,\mu{\rm Jy}}          
\newcommand{\mkJyb}{\,\mu{\rm mJy/beam}}    
\newcommand{\kW}{\,{\rm kW}}    
\newcommand{\MW}{\,{\rm MW}}    
\newcommand{\W}{\,{\rm W}}      

\newcommand{\rad}{\,{\rm rad}} 
\newcommand{\sterad}{\,{\rm sr}} 

\newcommand{\ave}[1]{{\left\langle{#1}\right\rangle}}
\newcommand{\avw}[1]{{\left\langle{#1}\right\rangle}_{\! W}}
\newcommand{\avwl}[1]{{\left\langle{#1}\right\rangle}_{\! W\times h}}


\newcommand{\ini}{{}_\mathrm{i}}
\newcommand{\f}{{}_\mathrm{f}}
\newcommand{\cl}{_\mathrm{c}}
\newcommand{\scl}{_\mathrm{sc}}
\newcommand{\phzero}{\phantom{0}}
\newcommand{\tphzero}{\phantom{0}\phantom{0}}
\newcommand{\phone}{\phantom{1}}

\index{gyration!frequency|see{Larmor frequency}}
\index{gyration!radius|see{Larmor radius}}
\index{material!derivative|see{Lagrangian derivative}}
\index{total!derivative|see{Lagrangian derivative}}

\review[Primordial magnetic fields]{The origin, evolution and signatures of 
primordial magnetic fields}

\author{Kandaswamy Subramanian}

\address{IUCAA, Post Bag 4, Ganeshkhind, Pune 411007, India}
\ead{kandu@iucaa.in}
\begin{abstract}
The universe is magnetized on all scales probed so far.
On the largest scales, galaxies and galaxy
clusters host magnetic fields at the micro Gauss level coherent
on scales up to ten kpc.  
Recent observational evidence suggests that even the 
intergalactic medium in voids could host a 
weak $\sim 10^{-16}$ Gauss magnetic field, 
coherent on Mpc scales.
An intriguing possibility is that these observed magnetic fields
are a relic from the early universe, albeit one which
has been subsequently amplified and maintained 
by a dynamo in collapsed objects. 
We review here the origin, evolution and signatures
of primordial magnetic fields. After a brief summary of
magnetohydrodynamics in the expanding universe, we turn to
magnetic field generation during inflation and phase transitions.
We trace the linear and nonlinear evolution of the generated 
primordial fields through the radiation era, including viscous effects. 
Sensitive observational signatures of
primordial magnetic fields on the cosmic microwave background,
including current constraints from Planck, are discussed.  
After recombination, primordial magnetic fields could 
strongly influence structure formation,
especially on dwarf galaxy scales. The resulting 
signatures on reionization, 
the redshifted 21 cm line, weak lensing and the
Lyman-$\alpha$ forest are
outlined. Constraints from radio and $\gamma$-ray astronomy 
are summarized. Astrophysical batteries and the role of dynamos 
in reshaping the primordial field are briefly considered. 
The review ends with some final thoughts on primordial magnetic fields.
\end{abstract}

\maketitle

\ioptwocol
\tableofcontents

\medskip

\hrule

\section{The magnetic universe}

Magnetic fields are ubiquitous on all scales probed so far,
from planets and stars to the large-scale magnetic fields 
detected in galaxies and galaxy clusters. 
The earths dipolar magnetic field of about a Gauss 
has been sustained for billions of years
by some form of dynamo action \citep{PO13}. Several other solar system planets
also display ordered fields \citep{St10}.
The Sun displays magnetic cycles with its dipolar magnetic field 
changing sign every 11 years \citep{H10} 
again possibly due to dynamo action \citep{BSS12,C14}.
Nearby spiral galaxies host
magnetic fields with a strength of
a few to tens of micro Gauss coherent on scales up to 
ten kpc \citep{B01,BW13}. Similar fields are also tentatively detected 
in higher redshift galaxies \citep{BMLKD08}. 
In clusters of galaxies, stochastic magnetic fields of a few micro 
Gauss strength are present, correlated on ten kpc scales 
\citep{CKB01,GF04,VE05}. 
Recent observational evidence suggests that even the 
intergalactic medium (IGM) in voids could host a 
weak $\sim 10^{-16}$ Gauss magnetic field, 
coherent on Mpc scales \citep{NV10}.
The origin and evolution of these magnetic fields is a 
subject of intense study. 

An intriguing possibility is that cosmic magnetic fields
are a relic from the early universe, albeit one which
has been subsequently amplified by a dynamo in collapsed
objects. 
Indeed any IGM field which volume fills the void regions would be
difficult to explain purely by astrophysical processes in the
late universe \citep{FL01,BVE06},
and would perhaps favour such a primordial origin.
Thus it is of great interest to ask if such a primordial field
can be generated in the early universe and also how they could
be detected and constrained. This forms the prime focus of the
present review, which considers
the origin, evolution and signatures of primordial magnetic fields.
Our guiding principle for the topics reviewed, is that the reader
gets a unified overview of primordial magnetic fields, right from its
generation, to its evolution, which then leads to observational signatures.

We will see that magnetic field strength
generally decreases (redshifts) as the universe expands as 
$B(t) \propto 1/a^2(t)$, where $B(t)$ is the field strength at epoch $t$,
and $a(t)$ is the expansion or scale factor of the universe 
(neglecting nonlinear and dissipative effects).
Thus the energy density in magnetic
fields generated in the early universe will scale as 
$\rho_B(t)=B^2(t)/(8\pi) \propto 1/a^4(t)$. 
This scaling also obtains for the energy density of 
any cosmic radiation present in the universe.
Indeed as discussed below, the universe is filled with a cosmic microwave
background radiation (CMB), a relic of its hot 'big bang' beginnings, 
with a thermal spectrum and present day
temperature of $T=2.725$ K \citep{mather_etal94}.
The energy density
of this radiation formed a dominant component of 
the energy density of the early universe, 
and dilutes as the universe expands as $\rho_\gamma(t) \propto 1/a^4(t)$.
Therefore, the ratio $r_B=\rho_B(t)/\rho_\gamma(t)$ is 
approximately constant
\footnote{Only approximate as during certain epochs, annihilation
of particles can increase the energy in photons.}
with epoch. 
It is then standard practice to characterize the primordial
field with either this ratio, or the present day value $B_0$
as a function of its coherence scale $L$. 
A present day magnetic field $B_0 \sim 3.2 \mu $G has 
an energy density equal to the present day CMB energy density,
or $r_B=1$.

A number of arguments suggest that a primordial 
field with a present day strength $B_0$ of order a nano Gauss (nG)
and coherent on Mpc scales, will have a significant effect on cosmology
(see below).
For such a field,
\EQ
r_B=\frac{B_0^2}{8\pi\rho_{\gamma 0}} \approx \frac{B^2(t)}{8\pi \rho_\gamma(t)}
 \approx 10^{-7} B_{-9}^2
\label{Bestimate}
\EN
where $\rho_{\gamma 0}$ is the present day energy density in radiation, and
$B_{-9} = B_0/(10^{-9} G)$ is the present-day magnetic field in units of
a nano Gauss.
So magnetic stresses are in general small compared to
the radiation energy density and its pressure, for nano Gauss fields.
The frozen in field assumption
breaks down at small scales; however the magnetic energy will only 
be smaller if there is decay.

An important question of course is how such a field can originate?
Likely scenarios include origin in various phase transitions which
may have occurred
in the early universe. The present day large scale structure 
in the universe is thought to be seeded by quantum fluctuations, which 
transit to classical density fluctuations, during an early inflationary 
(accelerated) expansion phase of the universe 
(cf. \cite{Kolb_Turner,Dodelson03,Padmanabhan02}. 
A possibility 
worth exploring is whether coherent large scale magnetic fields could 
also arise in this era 
\citep{TW88}? 
Or could a small fraction of the free energy
released during phase transitions like the electroweak or quark-hadron
transitions, be converted to large-scale magnetic fields 
\citep{Hogan83}? 
After all one requires only a small fraction $r_B$ to go into such 
long-wavelength modes.
These questions are discussed in \Sec{earlygen}.   

The further evolution of a primordial field generated during inflation 
or 
various
phase transitions, depends on its strength, spectrum and helicity
content. 
Large scales will have a frozen in evolution and simply dilute
with expansion as described above. Smaller scales will be subject to
nonlinear processing and damping
\citep{BJ04}.
The field coherence scale can increase
in the process, although its energy density will decrease. Conservation
of magnetic helicity plays an important role and leads to a larger
coherence scale than for a non helical field 
\citep{CHB01}.
The evolution of primordial 
fields in both the linear and nonlinear regime is taken up in \Sec{linearevol} 
and \Sec{nonlin} respectively.

A clean probe of primordial magnetic fields is to look for 
CMB anisotropies induced by such fields. 
The scalar, vector and tensor parts of the
perturbed stress tensor associated with primordial magnetic fields
lead to corresponding metric perturbations.
Further the compressible part of the Lorentz force leads to
compressible (scalar) fluid velocity and associated density
perturbations, while its vortical part leads to
vortical (vector) fluid velocity perturbation.
The magnetically induced compressible fluid perturbations,
for nano Gauss fields, are highly subdominant compared to
the fluid perturbations due to the scalar modes generated
during the inflationary era, and which are 
responsible for structure formation. 
For a CMB temperature anisotropy $\Delta T/T \sim 10^{-5}$
due to say the inflationary scalar modes, the scalar pressure
perturbations due to these modes are
$\delta p/p = 4\Delta T/T \sim 4 \times 10^{-5}$,
and so much larger than the magnetic pressure perturbation
$B^2/(8\pi p) \sim 10^{-7} B_{-9}^2$.
(Although scalar perturbations can still lead to
additional CMB anisotropies; see below).
Potentially more important are the vortical modes driven by
the rotational component of the Lorentz force, especially
since they survive damping due to radiative viscosity at scales much below
the scalar modes \citep{JKO98,SB98a}.

These perturbations due to primordial magnetic fields will induce 
temperature and polarization anisotropies in
the Cosmic Microwave Background (CMB)
The signals that could be searched for include
excess temperature anisotropies 
(from scalar, vortical and tensor perturbations), 
B-mode polarization (from tensors and vorticity), and non-Gaussian
statistics \citep{S06,Durrer07,WidSSR12,PlanckB15}. 
A field at a few nG level produces
temperature anisotropies at the $5\,\mu$K level, 
and B-mode polarization anisotropies 10 times smaller, and is 
therefore potentially detectable via the CMB anisotropies.
An even smaller field, with $B_0 \sim 0.1$ nG, if present on 
large scales, can lead to significant non-Gaussianity in the CMB
\citep{SS09,CFPR09,SNYIT11,TSS12,TSS14}. 
The CMB signatures are discussed in \Sec{cmb}.

After recombination, the baryons no longer feel the pressure
due to radiation but only their own pressure. Since the baryon to photon ratio
is  very small $\sim 10^{-9}$, the surviving inhomogeneous magnetic fields
can, if strong enough, induce compressible motions in the gas. 
For example nG fields which produced pressure perturbations
of order $3r_B \sim 3 \times 10^{-7}B_{-9}^2$, will 
just after recombination have a pressure a few hundred times
larger than the fluid pressure.
The gravitational influence of the resulting inhomogeneous
baryon distribution can seed density perturbations in the dark
matter. These perturbations will be amplified
due to gravitational instability, with the matter
power spectrum typically peaked on small scales,
for a scale invariant magnetic spectrum, 
and can lead to the
formation of the first dwarf galaxies.
The magnetic energy can also be dissipated by ambipolar
diffusion and decaying magneto hydrodynamic (MHD) turbulence
to heat and ionize the intergalactic medium (IGM)
\citep{Sethi_KS05}.
These processes leave signatures of primordial fields 
on reionization, the redshifted 21 cm line, 
and weak lensing.
We will see that a field with $B_0 \sim 0.1$ nG can
lead to structure formation at high redshift $z > 15$, 
impacting on the re-ionization of the Universe 
\citep{Sethi_KS05,CPFR15}
and significant weak lensing signatures \citep{PS12}.
The evolution and signatures of primordial fields post recombination
are discussed in \Sec{postrecomb}.
We also consider there constraints on primordial fields from a range of
other observational probes, like the gamma ray and
radio observations.

A $0.1$nG field in the IGM could also
be sheared and amplified due to flux freezing,
during the collapse to form a galaxy
to give $\mu$G strength fields observed in disk galaxies
(cf. \cite{K99}).
Of course, one will still need a dynamo to maintain such a field
against decay, unless it is helical \citep{Blackman_Sub13,BBS14} 
and/or explain the observed global structure of 
disk galaxy fields \citep{S07,CSS13a}.
Weaker primordial fields can 
still be sufficient to account for
the fields in voids which may have been detected 
in high energy gamma-ray observations \citep{NV10}, 
or to seed the first dynamos. 
In addition purely astrophysical processes can also
lead to coherent seed fields, albeit weaker than
required by gamma ray observations.
Batteries and dynamos are briefly discussed in \Sec{batdyn}.
The last section presents some 
final thoughts on the issues covered in the review.

There have been a number of excellent earlier reviews on primordial
magnetic fields \citep{GR01,W02,WidSSR12,DN13}, one of which
also included the present author. The current review differs 
from these in terms of perspective and emphasis, 
inclusion of new material and a somewhat
more pedagogical approach to some of the material.
In relation to the review of \citet{GR01,W02}, we
cover more recent material, particularly on the evolution and
signatures of primordial fields as presented in Chapters 5-9.
In relation to \citet{WidSSR12}, we give 
a more pedagogical discussion of several topics and our perspective 
and emphasis is somewhat different from \citet{DN13}.

We begin in the next section with a brief summary of
cosmology and the early universe, before describing
magnetohydrodynamics in the expanding universe.

\section{Cosmology and the Early Universe}
\label{cosmo}

Modern cosmology is based on a few basic observational keystones.
First is the discovery by Hubble that more distant a galaxy is from us
the faster it moves away from us. This discovery has been firmed up
considerably over the years and is known as the Hubble law. Combined with
the Copernican principle that we are not a special observer in the universe,
it leads to the concept of an expanding universe; that all observers move
away from each other due to an underlying expansion of the space, described
by an expansion or scale factor $a(t)$ (see below).

The second key input into cosmology arises from the discovery of the cosmic
microwave background radiation (CMB). 
The serendipitous discovery of the CMB
by \citet{PW65},
gave the first clear indication of an early hot "Big bang" stage of
the evolution of the universe. The subsequent verification by host
of experiments, culminating in the results of the COBE
satellite confirmed that its spectrum
is very accurately Planckian \citep{mather_etal94},
with a temperature $T=2.725$. This is the firmest evidence
that the universe was in thermal equilibrium at some early stage.

The dynamics of the universe on the largest scales is governed by
gravity. A study of cosmology necessarily entails
understanding and using a consistent theory of gravity, viz. 
general relativity, with of course simplifying assumptions.
The basic simplifying principle
known as the Cosmological principle, assumes that at each instant of {\it
time} the universe (the spatial geometry and matter) is {\it homogeneous} and
{\it isotropic}. Here we have to understand what is the "time" being referred to
as well as for which observer the universe is homogeneous and isotropic. This
is clarified by postulating that (i) the universe can be foliated by a regular
set of space like hyper surfaces $\Sigma$ and (ii) that there exist a set of
"fundamental observers" whose world lines $x^i$ are a set of non-intersecting
geodesics orthogonal to $\Sigma$. These assumptions are referred to as the
"Weyl postulate" (\cf\ \citet{Narlikar02}). The time $t$ is the parameter which
labels a particular space-like hypersurface, and it is for these fundamental
observers that the universe is assumed to be isotropic and homogeneous.

\subsection{The FRW models}

The geometry of spacetime in General Relativity is specified by the
metric tensor $g_{\mu \nu}$, which gives the spacetime interval $ds$
between two infinitesimally separated events, that is
$ds^2 = g_{\mu \nu}dx^\mu dx^\nu$.
Here and below the Greek indices $\mu,\nu$ etc run over the spacetime co-ordinates
and we assume that repeated indices are summed over all the co-ordinates.
For a universe whose constant time spatial slices are
isotropic and homogeneous, the metric is given by the Friedman-Robertson-Walker (FRW)
metric,
\EQ
ds^2 = -dt^2 + a^2(t) \left[ \frac{dr^2}{1 - kr^2} + r^2(d\theta^2
+ \sin^2\theta d\phi^2)\right],
\EN
where $0 < \theta < \pi$ and $0 \le \phi < 2\pi$ are the usual angular
co-ordinates on the sphere, whose comoving `radius' is $r$. 
The time coordinate $t$ is the proper time measured by a comoving
observer who is at rest with $(r,\theta,\phi)$ constant.
We also adopt units in which $c=1$ unless
otherwise stated. The spatial sections have flat, open or closed
geometry for $k=0,-1$ or $+1$ respectively. The expansion of the universe
is described by the scale factor $a(t)$.
Note that as the universe expands, particle momenta decrease as $1/a$,
and in particular the frequency of photons $\nu \propto 1/a$,
or wavelengths increase as $\lambda =c/\nu \propto a(t)$. Thus
one can also characterise the epoch $t$ or the scale factor $a(t)$
by redshift $z$, where $1+z(t) = a(t_0)/a(t)$. We can currently
detect galaxies to a redshift of about $10$ or so. 

The evolution of the scale factor $a(t)$ is determined
by Einstein equations
\EQ
R_{\mu\nu} - \frac{1}{2} g_{\mu\nu} R =  8 \pi G T_{\mu\nu}
\EN
where $R_{\mu\nu}$ is the Ricci tensor measuring the curvature of
space time, $R=g^{\mu\nu}R_{\mu\nu}$ is the scalar curvature, and $G$ is the
gravitational constant. The matter content is incorporated in the
energy-momentum tensor $T_{\mu\nu}$. For a perfect fluid with density
$\rho$ and pressure $p$, we have
\EQ
T_{\mu\nu} = (\rho +p)u_\mu u_\nu - p g_{\mu\nu}
\label{fluid_emt}
\EN
with $u^\mu \equiv (1,0,0,0)$ the 4-velocity of the
fundamental observers. For the FRW metric Einstein equations can be cast into the form,
\EQ
\left(\frac{\dot{a}}{a}\right)^2 +\frac{k}{a^2} = \frac{8\pi G}{3} \rho;
\qquad \frac{\ddot{a}}{a} = -\frac{4 \pi G}{3} \left( \rho + 3p \right).
\label{einstein}
\EN
The second of these equations shows that the universe accelerates
or decelerates, with $\ddot{a}$ positive or negative, when
$(\rho + 3p) < 0$ or $(\rho + 3p) > 0$, respectively.
For most of the evolution of the universe, its matter content has
positive $\rho$ and $p$ and so it decelerates. There are however
two very important epochs during its evolution, when the universe
is thought to be accelerating; during the epoch of inflation (see below) and
during the present epoch of dark energy domination.
The first of \Eq{einstein} can be cast into an equation for the constant $k$
\EQ
k = a^2 H^2 \left[ \frac{8\pi G\rho}{3H^2} - 1 \right]
= a_0^2 H_0^2 \left[ \frac{\rho_0}{\rho_{cr}} - 1 \right] \;,
\label{constantk}
\EN
where we have defined the Hubble rate $H(t) = \dot{a}(t)/a(t)$.
Since $k$ is a constant it can be evaluated at any epoch and in
the latter part of \Eq{constantk}, we have evaluated all the quantities at the
present epoch $t_0$. Thus $H_0=H(t_0)$ is the
Hubble constant giving the present rate of expansion,
$\rho_0$ is the present density and $a_0$ the present
expansion factor. We have also defined a critical density
$\rho_{cr} = 3H_0^2/(8\pi G)$. Whether the present density equals, exceeds
or is less than the critical density,
determines respectively, if the universe is flat with $k=0$ ($\rho_0 = \rho_{cr}$),
is closed with $k=1$ ($\rho_0 > \rho_{cr}$) or open with
$k=-1$ ($\rho_0 < \rho_{cr}$). Current observations indicate the universe
is very close to being spatially flat.

The two Einstein equations can also be combined to give the energy conservation
equation
\EQ
\frac{d}{dt} \left( \rho a^3 \right) + p \frac{d}{dt} \left( a^3 \right) = 0.
\label{energy}
\EN
To solve these equations requires one to specify
the relation between $p$ and $\rho$.
For a general equation of state of the form $p = w\rho$, \Eq{energy}
gives $\rho \propto a^{-3(1+w)}$.
Note that for 'dust' like matter $p=0$, and $\rho =\rho_M \propto a^{-3}$,
while for radiation $p=\rho/3$ leading to $\rho =\rho_R \propto a^{-4}$.

Consider the solutions for the spatially flat $k=0$ case. During
radiation domination, the solution of
\Eq{einstein} then gives $a(t) \propto t^{1/2}$,
while during the matter dominated
epoch we have $a(t) \propto t^{2/3}$.
For a general equation of state $p=w\rho$, we have $a(t) \propto t^{2/(3(1+w))}$,
as long as $w\ne -1$. For the case $w=-1$, with say $p=-\rho = -\rho_i < 0$,
one has accelerated expansion, with $a(t) = \exp( H_i t)$ where
$H_i = \sqrt{8\pi G\rho_i/3}$. Such epochs could be relevant
during inflation, or dark energy domination.

\subsection{Physics of the early universe}

We noted above
the discovery of expansion
of the universe and the CMB.
The spectrum of the CMB is to very high degree
of accuracy, a Planck spectrum with a present day temperature $T_0 = 2.725$ K.
This indicates that the radiation must have been in thermal equilibrium
with matter sometime in its history.
At the current epoch radiation is subdominant compared to matter.
However, from the fact that $\rho_R/\rho_M \propto 1/a$,
the universe will become radiation dominated in the past when
$a(t)$ is small enough. This happens according
to current estimates at a redshift $z = \rho_M(t_0)/\rho_R(t_0) \sim 3000$.
Since the frequency of photons redshift with expansion as
$\nu \propto 1/a$, the CMB will have
a Planck spectrum even in the past with a higher temperature $T=T_0/a(t)$.
These facts lead to the notion of the `hot big bang' model, whereby
the universe was radiation dominated and hot at some stage in its
evolution and subsequently cooled with expansion.

In the standard model of particle physics, it is thought that the
electromagnetic and weak interactions are unified into
the electroweak theory at energies higher than say 100 GeV.
Also the baryons are composite objects composed of quarks,
which would be revealed at high enough energies, greater than
about $150$ MeV. The theory
describing the strong interactions between quarks is
described by quantum chromodynamics (QCD).
All three interactions could be described by one grand unified theory
(GUT) at very high energies of about $10^{15}$ GeV. Even more
speculatively, gravity could also be unified with the other 3 interactions
at the Planck energy of $\sim 10^{19}$ GeV.
Since the universe can in principle attain
arbitrarily high energies, it is likely that in the beginning
all forces were unified and as the universe cooled it underwent
a series of symmetry breaking phase transitions; the GUT
phase transition at $T\sim 10^{15}$ GeV, the electroweak 
phase transition (EWPT)
at $T\sim 100$ GeV and finally the quark-hadron transition at
$T \sim 150$ Mev.
These phase transitions could be important for baryon number generation.
Equally they may be important for primordial magnetic field generation, 
as discussed in \Sec{earlygen}. 

As the universe cooled further
below a temperature of a few Mev, nucleosynthesis of light
elements occur. Also weak interaction rates become smaller
than the expansion rate and neutrinos decouple from the rest
of the matter and free stream to produce a neutrino background
analogous to the CMB. Just after this epoch of neutrino decoupling,
as $T$ drops below the electron rest mass, and electrons and positrons
annihilate and dump their energy into photons. This leads to a
slightly higher temperature for the CMB compared to that of the
neutrinos, with $T_\nu/T_0 = (4/11)^{1/3}$. When the
temperature drops below $T \sim 3000$ K,
the ions and electrons can combine to form atoms,
an epoch called the recombination epoch. This happens not at $T\sim 10^5$ K,
when the typical photon energy is $13.6$eV (the ionization potential
of hydrogen), but at lower temperatures $T \sim 3000$ K.
This is basically because the photon to baryon ratio
$n_\gamma/n_B \sim 10^9$, and so there are sufficient
number of photons even in the Planckian tail above the ionization potential,
to keep Hydrogen ionized,
until the temperature drops to $T \sim 3000$ K (much below $T\sim 10^5$
corresponding to the ionization potential).
After atoms form, the radiation decouples from matter
and free streams to give the radiation background
that we observe today as the CMB.
The above gives a brief thermal history of the universe.
More details can be found in many excellent cosmology textbooks
\citep{Kolb_Turner,Narlikar02,Padmanabhan02,M05,Weinberg08,Gor_Rub11}.

A number of puzzling features of the universe all find a plausible
explanation if one postulates an epoch of accelerated expansion in
the early universe, referred to as inflation.
There are extensive pedagogical discussions of the inflationary era 
in cosmology text books cited above (see also \citet{L90,L14,M15}).
Here we will give a very brief account of some relevant features
and point the interested reader to the above references for more
details.
For example, consider the comoving distance which light could have traveled in
a time interval from $t=0$ until time $t$. This is given by
\EQ
d_H(t) = \int_0^t \frac{dt'}{a(t')} = \int_0^a \frac{d\ln a}{aH}, 
\quad {\rm with} \quad H = \frac{1}{a} \ \frac{da}{dt},
\label{comove}
\EN
and gives the maximum comoving 'size'
of the region (called the particle horizon), 
which could have been in causal contact at any time $t$.
Note that the factor $(aH) = \dot{a}$ decreases with time in
a decelerating universe, or the comoving Hubble radius $R_H = (1/a) (1/H)$ 
increases with time. Thus the particle horizon $d_H$  
increases with time, in a decelerating universe. 
Its size today is much larger than 
at the time when the CMB photons decoupled from the baryonic matter.
This in turn implies that within our comoving horizon, there are
many causally unconnected patches at the time of CMB decoupling.
In fact locations on the CMB last scattering surface, which are separated
by more than a degree or so, were never in causal contact; their
past light cones never intersected before they hit the singularity.
How then is the CMB isotropic to 1 part in $10^5$ across 
the whole sky?
 
Moreover, the present day universe is inhomogeneous on scales
smaller than about $300$ Mpc; with structures like galaxies, 
galaxy clusters and super clusters. The matter in such structures, 
which presumably formed due a correlated collapse driven by 
gravitational instability, would
not be in causal contact at sufficiently early epochs, again if the
universe was always decelerating. 
How then were such correlated initial conditions set, between regions
which were apparently not in causal contact?

These features all appear to find
an explanation if the presently observable universe, was accelerating
at some early epochs. Such an acceleration leads to the the possibility 
that the comoving Hubble radius decreases with time, such that the
dominant contribution to the integral in \Eq{comove} is from early times. 
Then the elapsed conformal time to the last scattering surface (LSS) can
become sufficiently large, (if inflation lasts long enough), 
that light cones from all points on the LSS 
intersect sufficiently back in the past. This implies that the
whole observable universe was inflated
out of a region which was at some initial stage in causal contact.
The observed near isotropy of the CMB can then be accounted for 
and the needed correlated fluctuations to form
galaxies can arise from purely causal processes during inflation.
We will also see below that inflation provides 
ideal conditions for the generation
of primordial fields with large coherence scales.
Before coming to this, we consider first some general features
of how to formulate electrodynamics itself in the expanding universe.
  
\section{Electrodynamics in curved spacetime}
\label{EDcurved}

We 
first
discuss Maxwell equations in a general curved spacetime
and then focus on FRW models. Electrodynamics in curved spacetime
is most conveniently formulated by giving the action for electromagnetic
fields and their interaction with charged particles:
\EQ
S = -\int \sqrt{-g} \ d^4x \ \frac{F_{\mu \nu}F^{\mu \nu}}{16 \pi}
+ \int \sqrt{-g} \ d^4x \  A_\mu J^\mu
\label{emaction}
\EN
Here $F_{\mu \nu} = A_{\nu;\mu} - A_{\mu;\nu}
= A_{\nu,\mu} - A_{\mu,\nu}$ is the electromagnetic (EM) field tensor, with
$A_\mu$ being the standard electromagnetic 4-potential and $J^\mu$ the
4-current density.
Demanding that the action is stationary under the variation
of $A_\mu$, gives one half of the Maxwell equations.
And from the definition of the electromagnetic field tensor we also
get the source free part of the Maxwell equations. Thus
\EQ
F^{\mu \nu}_{\quad;\nu} = 4\pi J^\mu, \quad 
F_{[\mu\nu  ; \ \gamma]} = F_{[\mu\nu  , \ \gamma]} = 0.
\label{max1}
\EN
Here, the square brackets $[\mu\nu,\gamma]$
means adding terms with cyclic permutations of $\mu,\nu,\gamma$.
We can also define the dual electromagnetic field tensor
$^*F^{\mu \nu} =  (\epsilon^{\mu\nu\alpha\beta}/2) F_{\alpha\beta}$, and
write the latter half of \Eq{max1}, as  
$^*F^{\mu\nu}_{\quad ;\nu} = 0$.
Here $\epsilon^{\mu \nu \rho \lambda }=
({\cal A}^{\mu \nu \rho \lambda }/
{\sqrt{-g}})$, 
is the totally antisymmetric Levi-Civita tensor 
and ${\cal A}^{\mu \nu \rho \lambda }$,
the totally antisymmetric symbol such that ${\cal A}^{0123}=1$ and $\pm 1$ for
any even or odd permutations of $(0,1,2,3)$ respectively.
Note that we need to define ${\cal A}_{0123}=-1$.

We would like to cast these equations in terms of
electric and magnetic fields \citep{E73,Ts05,BMT07,S10}.
We closely follow the treatment of \citet{E73} as worked out
in \citet{S10}.
In flat spacetime
the electric and magnetic fields are written in terms of
different components of the EM tensor $F_{\mu\nu}$.
This tensor is antisymmetric, thus its diagonal components
are zero and it has 6 independent components, which can be thought
of the 3 components of the electric field and the 3 components of the
magnetic field. The electric field $E^i$ is given by time-space components
of the EM tensor, while the magnetic field $B^i$
is given by the space-space components
\[
F^{0i}=E^{i}\quad F^{12}=B^{3}\quad F^{23}=B^{1}\quad F^{31}=B^{2}\;.
\]
In a general spacetime, to define corresponding electric and magnetic fields
from the EM tensor, one needs to isolate a time direction. This
can be done by using a family of observers who measure the EM fields
and whose four-velocity is described by the 4-vector
$u^{\mu } = (dx^\mu/ds)$ with $u^\mu u_\mu =-1$.
Given this 4-velocity field, one can also define the 'projection tensor'
$h_{\mu\nu} = g_{\mu\nu} + u_\mu u_\nu$,
which projects all quantities into the 3-space orthogonal to $u^\mu$
and is also the effective spatial metric for these observers, i.e
\[
ds^2 = g_{\mu\nu} dx^\mu dx^\nu = - (u_\mu dx^\mu)^2
+ h_{\mu\nu} dx^\mu dx^\nu
\]
Using the four-velocity of these observers, the EM fields can be expressed
in a more compact form as a four-vector electric
field $E_\mu $ and magnetic field $B_\mu$ as
\begin{equation}
E_\mu =F_{\mu \nu }u^{\nu} , \quad
B_\mu ={\frac 12}\epsilon _{\mu \nu \rho \lambda }u^{\nu}F^{\rho \lambda }
= ^*F_{\mu\nu} u^\nu  \ .
\label{ebnudef0}
\end{equation}
From the definition of $E_\mu$ and $B_\mu$, we have
$E_\mu u^\mu = 0$ and $B_\mu u^\mu =0$.
Thus the four-vectors $B_\mu$ and $E_\mu$ have purely spatial components
and are effectively 3-vectors in the space orthogonal to $u^\mu$.
They generalize the flat space-time notion of electric field as the time-space
component, and the magnetic field as the space-space component 
of the electromagnetic field tensor.
One can also invert \Eq{ebnudef0} to write the EM tensor and its dual in
terms of the electric and magnetic fields
\EQ
F_{\mu\nu} =  u_\mu E_\nu - u_\nu E_\mu
+ \epsilon_{\mu\nu\alpha\beta}B^\alpha u^\beta
\label{feb}
\EN
\EQ
^*F^{\alpha\beta} = \frac{\epsilon^{\alpha\beta\mu\nu}}{2}F_{\mu\nu}=
\epsilon^{\alpha\beta\mu\nu}
u_\mu E_\nu
+ (u^\alpha B^\beta - B^\alpha u^\beta ) \ .
\label{febdual}
\EN
We can now use the time-like vector $u^\mu$ and the spatial
metric $h^\mu_\nu$ to decompose the Maxwell equations
into timelike and spacelike parts.
The details of this procedure is given in for example \citet{S10}.
We merely state the results here.

For this it will also be useful to
define the spatial projection of the covariant derivative
as $D_\beta B^\alpha = h^\mu_\beta h^\alpha_\nu B^\nu_{;\mu}$.
And also split the covariant velocity gradient tensor,
$u_{\alpha;\beta}$, in the following manner:
\EQ
u_{\alpha;\beta} 
= \frac13 \Theta h_{\alpha\beta} +
\sigma_{\alpha\beta} 
+\omega_{\alpha\beta}
- \dot{u}_\alpha u_\beta
\label{sxva}
\EN
Here $\Theta = u^\alpha_{;\alpha}$ is called 
the expansion scalar and $\sigma_{\alpha\beta}$ is
the shear tensor, which is symmetric, traceless 
($\sigma^\alpha_\alpha = 0$) part of the velocity gradient, and is 
purely spatial as $\sigma_{\alpha\beta} u^\beta =0$.
The antisymmetric part of the velocity gradient, 
$\omega_{\alpha\beta}$ is called vorticity and
the `time' derivative of
$u_\beta$, defined by 
$\dot{u}_\beta = u^\alpha u_{\beta:\alpha}$ is the acceleration
of the observer. We also define 
the vorticity vector,
$\omega_\nu = -\omega_{\alpha;\beta}\epsilon^{\alpha\beta\mu\nu}u_\mu/2$
Then the projection of the second part of \Eq{max1} on $u_\alpha$ gives,
\EQ
D_\beta B^\beta = 
h^\mu_\beta h^\beta_\nu B^\nu_{;\mu} =
2 \omega^\beta E_\beta \; .
\label{divbeq}
\EN
This equation generalizes the flat space equation
$\nab\cdot\BBB =0$, to a general curved spacetime. We see
that $2 \omega^\beta E_\beta$ acts as an effective magnetic
charge, driven by the vorticity of the relative motion
of the observers measuring the electromagnetic field.

The spatial projection of the second part of
Maxwell equations in \Eq{max1}, on $h^\kappa_\alpha$ gives the
generalization of Faraday law to curved spacetime,
\EQA
h^\kappa_\alpha\dot{B}^\alpha
= \left[\sigma^\kappa_\beta +\omega^\kappa_\beta - \frac23 \Theta \delta^\kappa_\beta \right]B^\beta
&-& \bar{\epsilon}^{\kappa\mu\nu} \dot{u}_{\mu} E_\nu
\nonumber \\
&-& {\rm Curl}(E^\kappa).
\label{faradayg}
\ENA
Here we have defined $\dot{B}^\alpha = u^\beta B^\alpha_{;\beta}$ and 
a `Curl' operator
${\rm Curl}(E^\kappa) = \bar{\epsilon}^{\kappa\beta\nu} E_{\nu;\beta}$, 
where $\bar{\epsilon}^{\kappa\beta\nu}
= \epsilon^{\kappa\beta\nu\mu}u_\mu$
is a 3-d fully antisymmetric tensor. The 'time' component
of $\bar{\epsilon}^{\kappa\beta\nu}$ got by its projection
on to $u^\alpha$ vanishes.

The other two Maxwell equations, involving source terms,
can be derived from the following symmetry argument.
If we map $\EE \to -\BBB$, and $\BBB \to \EE$, then
the dual EM tensor is mapped to the EM tensor, that is
$^*F^{\mu\nu} \to F^{\mu\nu}$. 
Also in deriving \Eq{divbeq} and \Eq{faradayg}, one needs to 
change the sign of all the terms appearing in \Eq{max1}.
Thus mapping $\EE \to -\BBB$, and $\BBB \to \EE$ in \Eq{divbeq} and \Eq{faradayg}
respectively, and also changing the sign of the source term
$4\pi J^\mu \to -4\pi J^\mu$, the Maxwell equations
$F^{\mu\nu}_{;\nu} = 4\pi J^\mu$,
in terms of the $E^\mu$ and $B^\mu$ fields, become
\EQ
D_\beta E^\beta = 4\pi \rho_q - 2 \omega^\beta B_\beta  \; ,
\label{diveeq}
\EN
\EQA
h^\kappa_\alpha\dot{E}^\alpha
&=& \left[\sigma^\kappa_\beta +\omega^\kappa_\beta - \frac23 \Theta \delta^\kappa_\beta \right]E^\beta
+ \bar{\epsilon}^{\kappa\mu\nu} \dot{u}_{\mu} B_\nu
\nonumber \\
&+& {\rm Curl}(B^\kappa) - 4\pi j^\kappa.
\label{amphereg}
\ENA
Here we have defined the charge and 3-current densities as perceived by the
observer with 4-velocity $u^\alpha$ by projecting the 4-current density
$J^\mu$, along $u^\alpha$ and orthogonal to $u^\alpha$. That is
\[
\rho_q = -J^\mu u_\mu  \; , \quad j^\kappa = J^\mu h^\kappa_\mu.
\]
Note that $j^\kappa u_\kappa = 0$.
To do MHD in the expanding universe,
we also need the relativistic generalization to Ohm's law. This is given by
\EQ
h^\alpha_{(f) \beta} J^\beta = \sigma F^{\alpha\beta} w_\beta, \quad
{\rm or} \quad J^\alpha = \rho_{(f) q} w^\alpha
+ \sigma E_{(f)}^\alpha \; .
\label{relohm}
\EN
Here the symbol $(f)$ stands for a fluid variable, that is
$w^\alpha$ is the mean 4-velocity of the fluid,
$h^\alpha_{(f) \beta}= (\delta^\alpha_\beta + w^\alpha w_{\beta})$
and $E_{(f)}^\alpha = F^{\alpha\beta} w_{\beta}$ is the electric
field as measured in the fluid rest frame. Also
$\rho_{(f) q}$ and $\sigma$ are the fluid charge densities and
conductivity as measured in its rest frame.
Note that the fluid 4-velocity $w^\alpha$,
need not be the 4-velocity $u^\alpha$
of the family of fundamental observers used to define the EM fields
in Maxwell equations; indeed the conducting fluid will in general
have a peculiar velocity in the rest frame of the fundamental observers.

\subsection{Electrodynamics in the expanding universe}

Let us now consider Maxwell equations for the particular case
of the spatially flat FRW spacetime. We choose $u^\alpha$ corresponding
to the fundamental observers of the FRW spacetime, that is
$u^\alpha \equiv (1,0,0,0)$. For such a choice and in
the FRW spacetime, we have
$\dot{u}^\alpha = 0$,
$\omega_{\alpha\beta}=0$, $\sigma_{\alpha\beta} =0$ and
$\Theta = 3 \dot{a}/a$. Further, we can simplify
$h^\kappa_\alpha\dot{B}^\alpha$ as follows:
\EQA
&h^\kappa_\alpha & \dot{B}^\alpha  =
(\delta^\kappa_\alpha + u^\kappa u_\alpha) u^\gamma B^\alpha_{;\gamma}
\nonumber \\
&=&u^\gamma B^\kappa_{;\gamma} + u^\kappa u^\gamma[(u_\alpha B^\alpha)_{;\gamma}
- u_{\alpha;\gamma}B^\alpha]
= u^\gamma B^\kappa_{;\gamma}
\ENA
Thus the Maxwell equations reduce to,
\EQA
B^\beta_{;\beta} = 0, \quad
&&
u^\gamma B^\kappa_{;\gamma}
+ \frac23 \Theta B^\kappa =
-{\rm Curl}(E^\kappa),
\nonumber \\
E^\beta_{;\beta} = 4\pi \rho_q, \quad
&&
u^\gamma E^\kappa_{;\gamma}
+ \frac23 \Theta E^\kappa =
{\rm Curl}(B^\kappa) - 4\pi j^\kappa.
\label{maxfrw}
\ENA
In the spatially flat FRW metric the connection co-efficients
take the form
\EQ
\Gamma^0_{0 0} =0 = \Gamma^0_{0 i} = \Gamma^{i}_{j k},
\quad \Gamma^0_{i j} = \delta_{ij} a \dot{a},
\quad \Gamma^i_{0 j} = \delta_{i j} \frac{\dot{a}}{a}.
\label{connection}
\EN
Using these \Eq{maxfrw} can be further simplified as,
\EQA
\frac{\partial B^i}{\partial x^i} = 0,
&&
\frac{1}{a^3} \frac{\partial}{\partial t} \left[a^3 B^i\right]
= -\frac{1}{a} \epsilon^*_{ilm} \frac{\partial E^m}{\partial x^l},
\nonumber \\
\frac{\partial E^i}{\partial x^i} =
4\pi \rho_q, \ 
&&
\frac{1}{a^3} \frac{\partial}{\partial t} \left[a^3 E^i\right]
= \frac{1}{a} \epsilon^*_{ilm} \frac{\partial B^m}{\partial x^l}
- 4 \pi j^i \;.
\label{maxsimp}
\ENA
Here we have defined the 3-d fully antisymmetric symbol
$\epsilon^*_{ijk}$, where as usual $\epsilon^*_{123} = 1$.

The electric and magnetic field 4-vectors we have used above
are referred to a co-ordinate basis, where the spacetime metric
is of the FRW form. 
They have the following curious property.
Consider for example the case when the plasma in the universe has
no peculiar velocity, that is $w^\alpha = u^\alpha$,
and also highly conducting 
with $\sigma \to \infty$. Then from \Eq{relohm}, we have 
$E_{(f)}^\alpha = 0 = E^\alpha$, and from Faraday's law
in \Eq{maxsimp}, $B^i \propto 1/a^3$.
There is however a simple result derivable in flat space time that
in a highly conducting fluid, the magnetic flux through
a surface which co-moves with the fluid is constant (see below).
Since in the expanding universe all proper surface areas
increase as $a^2(t)$, one would expect the strength
of a 'proper' magnetic field to go down with expansion as $1/a^2$.
This naively seems to be at variance with
the fact that $B^i \propto 1/a^3$ and $B_i = g_{i\mu} B^\mu \propto 1/a$. 
Of course,
if we define the magnetic field
amplitude, say $B$, by looking at the norm of the four vector $B^\mu$,
that is let $B^2 = B^\mu B_\mu = B^iB_i \propto 1/a^4$, then we do get
$B \propto 1/a^2$. This procedure however does not appear
completely satisfactory as one would prefer to deal with the field
components themselves. 

In this context, we note that laboratory measurements
of the EM fields would use a locally inertial co-ordinates around
the observer. Thus it would be interesting to set up a coordinate
system around any event ${\cal P}$, where the metric is is
flat ($\bar{g}_{ \mu\nu} = \eta_{\mu\nu}$) and the connection co-efficients vanish
($\bar{\Gamma}^\mu_{\alpha\beta} = 0$). We have used a `bar' over physical quantities
to indicate they are evaluated in the locally inertial frame.
Such a locally inertial co-ordinate system can be conveniently defined using
a set of orthonormal basis vectors, more generally referred
to as tetrads. 

Any observer can be thought to be carrying along
her/his world line a set of four orthonormal vectors ${\bf e}_{(a)}$
($a=0,1,2,3$), which satisfy the relation
\EQ
g_{\mu\nu} e^\mu_{(a)} e^\nu_{(b)} = \eta_{ab},
\quad \eta^{a b} e^\mu_{(a)} e^\nu_{(b)} = g^{\mu \nu}
\label{tetradprop}
\EN
Here $\eta_{a b}$ has the form of the flat space metric.
The observer's 4-velocity itself is the tetrad with $(a)=0$, i.e
$e^\mu_{(0)} = u^\mu$. The other three tetrads are orthogonal to the
observer's 4-velocity. In the present case, we consider the observer to
be the fundamental observer of the FRW space time, and the
components of the tetrads, which satisfy \Eq{tetradprop} are given by
\[
e^\mu_{(0)} = \delta^\mu_0, \quad e^\mu_{(i)} = \frac{1}{a} \delta^\mu_i, \quad i = 1,2,3
\]
Note that the fundamental observers move along the geodesics, and
as we noted earlier, do not have either relative
acceleration or rotation.
Such observers parallel transport their tetrad along their
trajectory, i.e $u^\mu e^\alpha_{(a);\mu} = 0$, as can be easily
checked dy direct calculation using the connection co-eficients
given in \Eq{connection}.

Given the set of tetrads one can set up a local
co-ordinate system around any event ${\cal P}$ by using geodesics
emanating from ${\cal P}$ and whose tangent vectors at ${\cal P}$ are the
four tetrads ${\bf e}_{(a)}$. This co-ordinate frame, is a locally inertial frame; that
is the spacetime is locally flat with the metric in the form of
$\eta_{a b}$ and the connection co-efficients in these co-ordinates
vanishing, all along the geodesic world line \citep[see section 13.6][for a proof]{MTW}.
In this co-ordinate system
(called Fermi-Normal co-ordinates), the metric differs from
flat space-time metric only to the second order (due to finite
space-time curvature). The metric $\eta_{a b}$ can also be used to
raise and lower the index of the tetrad to define
$e^{\mu (a)} = \eta^{a b} e^\mu_{(b)}$.
This co-ordinate system is the natural co-ordinate
system where one measures the EM fields in the Laboratory.
For example the physical magnetic field components can be
represented as its projection along the four tetrads using,
\EQA
\bar{B}^{a} &=& g_{\mu\nu} B^\mu e^{\nu (a)}=B^{\mu}e_{\mu}^{(a)}, 
\nonumber \\ 
\bar{B}^0 &=& 0, \quad \bar{B}^a = a(t) B^a,  \
{\rm for} \ a=1,2,3.
\label{BPhys}
\ENA
Note that this is still a vector as far as Lorentz transformation is concerned
(which preserves the orthonormality conditions in \Eq{tetradprop}). If we
define $\bar{B}_{a} = \eta_{ab} \bar{B}^b$, then numerically $\bar{B}^{i} = \bar{B}_i$ and
$\bar{B}^{0} = - \bar{B}_0 = 0$.
A similar relation $\bar{E}^b = a(t) E^b$  
is obtained for 
the electric field components.
In the FRW universe, as $B^i \propto 1/a^3$, we see that
$\bar{B}^{i} = \bar{B}_i \propto 1/a^2$, as one naively expects
from flux freezing of the magnetic field. Thus the magnetic field
components projected onto the orthonormal tetrads seem to be
the natural quantities to be used as the 'physical' components of the
magnetic field. 
Note that this is similar to using the Cartesian components of
a vector as the physical components in 3 dimensional vector analysis.

Let us now define the vectors $\BBB \equiv (\bar{B}^1, \bar{B}^2, \bar{B}^3)$
and $\EE \equiv (\bar{E}^1, \bar{E}^2, \bar{E}^3)$
and $\JJ = (\bar{j}^1, \bar{j}^2, \bar{j}^3)$ . 
Let us also define a new set of variables,
\EQ
\BBB^* = a^2\BBB, \quad \EE^* = a^2 \EE, \quad \rho_q^* = a^3\rho_q, \quad
\JJ^* = a^3 \JJ,
\EN
and transform to conformal time $d\tau = dt/a$ and continue to use 
co-moving space co-ordinates $x^i$.
Then the Maxwell equations \Eq{maxsimp} in terms of the starred variables become,
\EQA
{\bf \nabla }\cdot{\bf B}^{*}&=&0, \quad
{\bf \nabla }\times {\bf E}^{*}=-{\frac{\partial {\bf B}^{*}}{\partial \tau}},
\nonumber \\ 
{\bf \nabla}\cdot{\bf E}^{*}&=&4\pi \rho_q^{*}, \quad 
{\bf \nabla} \times {\bf B}^{*}=4\pi {\bf J}^{*}+{\frac{\partial {\bf E}^{*}%
}{\partial \tau}}, 
\label{maxstar}
\ENA
and Ohm's law becomes,
\EQ
\JJ^* = \rho_q^* \vv + \sigma^* \left(\EE^{*}+\vv^*\times \BBB^{*} \right)
\label{ohmstar}
\EN
where we define $\sigma^* = a \sigma$.
These are exactly the Maxwell equations and Ohm's law in flat spacetime.
This result also follows quite generally
from the conformal invariance of electrodynamics.

\subsubsection{The induction equation}

One can derive an evolution equation for the magnetic field,
by using Ohm's law in the Maxwell equations. 
Introducing the magnetic diffusivity
$\eta^*=(4\pi\sigma^*)^{-1}$ in cgs units (and with $c=1$), we get
\EQ
\eta\frac{\partial\EE^*}{\partial t} + \eta \vv
(\nab\cdot\EE^*) + \EE^*
=\eta^*\nab \times \BBB^* -\vv\times\BBB^*.
\label{FaradayOhm}
\EN
Here we have also used \Eq{maxstar} to eliminate 
$\rho_q^*$ in terms of the electric field.
We can generally neglect the time derivative (arising from the displacement 
current) and the space derivative of $\EE^*$, 
as the Faraday time $\tau_F = \eta^* = (4 \pi \sigma^*)^{-1}$ 
is much smaller than other
relevant time scales \citep{BS05a}. Then taking curl of \Eq{FaradayOhm}, 
the magnetic field evolution is
governed by the induction equation,
\EQ
\frac{\partial\BBB^*}{\partial \tau} 
= \nab \times \left [ \vv \times \BBB^* - \eta^* \nab \times \BBB^* \right].
\label{expind}
\EN
Thus we see that in the absence of resistivity ($\eta =0$)
and peculiar velocities ($\vv =0$), $\BBB^*$ is constant, or 
the magnetic field defined in the
local inertial frame, decays with expansion factor
as $\BBB \propto 1/a^2$. This decays is as expected, when the magnetic
flux is frozen to the plasma (see below), since all proper areas 
in the FRW spacetime increase with expansion as $a^2$.

\subsubsection{Magnetic flux freezing}

The $\vv\times\BBB^*$ term in \Eq{expind} is usually referred to as
the induction term.
This term more generally implies
that the magnetic flux through a surface moving with the
fluid remains constant in the high-conductivity limit.
To prove this consider a comoving surface $S$, bounded by a curve $C$,
moving with the fluid.
Note that the surface $S$ need not lie in a plane.
Suppose we define the magnetic flux through this surface,
$\Phi = \int_S \BBB^*\cdot\dd\SSS$. 
After a time $\dd \tau$, let the surface move to a new surface 
$S'$. Then the change in flux is given by
\EQ
\Delta\Phi = \int_{S'} \BBB^*(\tau+\dd \tau)\cdot\dd\SSS -
\int_S \BBB^*(\tau)\cdot\dd\SSS.
\label{Delphi}
\EN
Applying $\int\nab\cdot\BBB^*\,\dd V=0$ at time $\tau+\dd \tau$, to
the `tube'-like volume swept up by the
moving surface $S$, 
we also have
the flux $\int_{S'} \BBB^*(\tau+\dd \tau)\cdot\dd\SSS$ leaving $S'$, 
is that entering $S$, $\int_S \BBB^*(\tau+\dd \tau)\cdot\dd\SSS$, minus 
that leaving the sides of the tube 
($\oint_C \BBB^*(\tau+\dd \tau)\cdot (\dd\lv \times\vv \dd \tau)$). 
Here $C$ is the curve bounding the surface $S$,
and $\dd\lv$ is the line element along $C$.
(In the last term, to linear order in $\dd \tau$, it does not matter
whether we take the integral over the curve $C$ or $C'$ the bounding
curve of $S'$.)
Using the above condition in \Eq{Delphi}, we obtain
\EQA
\Delta\Phi &=& \int_S [\BBB^*(\tau+\dd \tau) - \BBB^*(\tau)]\cdot\dd\SSS 
\nonumber \\
&-& \oint_C \BBB^*(\tau+\dd \tau)\cdot (\dd\lv \times\vv) \dd \tau.
\ENA
Taking the limit of $\dd \tau\to 0$, and noting that
$\BBB^*\cdot(\dd\lv\times\vv)=(\vv\times\BBB^*)\cdot\dd\lv$, we have
\EQA
\frac{\dd\Phi}{\dd \tau} &=& \int_S \deriv{\BBB^*}{\tau}\cdot\dd\SSS
- \oint_C (\vv\times\BBB^*)\cdot\dd\lv \nonumber \\
&=& -\int_S (\nabla \times (\eta^* \nab\times\BBB^*))\cdot\dd\SSS.
\ENA
In the second equality we have used
$\oint_C(\vv\times\BBB^*)\cdot\dd\lv
=\int_S\nab\times(\vv\times\BBB^*)\cdot\dd\SSS$
together with the induction equation (\ref{expind}).
One can see that,
when $\eta \to 0$, $\dd\Phi/\dd \tau \to 0$ and so $\Phi$ is constant.
Thus $\int_S \BBB\cdot\dd\SSS \propto 1/a^2$ even in the presence
of the peculiar velocity $\vv$, when $\eta^* \to 0$..

Now suppose we consider a small segment of a thin flux
tube of comoving length $l$ and cross-section $A$, in a highly
conducting fluid. Then, as the fluid moves about, conservation
of flux implies $B^*A$ is constant. Thus a decrease in $A$ leads
to an increase in $B^*$. 
Any 'incompressible' shearing motion which increases
$l$ will also amplify $B^*$; an increase in $l$ leading to
a decrease in $A$ (because of incompressibility) and hence
an increase in $B^*$ (due to flux freezing). This effect,
plays a crucial role in all scenarios involving
turbulent dynamo amplification of magnetic fields, 
from seed fields.

\subsubsection{Magnetic diffusion and the Reynolds number}

Let us now consider the opposite limit when $\vv =0$. 
Then for a constant $\eta^*$ the induction equation \Eq{expind},
reduces to the diffusion equation
\EQ
\frac{\partial\BBB^*}{\partial \tau}
=\eta^* \nab^2\BBB^*\;.
\label{Diffusion}
\EN
The field $B^*$ decays on the (comoving) 
diffusion timescale $\tau_d \sim L^2/\eta^*$, where
$L$ is the co-moving scale over which the magnetic field varies.

The importance of the magnetic induction
relative to magnetic diffusion, in the induction equation
is characterized by the magnetic Reynolds number, which
is defined as
\EQ
\Rm=\frac{v L}{\eta^*} = \frac{v l}{\eta}\;,
\label{Rmdefinition}
\EN
where $v$ gives the typical fluid velocity on the comoving 
scale $L$ (or the proper scale $l=aL$).
In some applications, it may be more
convenient to define the magnetic Reynolds number based
on the wavenumber, $k= 2\pi/L$, using $\Rm = v/(\eta^* k)$,
which is smaller by a factor $2 \pi$ compared to the definition
given in \Eq{Rmdefinition}.

\subsubsection{Magnetic helicity}
\label{helicity_basic}

The equations of magnetohydrodynamics imply a very useful
conservation law, that of magnetic helicity, which constrains
the dynamics of cosmic magnetic fields.
We define magnetic helicity by the volume integral
\EQ
H=\int_V\AAA\cdot\BBB\,\dd^3\rr = \int_{V^*}\AAA^*\cdot\BBB^*\,\dd^3\xx
\label{Hgaugedep}
\EN
over a closed or periodic volume proper $V$ (or comoving volume $V^*$.
Here $\BBB^* = {\bf \nabla}_{\bf x} \times \AAA^*$ and 
$\BBB = {\bf \nabla}_{\bf r} \times \AAA$, with ${\bf r} = a(t) {\bf x}$.
Since $\BBB^* = \BBB a^2$, we also have $\AAA^* = a(t)\AAA$ 
and so interestingly, the helicity is the same whether defined
in terms of the comoving or proper fields.
Also by a closed volume we mean one in which the magnetic field lines
are fully contained, so the field has no component normal to the
boundary, \ie$\BBB\cdot\nnn=0$. The volume $V$ could also
be an unbounded volume with the fields falling off sufficiently
rapidly at spatial infinity.
In these particular cases, $H$ is invariant under the gauge
transformation $\AAA'=\AAA + {\bf \nabla}\Lambda$.

Magnetic helicity has a simple topological interpretation
in terms of the linkage and twist of isolated (non-overlapping) flux tubes.
For example consider the magnetic helicity for an interlocked, but untwisted,
pair of thin flux tubes,
with $\Phi_1$ and $\Phi_2$ being the fluxes in the tubes around $C_1$ and
$C_2$ respectively (with the field $\BBB$ in the tubes going around 
in an anti-clockwise direction;
For example see Fig 3.2 in \citet{BS05a}
). For this configuration
of flux tubes, $\BBB\,\dd^3r$ can be replaced by
$\Phi_1 \dd \lv$ on $C_1$ and $\Phi_2 \dd \lv$ on $C_2$.
The net helicity is then given by the sum
\EQ
H = \Phi_1 \oint_{C_1} \AAA\cdot\dd\lv + \Phi_2 \oint_{C_2} \AAA\cdot\dd\lv,
= 2\Phi_1\Phi_2.
\EN
For a general pair of non-overlapping thin flux tubes, the
helicity is given by $H=\pm2\Phi_1\Phi_2$;
the sign of $H$ depending on the relative orientation of the two tubes
\citep{Mof78}.

The evolution equation for $H$ can be derived from Faraday's law 
in \Eq{maxstar} and
its uncurled version, $\partial\AAA^*/\partial \tau = -\EE^* -\nab\phi^*$, 
where $\phi^*$ is a scalar potential.
We have
\EQA
{\partial \over \partial \tau}(\AAA^*\cdot\BBB^*)
&=& (-\EE^* -\nab\phi^*)\cdot\BBB^* + \AAA^*\cdot(-\nab \times \EE^*)
\nonumber \\
= -2\EE^*\cdot\BBB^* &-& \nab\cdot(\phi^*\BBB^* + \EE^* \times \AAA^* ).
\ENA
Integrating this over the volume $V^*$, and using 
Ohm's law, $\EE^* = - (\vv \times \BBB^*) + \JJ^*/\sigma^*$,
in the volume integral,
the magnetic helicity satisfies the evolution equation
\EQA
{\dd H\over\dd \tau}&=&
-2\eta^* \int_{V^*} 4\pi \JJ^*\cdot\BBB^* \ \dd^3\xx
\nonumber \\
&&- \oint_{\partial V^*}(\phi^*\BBB^* + \EE^* \times \AAA^*)
\cdot\nnn\dd S \nonumber \\
&=&  
-2\eta^* \int_{V^*} \BBB^*\cdot(\nab \times \BBB^*) \ \dd^3\xx,
\label{magn_hel_evol}
\ENA
where the last equality holds for closed
domains, when the surface integral vanishes.

In the non-resistive case, $\eta^*=0$, and assuming a closed domain,
the magnetic helicity is conserved, \ie \ $\dd H/\dd \tau=0$ and so
also $\dd H/\dd t=0$.
However, this does not guarantee
conservation of $H$ in the limit $\eta^*\rightarrow0$,
because the current helicity, $\int\JJ^*\cdot\BBB^*\,\dd V^*$, 
may in principle still
become large. For example, the Ohmic dissipation rate of magnetic energy
$Q_{\rm Joule}\equiv (4\pi\int \eta^* {\bf J}^{*2} \dd V^*)$
can be finite and balance magnetic energy input by motions,
even when $\eta^* \rightarrow0$. This is
because small enough scales develop in the field (current sheets)
where the current density increases with decreasing $\eta^*$ as
$\propto\eta^{*-1/2}$ as $\eta^*\rightarrow0$,
whilst the rms magnetic field strength, $B_{\rm rms}$, remains
essentially independent of $\eta^*$. Even in this case, however,
the rate of magnetic helicity dissipation {\it decreases} with $\eta^*$,
with an upper bound to the dissipation rate
$\propto\eta^{*+1/2}\rightarrow0$, as $\eta^*\rightarrow0$.
Thus, under many astrophysical conditions where $\Rm$ is
large ($\eta^*$ small), the magnetic helicity $H$, is almost independent
of time, even when the magnetic energy is dissipated at finite
rates.

We note the very important fact that the fluid velocity
completely drops out from the volume generation term of the 
helicity evolution equation
\Eq{magn_hel_evol}, since $(\vv \times \BBB^*)\cdot\BBB^* = 0$.
Therefore, any change in the nature of the fluid velocity,
for example due to turbulence
(turbulent diffusion), the Hall effect,
or ambipolar drift (see below), does not affect the
volume generation/dissipation of magnetic helicity.

We should point out that it is also possible to define
magnetic helicity as linkages of flux analogous to the
Gauss linking formula for linkages of curves
\citep{berger_field84,moffatt69}.
This approach can be used to formulate the concept of a gauge invariant
magnetic helicity {\it density} in the case of random fields, 
as the density of correlated links \citep{SB06}. Such a concept
would be especially useful in the context of early universe magnetogenesis,
where the field is generally random and has a finite correlation length.

\subsubsection{Resistivity in the early universe}
\label{resist}

A simple physical picture for the conductivity
in a plasma is as follows: The force due to an electric field 
$\EE$ accelerates negative charges (electrons in the current universe) 
relative to the positive charges (ions at present); 
but they cannot move freely
due to friction caused by collisions between these drifting components.
A `terminal' relative drift velocity ${\bf u}$ would result
obtained by balancing the Lorentz force with friction. This
velocity can be estimated by assuming that after a collision
time $\tau_c$ the drift velocity is randomized.
During the radiation dominated phase, assume that the
currents are carried by charged particles with charge $e$,  
inertia of order $T$ (the Boltzmann
constant $k_B$ is set to 1) and number density $n$. Then 
during the time $\tau_c$ they would acquire from the action of
an electric field $\EE$ an drift velocity 
${\bf u} \sim \tau_c e \EE/T$, which will correspond to 
a current density $\JJ \sim e n {\bf u} \sim
(n e^2 \tau_c/T) \EE$. This leads to an estimate
$\sigma \sim n e^2 \tau_c/T$.

The collision time scale $\tau_c$ can itself be estimated as follows.
For a strong collision one
needs an impact parameter $b$ which satisfies the condition
$e^2/b > T$ (potential energy greater than kinetic). 
This gives a cross section
for strong scattering of $\sigma_t \sim \pi b^2 $. If the
scattering is due to a long range force, the larger number
of random weak scatterings add up to give an extra `Coulomb
logarithm' correction to make
$\sigma_t \sim \pi (e^2/T)^2 \ln\Lambda$,
where $\ln\Lambda$ is to be determined.
The corresponding mean free time between collisions is
$\tau_c \sim 1/(n \sigma_t)$
giving an estimate of the conductivity
\EQ
\sigma \sim \frac{n e^2 \tau_c}{T} = 
\frac{n e^2}{T} \frac{T^2}{n e^4 \ln\Lambda} \sim \frac{T}{\alpha \ln\Lambda},
\label{cond}
\EN
where $\alpha = e^2$ is a dimensionless `fine structure constant' 
and most importantly the dependence on the particle 
density has canceled out. A more careful calculation by \citet{BH97} 
bears out the above qualitative estimate and gives $\Lambda \sim 1/\alpha$
at temperatures well below the electroweak scale of $T \sim 100$ GeV.
Above this temperature, \citet{BH97} argue that $W^{\pm}$ charge-exchange
processes effectively stop left handed charged leptons, but right handed
ones can still carry the current, and $\sigma$ is reduced by a 
$\cos^4\theta_W$ factor, where $\theta_W$ is the weinberg angle.

We can gauge the importance of resistive decay by
estimating $\Rm = vl/\eta = 4\pi vl \sigma$. 
A characteristic velocity will be the Alfv\'en velocity $V_A$ and consider
a scale $l = V_A t$ which we will see to be
the coherence scale for causally generated
field. Then $\Rm = 4\pi V_A^2 t\sigma$. From Einstein equation
$t= H^{-1} \approx m_{pl}/T^2$ during the radiation dominated era,
where $m_{pl}= 1/\sqrt{G}$ is the Planck mass. Then we have
$ \Rm \sim (10^{-6}B_{-9}^2/\alpha\ln\Lambda) (m_{pl}/T)$,
and for the electroweak era ($T\sim 100$ GeV) 
or the QCD era ($T \sim 150$ MeV), we see that $\Rm \gg1$
for most field strengths. Of course,
after inflation and reheating,
$T$ will be larger, but the relevant scales $l$ will now be 
super Hubble and will not have associated motion. Then the
relevant quantity to estimate is the resistive decay time
$l^2/\eta$ compared to the Hubble time $t$. This ratio
is given by $(lH)^2 (m_{pl}/T) (1/\alpha\ln\Lambda) \gg 1$, 
and so long wavelength magnetic fields will decay negligibly.
All in all, the early universe is an excellent conductor.
(see also the appendix in \citet{TW88}).

\subsection{The fluid equations}

In the early universe during the radiation dominated era,
the fluid equations including the effect of the shear
viscosity can also be transformed to a simple form that
it takes in flat space time. The detailed derivation is given in
\citet{SB98a}, 
using the conformal invariance
of the relativistic fluid, electromagnetic and shear viscous
energy momentum tensors.
Transforming the fluid pressure ($p$), energy density ($\rho$) and
the dynamic shear viscosity ($\mu$) 
to a set of new variables,
\EQ
p^{*} = a^4 p,\quad \rho ^{*} = a^4 \rho, \quad \mu^* = a^3 \mu . 
\label{tranfl}
\EN
and using the conservation of the total energy momentum tensor, 
one gets in the non-relativistic limit,
\begin{equation}
{\bf v}]-{\bf E}^{*}\cdot{\bf J}^{*}-\mu ^{*}{\bf \nabla }\cdot{\bf f}=0.
\label{energ}
\end{equation}
\begin{eqnarray}
{\frac \partial {\partial \tau }}&&\left[ (\rho ^{*} + p^{*}){\bf v}\right] 
+ ({\bf v}.{\bf \nabla })\left[ (\rho ^{*}
+p^{*}){\bf v}\right] 
\nonumber \\
&&+{\bf v}{\bf %
\nabla }.\left[ (\rho ^{*}+p^{*}){\bf v}\right] 
=-{\bf \nabla }p^{*}+{\bf J}^{*}\times {\bf B}^{*} \nonumber \\
&& +(\rho^*+p^*)\nu ^{*}\left[
\nabla ^2{\bf v}+{\frac 13}{\bf \nabla }({\bf \nabla }.{\bf v})\right] .
\label{eulers}
\end{eqnarray}
where ${\bf f}={\bf \nabla }(\vv^2/2)-(2/3)\vv{\bf \nabla }\cdot\vv$. 
We have also defined the kinematic viscosity, $\nu = \mu/(\rho +p)$, such that
$\nu^* = \nu/a$. In the radiation dominated era this is 
given by (see \cite{weinb}, section 2.11 and 15.8);
\EQ
\nu = \frac{(4/15)\rho_d l_d}{\rho+p} = \frac{l_d}{5} \frac{g_d}{g_f}.
\label{earlynu}
\EN
where $\rho_d$ and $l_d$ are the energy density,
mean-free-path of the diffusing particle. In the latter
equality, $g_d$ and $g_f$ are respectively the statistical weights contributed
to the energy density by the diffusing particle and the total fluid.
After neutrino decoupling, when photons dominate the energy density,
coupled to the field and the baryons, and we have $g_d/g_f =1$.
In the radiation-dominated epoch, the bulk 
viscosity is zero and we have
neglected the thermal conductivity term since it does not affect the
nearly incompressible fluid motions 
that we will mostly focus upon.

We considered in \Eq{expind} how the peculiar 
velocity field influences the evolution of
the magnetic field.
The magnetic field in turn influences the fluid velocity
through the Lorentz force, as 
given by the term $({\bf J}^{*}\times {\bf B}^{*})$ 
in \Eq{eulers}.
In general there would also be an electric part to the Lorentz force.
But for highly conducting fluid moving with non-relativistic velocities,
this part turns out to be negligible compared to the magnetic part.
These equations will be used below to follow the evolution of primordial
magnetic fields.
We had defined the magnetic Reynolds number above. In a similar vein,
the relative importance of the nonlinear term
in the velocity to the viscous dissipation is also given by a
dimensionless number, called fluid Reynolds number 
\EQ
\Rey = \frac{vL}{\nu^*} = \frac{vl}{\nu}.
\label{Re}
\EN
The ratio $\Prm=\nu^*/\eta^*= \nu/\eta = \Rm/\Rey$ 
is called the magnetic Prandtl number.

The frictional drag in \Eq{eulers} assumes that the mean free path $l_d \ll l$,
the proper wavelength of a perturbation. In general $l_d$ is a rapidly
increasing function of time as the universe expands and 
indeed increase faster than $l$.  If it increases such that
$l_d > l$, then the diffusion approximation for describing
viscosity will break down and in principle one has to
use a full Boltzmann treatment for the drag force. 
A simpler approximation, that is adequate for most purposes,
is to assume that the particle responsible for the drag can free
stream on the scale $l$, and its occasional scattering
induces a viscous force $\ff_v = - \kappa \vv$.
In this case one can also re define a fluid Reynolds number as
\EQ
\Rey = \frac{(v^2/l)}{\kappa v} = \frac{v}{\kappa l}.
\label{freerey}
\EN
A nice discussion of viscosity at various epochs in the universe 
is given in the appendix of \citet{DN13}. 

\section{Generation of primordial magnetic fields}
\label{earlygen}

\subsection{Generation during Inflation}
\label{geninf}

As mentioned earlier, inflation provides an ideal setting 
for the generation of primordial field with large coherence scales \citep{TW88}.
First the rapid expansion in the inflationary era provides the kinematical
means to produce fields correlated on very large scales by just the
exponential stretching of wave modes. Also vacuum fluctuations
of the electromagnetic (or more correctly the hypermagnetic) field
can be excited while a mode is within
the Hubble radius and these can be transformed
to classical fluctuations as it transits outside the Hubble radius.
Finally, during inflation any existing charged particle densities
are diluted drastically by the expansion, so that the universe is
not a good conductor; thus magnetic flux conservation then does not
exclude field generation from a zero field.
There is however one major difficulty, which arises because
the standard electromagnetic
action is conformally invariant, and the universe metric is conformally flat.

Consider again the electromagnetic action
\EQ
S 
= - \int \sqrt{-g} \ d^4x \ \frac{1}{16\pi}
g^{\mu \alpha}g^{\nu \beta} F_{\mu \nu}F_{\alpha \beta}
\EN
Suppose we make a conformal transformation of the metric given by
$g^*_{\mu \nu} = \Omega^2 g_{\mu \nu}$.
This implies $\sqrt{-g^*} =\Omega^4\sqrt{-g}$ and
$g^{* \mu \alpha}= \Omega^{-2} g^{\mu \alpha}$. Then
taking $A_\mu^* = A_\mu \Rightarrow S^* = S$.
Thus the action of the free electromagnetic field is
invariant under conformal transformations.
Note that the FRW models are all conformally flat;
that is the FRW metric can be written as
$g^{FRW}_{\mu \nu} = \Omega^2 \eta_{\mu \nu}$, where
$\eta_{\mu \nu}$ is the Minkowski, flat space metric.
As we will see below this implies that one can transform
the electromagnetic wave equation and Maxwell equations in general
into their flat space versions.
It turns out that one cannot then amplify electromagnetic
wave fluctuations in a FRW universe and
the field then always decreases
with expansion as $1/a^2(t)$.
\footnote{
This decay of the magnetic field can made to slow down in the
case of open models for the universe for super curvature modes \citep{BTY12}. 
In these $k=-1$ models the effect is purely due to geometric reasons. 
It is not clear however if inflation can lead to the generation of 
the required super curvature modes \citep{AdD12,YFM14} or the conformal
time interval available during inflation or later is
sufficient to make the decay much slower, if curvature is small 
(see \citet{ShSa13} versus \citet{T14}).
}

Therefore mechanisms for magnetic field generation need to invoke
the breaking of conformal invariance of the electromagnetic action,
which changes the above behaviour to $B \sim 1/a^{\epsilon}$ with typically
$\epsilon \ll 1$ for getting a strong field.
A multitude of ways have been considered for breaking conformal invariance  of the
EM action during inflation. Some of them are illustrated in the
action below (see \citet{TW88,Ratra92,Dol93,GGV95,Gio00,Gio07,MY08,S10,KKT11,
DN13,APSS14,FNTTT15} 
and references, for some early and some recent range of models).
\EQA
S = &\int& \sqrt{-g} \ d^4x b(t)[-\frac{f^2(\phi,R)}{16\pi} F_{\mu \nu}F^{\mu \nu}
-g_1 R A^2 \nonumber \\
&+& g_2 \theta F_{\mu \nu} \tilde{F}^{\mu \nu}
- D_{\mu}\psi (D^\mu \psi)^* \ ]
\label{cibreak}
\ENA
They include coupling of EM action to scalar fields ($\phi$) like the
inflaton and the dilaton, coupling to the evolution of an extra-dimensional
scale factor ($b(t)$), to curvature invariants ($R$),
coupling to a pseudo-scalar field like the axion ($\theta$),
having charged scalar fields ($\psi$) and so on.
(Note that models involving a non zero $RA^2$ term 
are strongly disfavored as they
imply 'ghosts' in the theory \citep{HCP09}).
If conformal invariance of the EM action
can indeed be broken, the EM wave can amplified from vacuum fluctuations,
as its wavelength increases from sub-Hubble to super-Hubble scales.
After inflation ends, the universe reheats and leads to the production
of charged particles leading to a dramatic increase in the plasma
conductivity. Then the electric field $\EE$ would get
shorted out while the magnetic field $\BBB$ of the EM wave gets frozen in.
This is the qualitative picture for the generation
of primordial fields during the inflationary era.

There is however another potential difficulty;
since $a(t)$ is almost exponentially
increasing during slow roll inflation, the predicted field amplitude,
which behaves say as $B \sim 1/a^{\epsilon}$ is
exponentially sensitive to any changes of the parameters of the model
which affects $\epsilon$. Therefore models of magnetic
field generation can lead to fields as large as $B\sim 10^{-9}$ G
(as redshifted to the present epoch) down to
fields which much smaller than that required for even seeding
the galactic dynamo. For example in  model considered by
\citet{Ratra92} with $ f^2(\phi) \sim e^{\alpha\phi}$,
with $\phi$ being he inflation, one gets $B\sim 10^{-9}$ to $10^{-65}$ G,
for $\alpha \sim 20-0$.
Note that the amplitude of scalar perturbations generated
during inflation is also dependent on the parameters of the theory and
has to be fixed by hand. But the sensitivity to parameters seems
to be stronger for magnetic fields than for scalar perturbations due
to the above reason. Nevertheless one may hope that there would
arise a theory where the parameters are naturally such as to produce
interesting primordial magnetic field strengths. 

Indeed, since the seminal papers of \citet{TW88,Ratra92}, 
there has been an extensive exploration of different models
of inflationary magnetogenesis. However, 
there is as yet no compelling model, where the above issue is resolved,
and which solves a number of other 
problems, like the back reaction and strong coupling problems 
discussed below.
In this situation, we discuss below one of the simpler frameworks for 
inflationary magnetogenesis, discussed extensively in the literature, 
where the above issues can also be brought out. 
This scenario also encompasses the \citet{Ratra92} model, 
one example of the 
\citet{TW88} models, and several models which can arise from particle
physics theories \citep{MY08}. 
Our discussion closely follows \citet{MY08,S10}, where the detailed
derivations can be found.

Let us assume that the scalar field $\phi$ in \Eq{cibreak}
is the field responsible for inflation and also assume that this
is the sole term which breaks the conformal invariance of the
electromagnetic action. 
We assume that the electromagnetic field is a `test' field
which does not perturb either the scalar field evolution or
the evolution of the background FRW universe. We take the metric
to be spatially flat with $k=0$.
It is convenient to adopt the Coulomb gauge by adopting
$A_0(t,{\bf x}) = 0$ and ${\partial}_jA^j(t,{\bf x}) =0$.
In this case the time component of
Maxwell equations 
becomes a trivial identity, while
the space components give
\EQ
A_i{''}
+ 2\frac{f'}{f}A_i' - a^2\partial_j\partial^j A_i = 0
\label{Aevol}
\EN
where we have defined
$\partial^j = g^{jk}\partial_k = a^{-2}\delta^{jk}\partial_k$,
and a prime denotes derivative with respect to $\tau$.

We can also use \Eq{ebnudef0} to write the electric and magnetic
fields in terms of the vector potential. Note that the
four velocity of the fundamental observers used to define
these fields is given by $u^\mu \equiv (1/a, 0,0,0)$.
The time components of $E_\mu$ and $B_\mu$ are zero, while
the spatial components are given by
\EQ
E_i = -\frac{1}{a} A_i', \ 
B_i = \bar{\epsilon}_{ijk}\delta^{jl}\delta^{km}(\partial_l A_m)
\label{EandB}
\EN

\subsubsection{Quantizing the EM field}

We would like quantize the electromagnetic field in the FRW background.
For this we treat $A_i$ as the co-ordinate, and find the conjugate
momentum in the standard manner by varying the EM Lagrangian density ${\cal L}$
with respect to $A_i'$. We get
\[
\Pi^i = \frac{\delta{\cal L}}{\delta A_i'}
=\frac{1}{4\pi} f^2 a^2 g^{ij} A_j', \quad \Pi_i
= \frac{1}{4\pi} f^2 a^2 A_i'
\]
To quantize the electromagnetic field, we promote $A^i$ and $\Pi_i$
to operators and impose the canonical commutation relation between them,
\EQA
\left[A^i(\xx,\tau),\Pi_j(\yy,\tau) \right]
&=& i \int \frac{d^3k}{(2\pi)^3}  \  e^{i\kk\cdot(\xx-\yy)} P^i_j(\kk)
\nonumber \\
&=& i \delta_{\perp \ j}^i(\xx-\yy).
\label{quant1}
\ENA
Here the term $P^i_j(\kk) = (\delta^i_j -\delta_{jm} (k^i k^m/k^2))$
is introduced to ensure that the Coulomb gauge condition is
satisfied and $\delta_{\perp \ j}^i$ is the transverse delta function.
This quantization condition is most simply incorporated
in Fourier space. We expand the vector potential in
terms of creation and annihilation operators,
$b_\lambda^{\dagger}(\kk)$ and $b_\lambda(\kk)$, with $\kk$ the
co-moving wave vector,
\EQA
&&A^i(\xx,\tau) = \sqrt{4\pi}
\int \frac{ d^3 k}{(2\pi)^3}\sum _{\lambda =1}^2
{\cal \ee}^i_{\lambda }(\kk) \times
\nonumber \\
&&\left[
b_\lambda(\kk) A(k,\tau)e^{i \kk\cdot\xx }
+ b_\lambda^{\dagger}(\kk) {A}^*(k,\tau)e^{-i\kk\cdot\xx} \right] \;.
\label{creation_exp}
\ENA
Here the index $\lambda=1,2$ and ${\cal \ee}^i_{\lambda }(\kk)$
are the polarization vectors, which form part of
an orthonormal set of basis four-vectors,
\EQ
{\cal\ee}^\mu_0 = \left( \frac{1}{a}, {\bf 0} \right), \quad
{\cal\ee}^\mu_\lambda = \left( 0, \frac{\bar{\cal\ee}^i_{\lambda}}{a} \right), \quad
{\cal\ee}^\mu_{3} = \left( 0, \frac{\uvec{k}}{a}\right) \; .
\label{orthodef}
\EN
The 3-vectors $\bar{\cal\ee}^i_{\lambda}$ are unit vectors,
orthogonal to $\kk$ and to each other. The
expansion in terms of the polarization vectors incorporates
the Coulomb gauge condition in Fourier space. It also shows that
the free electromagnetic field has two polarization degrees of freedom.
Substitution of \Eq{creation_exp} into \Eq{Aevol}, shows that the Fourier
coefficients $\bar{A} = (aA(k,\tau))$ satisfies,
\EQ
\bar{A}{''} + 2\frac{f'}{f}\bar{A}' + k^2 \bar{A} = 0. 
\label{barAevol}
\EN
One can also 
define a new variable
${\cal A} = a(\tau )f(\tau )A(\tau ,k) $ to eliminate
the first derivative term.
 to get
\EQ
{\cal A}''(\tau ,k)+\left(k^2-\frac{f''}{f}\right){\cal A}(\tau ,k)=0
\label{calAeq}
\EN
We can see that the mode function ${\cal A}$ satisfies the
equation of a harmonic oscillator with a time dependent mass term.
The case $f=1$ corresponds to the standard EM action where ${\cal A}$
oscillated with time. 

The quantization condition given in
\Eq{quant1} can be implemented by imposing
the following commutation relations between the creation and
annihilation operators,
\EQA
\left[ b_\lambda(\kk), b_{\lambda'}^\dagger(\kk')\right] &=&
(2\pi)^3 \ \delta^3(\kk -\kk') \ \delta_{\lambda\lambda'}, \nonumber \\
\left[ b_\lambda(\kk), b_{\lambda'}(\kk')\right] &=&
\left[ b_\lambda^\dagger(\kk), b_{\lambda'}^\dagger(\kk')\right] =
0 \;.
\label{bcommutation}
\ENA
We also define the vacuum state $\vert0>$ as one which is annihilated
by $b_\lambda(\kk)$, that is $b_\lambda(\kk)\vert0> = 0$.

Once we have set up the quantization of the EM field, it
is of interest to ask how the energy density of the EM
field evolves. The energy momentum tensor is given by
varying the EM Lagrangian density with respect to the metric.
The energy density $T^0_0$ can be written as the sum of a magnetic
contribution,  $T^{0 B}_0 = -f^2 (B_iB^i)/(8\pi)$ and 
electric contribution $T^{0 E}_0 = -f^2 (E_iE^i)/(8\pi)$.
We substitute the Fourier expansion of $A_i$
into the magnetic and electric energy densities,
and take the expectation value in the vacuum state
$\vert 0 >$. Let us define
$\rho_B = <0\vert T^{0 B}_0\vert0>$ and 
$\rho_E = <0\vert T^{0 E}_0\vert0>$.
 Using the properties
$b_\lambda(\kk)\vert0> = 0$, 
and $<0\vert b_\lambda(\kk) b_{\lambda'}^\dagger(\pp)\vert 0 > =
(2\pi)^3 \delta(\kk - \pp)\delta_{\lambda\lambda'}$,
we get for the spectral energy densities in the magnetic and electric fields,
\EQA
\frac{d\rho_B}{d\ln{k}} &=& \frac{1}{2\pi^2} \left(\frac{k}{a} \right)^4
k \left\vert {\cal A}(k,\tau) \right\vert^2 , \nonumber \\
\frac{d\rho_E}{d\ln{k}} &=& \frac{f^2}{2\pi^2} \frac{k^3}{a^4}
\left\vert \left[\frac{{\cal A}(k,\tau)}{f} \right]' \right\vert^2.
\label{rhoB}
\ENA
Once we have calculated the evolution of the mode function
${\cal A}(k,\tau)$, then the evolution of energy densities in the
magnetic and electric parts of the EM field can be calculated.

\subsubsection{The generated magnetic and electric fields}

Consider for example a case where the scale factor evolves with conformal
time as
$a(\tau) = a_0 \left\vert (\tau/\tau_0)\right\vert^{1 + \beta}$.
The case when $\beta = -2$ corresponds to de Sitter space-time of
exponential expansion in cosmic time, or $a(t) \propto \exp{(Ht)}$.
On the other hand for an accelerated power law expansion with
$a(t) = a_0(t/t_0)^p$ and $p > 1$, integrating $dt = ad\tau$, we have
$\tau \propto -(t/t_0)^{1/(p-1)}$ and $a(\tau) \propto \tau^{-p/(p-1)}$.
Here we have assumed that $\tau \to 0_{-}$ as $t \to \infty$, such that
during inflation, the conformal time lies in the range $-\infty < \tau < 0$.
In the limit of $p \gg 1$, one goes over to an almost
exponential expansion with $\beta \to -2 -1/p$.

Let us also consider a model potential where the gauge coupling
function $f$ evolves as a power law,
$f(\tau) \propto a^\alpha$.
This could obtain for example for exponential form of $f(\phi)$ and
a power law inflation. We then have
$f''/f = \gamma(\gamma -1)/\tau^2$
and $\gamma = \alpha(1+\beta)$.
Then the evolution of the mode function ${\cal A}$ is given by
\EQ
{\cal A}''(k,\tau)
+\left(k^2-\frac{\gamma(\gamma -1)}{\tau^2}\right){\cal A}(k,\tau)=0,
\label{Aevol2}
\EN
whose solution can be written in terms of Bessel functions,
\EQ
{\cal A} = \sqrt{(-k\tau)} [ C_1J_{\gamma -1/2}(-k\tau)
+ C_2J_{-\gamma + 1/2}(-k\tau)],
\label{Asoln}
\EN
where $C_1(k)$ and $C_2(k)$ are 
fixed by the initial conditions.

The initial conditions are specified for each mode (or wavenumber $k$),
when it is deep within the Hubble radius, where one can assume the mode
function goes over to that relevant for the Minkowski space vacuum.
Note that the expansion rate is given by $H(t) = \dot{a}/a = a'/a^2$.
For the expansion factor given above, we have
$a'/a = -(p/(p-1))(1/\tau)$, and for $p \gg 1$, $aH \to -1/\tau$.
Thus the ratio the Hubble radius to the proper scale of a perturbation is
given by $(1/H)(a/k)^{-1} = k/(aH) = -k\tau$. A given mode is
therefore within the Hubble radius for $-k\tau > 1$ and outside the Hubble
radius when $-k\tau < 1$.

In the short wavelength limit, $(k/a)/H = (-k\tau) \to \infty$,
the solutions of \Eq{Aevol2} are simply ${\cal A} \propto \exp{(\pm ik\tau)}$,
and we choose the solution which reduces to that relevant for
the Minkowski space vacuum. Thus we assume
as initial condition that as $(-k\tau) \to \infty$,
${\cal A} \to (1/\sqrt{2k})  \exp^{-ik\tau}$.
This fixes the constants in \Eq{Asoln}.
In the other limit of modes well outside the Hubble radius,
or at late times, with $(-k\tau) \to 0$
\EQ
{\cal A} \to k^{-1/2} [ c_1(\gamma) (-k\tau)^{\gamma}
+ c_2(\gamma) (-k\tau)^{1-\gamma}],
\label{finalA}
\EN
where
\EQA
c_1 &=& {\frac{\sqrt{\pi}}{2^{\gamma+1/2}}}
\frac{e^{-i\pi\gamma/2}}{\Gamma(\gamma +\frac12)\cos(\pi\gamma)},
\nonumber \\
c_2 &=& {\frac{\sqrt{\pi}}{2^{3/2-\gamma}}}
\frac{e^{i\pi(\gamma+1)/2}}{\Gamma(\frac32-\gamma)\cos(\pi\gamma)},
\label{confin}
\ENA
From \Eq{finalA} one sees that, as $(-k\tau) \to 0$,
the $c_1$ term dominates for $\gamma \le 1/2$,
while $c_2$ term dominates for $\gamma \ge 1/2$.
We can substitute \Eq{finalA} into \Eq{rhoB}
to calculate the spectrum of $\rho_B$ and $\rho_E$.
We get for the magnetic spectrum,
\EQ
\frac{d\rho_B}{d\ln{k}} 
\approx
\frac{{\cal F}(n)}{2\pi^2} H^4
\left(-k\tau\right)^{4+2n},
\label{rhoBn}
\EN
where $n=\gamma$ if $\gamma \le 1/2$ and $n =1-\gamma$ for
$\gamma \ge 1/2$, and
${\cal F}(n) = \pi/(2^{2n+1}\Gamma^2(n +\frac12)\cos^2(\pi n))$.
We have also taken $(k/aH) \approx -k\tau$, valid for
nearly exponential expansion with $p\gg1$.
During slow roll inflation, the Hubble parameter $H$
is expected to vary very slowly, and thus most of the
evolution of the magnetic spectrum is due to the
$(-k\tau)^{4+2n}$ factor.
One can see that the property of
scale invariance of the spectrum (with
$4 + 2n = 0$), and having $\rho_B \sim a^0$ go together,
and they require either $\gamma = 3$ or $\gamma = -2$.

We can calculate the electric field spectrum in a very
similar manner by first calculating $({\cal A}/f)'$ from
\Eq{Asoln} in the limit $(-k\tau) \to 0$, and the using \Eq{rhoB}.
We get
\EQ
\frac{d\rho_E}{d\ln{k}} 
\approx
\ \frac{{\cal G}(m)}{2\pi^2} \ H^4
\ \left(-k\tau\right)^{4+2m},
\label{rhoEn}
\EN
where now $m=\gamma+1$ if $\gamma \le -1/2$ and $m =-\gamma$ for
$\gamma \ge -1/2$, and
${\cal G}(m) = \pi/(2^{2m+3}\Gamma^2(m +\frac32)\cos^2(\pi m))$.
Thus having a scale invariant magnetic spectrum implies
that the electric spectrum is not scale invariant, and
in addition varies strongly with time. For example
if $\gamma =3$, then $(4+2m) = -2$, although $(4+2n) =0$.
In this case as $(-k\tau) \to 0$, the electric field
increases rapidly and there is the danger of its energy
density exceeding the energy density in the universe,
unless the scale of inflation (or the value of $H^4$ is
sufficiently small. Such values of $\gamma$ 
(and the associated magnetogenesis models) 
are strongly
constrained by the back reaction on the background expansion
they imply \citep{MY08,FM12}.

On the other hand consider the case near $\gamma=-2$. In this case
the magnetic spectrum is scale invariant, and at the same time
$(4+2m) = 2$, and so the electric field energy density goes as
$(-k\tau)^2 \to 0$ as $(-k\tau) \to 0$. Thus these values
of $\gamma$ are acceptable for magnetic field generation
without severe back reaction effects.

For $\gamma \le 1/2$, and using $(k/aH) = (-k\tau)$,
\EQ
\frac{d\rho_B}{d\ln{k}} = \frac{C(\gamma)}{2\pi^2}
H^4 (-k \tau)^{4 +2\gamma}
\approx \frac{9}{4\pi^2} H^4
\ ({\rm for \ \gamma=-2 })
\label{finamp}
\EN

In the above scenario, one generates basically a non-helical field.
It is possible to also generate a helical field if the action also
contains a term of the form $F_{\mu\nu}\tilde{F}^{\mu\nu}$ as
in \Eq{cibreak}, with a time dependent co-efficient \citet{DHJ11}. 
If the same time dependent function couples
both  $F_{\mu\nu}F^{\mu\nu}$ and $F_{\mu\nu}\tilde{F}^{\mu\nu}$,
then one can also generate helical fields with 
a scale invariant spectrum \citep{ASS15,CS14}. Such a situation
naturally obtains in  higher dimensional cosmology with
say the extra dimensional scale factor $b(t)$ multiplying the
whole action \citep{ASS15}.

\subsubsection{Post inflationary evolution}
\label{postinf}

Post inflationary reheating is expected to
convert the energy in the inflaton field to radiation
(which will include various species of relativistic charged particles).
For simplicity let us assume this reheating to be instantaneous.
After the universe becomes radiation dominated its conductivity
($\sigma$) becomes important. From \Sec{resist}, 
we  find that the ratio $\sigma/H \sim (1/\alpha\ln\Lambda)(m_{pl}/T) \gg 1$. 
Thus the time scale for conductivity to operate is much smaller than
the expansion time scale. 
In order to take into account this
conductivity, one has to reinstate the interaction term in the
EM action, given in \Eq{emaction}. Further, as the inflaton has
decayed, we can take $f$ to have become constant with time
and settled to some value $f_0$.
Varying the action with respect to $A_\mu$ now gives
\[
F^{\mu\nu}_{; \ \nu} = \frac{4\pi J^\mu}{f_0^2}
\]
The value of $f_0$ in this model thus
goes to renormalize the value of electric charge 
$e$ to be $e_N = e/f_0^2$.

Let us proceed for now by assuming that we have absorbed $f_0^2$
into $e$. In the conducting plasma which 
is present
after reheating,
the current density will be given by the Ohm's law of \Eq{relohm}.
The fluid velocity at this stage is expected to be that of
the fundamental observers, i.e. $w^\mu = u^\mu$.
Thus the spatial components $J^i = \sigma E^i = -g^{ij}\dot{A_j}$.
Let us assume that the net charge density is negligible and
thus neglect gradients in the scalar potential $A_0$.
Then the evolution of the spatial components of the 
vector potential is given by
\EQ
\ddot{A}_i +(H + 4\pi\sigma) \dot{A}_i
 -\partial_j\partial^j A_i = 0 \; .
\label{Aevolsig}
\EN
We see that any time dependence in $A_i$ is damped out on
the inverse conductivity time-scale. To see this explicitly,
consider modes which have been amplified during inflation
and hence have super Hubble scales $k/(aH) \ll 1$.
Also let us look at the high conductivity limit
of $\sigma/H \gg 1$. Then \Eq{Aevolsig}
reduces to 
\[
\ddot{A}_i + 4\pi\sigma \dot{A}_i = 0; \quad {\rm or} \quad 
A_i = \frac{D_1(\xx)}{4\pi \sigma} e^{-4\pi\sigma t}
+ D_2(\xx)\; .
\]
We see that the $D_1$ term decays exponentially on a time-scale of
$(4\pi\sigma)^{-1} \ll (1/H)$. This leaves behind a constant
(in time) $A_i = D_2(\xx)$. Thus the electric field $E_i =0$, 
and the high conductivity of the plasma has led to the shorting out
of the electric field. Note that the time scale 
in which the electric field decays does not depend on the scale 
of the perturbation, that is the $\sigma$ dependent damping term
in \Eq{Aevolsig} has no dependence on spatial derivatives.
As far as the magnetic field is concerned, \Eq{EandB} shows that
$B_i \sim 1/a$ when $A_i = D_2(\xx)$. Therefore $\bar{B}_i \sim 1/a^2$,
as expected when the magnetic field is frozen into the highly conducting plasma.

Let us now make a numerical estimate of the 
strength of the magnetic fields generated in the scale invariant case.
For both $\gamma =-2$ and $\gamma = 3$, we have
from \Eq{rhoBn} and \Eq{finamp},  
$d\rho_B/d\ln{k} \approx (9/4\pi^2) H^4$. 
Cosmic Microwave Background 
limits on the amplitude of scalar perturbations
generated during inflation, give an upper limit on $H/M_{pl} \sim 10^{-5}$
(cf. \citet{BTW06}). Here
$M_{pl} = 1/\sqrt{G}$ is the Planck mass. The magnetic energy
density decreases with expansion 
as $1/a^4$, and so its present day value
$\rho_B(0) = \rho_B (a_f/a_0)^4$, where $a_f$ is the scale factor
at end of inflation, while $a_0$ is its present day value.
Let us assume that the universe transited to radiation domination
immediately after inflation and use entropy conservation,
that is the constancy of $g T^3 a^3$ during its evolution, where
$g$ is the effective relativistic degrees of freedom and $T$ the temperature
of the relativistic fluid. 
Then $(a_0/a_f) = (g_f/g_0)^{1/3} (T_f/T_0)$. To find $T_f$, we assume
instantaneous reheating at the end of inflation, to generate
relativistic plasma. Then Einstein equation gives
$H^2 = (8\pi G/3 M_{pl}^2) [g_f (\pi^2/30) T_f^4]$.
We then get
\EQA
\frac{a_0}{a_f} &=& 
\frac{g_f^{1/12}}{g_0^{1/3}} \frac{H^{1/2} M_{pl}^{1/2}}{T_0}
\left(\frac{90}{8\pi^3}\right)^{1/4} \nonumber \\
&\approx& 0.9 \times 10^{29}
\left(\frac{H}{10^{-5} M_{pl}}\right)^{1/2},
\label{a0af}
\ENA
where we have taken $g_f \sim 100$ and $g_0 = 2.64$.
(for two neutrino species being non-relativistic today),
This leads to an estimate the present day value of
the magnetic field strength, $B_0$ at any scale,
\EQ
B_0 \sim 0.6 \times 10^{-10} {\rm G} 
\left(\frac{H}{10^{-5} M_{pl}}\right)
\; .
\label{Bstrength}
\EN
Thus interesting field strengths can in principle be created
if the parameters of the coupling function $f$ are
set appropriately.

\subsubsection{Constraints and Caveats}

We have already described above one possible difficulty which
needs to be addressed in models of inflationary magnetogenesis,
that of avoiding strong back reaction due to the 
generated electric fields.
Another interesting problem was raised
by \citet{DMR09} (DMR), which has come to be known as the strong coupling
problem.
Suppose the inflationary expansion is almost exponential with
$\beta =-2$, then for $\gamma \approx -2$, we have 
$\alpha = \gamma/(1+\beta) \approx 2$.
This implies that the function $f =f_i (a/a_i)^2$ increases
greatly during inflation, from its initial value of $f_i$ at $a=a_i$.
Thus if we want $f_0 \sim 1$ at the end of inflation, 
then at early times $f_i \ll f_0$ and the renormalized charge
at these early times $e_N  = e/f_i^2 \gg e$. DMR argue 
that one is then in a strongly coupled
regime at the beginning of inflation where such a theory
is not trustable. There is however the following naive caveat to the
above argument: Suppose one started with a weakly coupled theory where
$f_i \sim 1$. Then at the end of inflation $f_0 \gg f_i$, and so
the renormalized charge $e_N \ll e$. Such a situation
does not seem to have the problem of strong coupling raised by DMR;
however it does leave the gauge field extremely weakly coupled
to the charges at the end of inflation.
This also means that even if $\rho_B$ is large, the magnetic
field strength itself as deduced from the expression for $T^{0B}_0$ 
is $B^iB_i = 8\pi \rho_B/f_0^2 \ll 8\pi\rho_B$. 

Possible ways of getting around the strong coupling problem
have been explored \citep{FJS13,Camp15,Tas15}.
A particularly simple possibility 
arises if conformal invariance is broken by extra dimensional scale factor, 
like $b(t)$ in \Eq{cibreak}, which is outside all parts of the action 
(Subramanian et al., 2015 (in Preparation) and also \citet{ASS15}).
Then when $b(t)$ stops evolving and settles down to a constant value $b_0$,
this constant may be absorbed in to the re-definition of the 4-d metric,
instead of renormalizing the coupling or the fields. The details
of this possibility remains to be fully explored.
It has also been argued that some of these problems can be 
circumvented if magnetic field generation takes place in
a bouncing universe \citep{Mem14}, and also scale
invariant spectra can be generated in such models \citep{SAJ15}.
An additional potential problem raised recently is that the
generated electric fields can lead to generation of
light charged particles due to the Schwinger effect, whose
conductivity freezes the magnetic field generation \citep{KA14}.
This effect is derived in pure de Sitter space assuming 
adiabatic regularization, and needs to be examined in greater detail.

Before leaving this section, it is also of interest to mention
a possible origin of magnetic fields during inflation, suggested
by \citet{Camp13}, which does not require explicit breaking
of conformal invariance. \citet{Camp13} calculated the renormalized
expectation value of the two-point magnetic correlator in de-Sitter
space time (which is mimicked by exponential inflation) using
adiabatic regularization, and finds
it to have a value
$<0\vert \BBB(\xx)\BBB(\yy)\vert0>_{phys} = (19 H^4)/(160 \pi^2)$,
independent of scale. This is a factor $(19/360)$ times smaller than
the value in \Eq{finamp}, obtained for 
models which explicitly break conformal invariance
and generate scale invariant magnetic fields. 
It is not clear whether such a value obtains if the 
spacetime is not eternally de-Sitter, but one where inflation
starts and ends at finite times (\citet{DuMR13} and reply by \citet{Campb13}).
\citet{ALN14} have pointed out that duality symmetry and
conformal invariance of free electromagnetism, could 
both break at the quantum level in the presence of a classical background,
and for de-Sitter recovered the result of \citet{Camp13},
for the single-point $\bra{\BBB^2(\xx)}$,
also showing that the electric field has $\bra{\EE^2} = - \bra{\BBB^2}$.
This also implies that the energy density $\propto 
\bra{\EE^2}+\bra{\BBB^2}=0$ in pure de-Sitter!
It is of interest for more workers to examine these issues
critically.

\subsection{Generation during Phase transitions}
\label{genpt}

Primordial magnetic fields could also be generated in various 
phase transitions, like
the electroweak phase transition (EWPT) or 
the QCD transition due to causal processes (for some reviews see
\citet{GR01,KKT11,DN13}).
However these will lead to a correlation scale of the field
smaller than the Hubble radius at that epoch. Hence
very tiny fields on galactic scales obtain,
unless helicity is also generated; in which case
one can have an inverse cascade of energy to larger scales
\citep{BEO96,BJ04}.

Magnetic fields can optimally arise if the phase transition is a first
order phase transition. The idea is that in such a transition, bubbles
of the new phase nucleate in a sea of the old phase, then expand, collide
until the new phase occupies the whole volume. 
Such a process also provides non-equilibrium conditions 
for processes like baryogenesis \citep{Shap87,TZ90} 
and leptogenesis, which in turn
could also lead to magnetic field generation. 
The process of collisions
of the bubbles is likely to be "violent" and 
generate turbulence. This can in turn 
amplify magnetic fields further by dynamo action (cf. \citet{BS05a} and 
\Sec{fd}).

The QCD transition occurs as the universe cools below a temperature 
$T_c\sim150$ MeV \citep{Bazavov14}, when the universe transits from 
a quark-gluon plasma to a hadronic phase. For a universe where chemical 
potentials are small, that is the excess of matter over antimatter is small, 
lattice calculations show that the transition occurs 
smoothly as the universe cools in what is referred to as an
'analytic crossover' \citep{Aoki06}. This is not a real phase transition, 
but rather many thermodynamic variables change dramatically but continuously 
around a narrow range of temperature, as the universe cools below $T_c$.
However, if the lepton chemical potential (say in neutrinos)
is sufficiently large, but within cosmologically allowed values,
the nature of the QCD phase transition need not be a crossover, 
and could be first order \citep{SchStu09}.

Similarly, the EWPT in case of the standard model of particle physics  
is also not a first order transition \citep{Kaj_etal96,CFH98}, 
for the observed high Higgs mass $M_H \sim 125$ GeV 
\citep{HiggsATLAS,HiggsCMS}, but again a crossover.
However supersymmetric extensions to the standard model (like the MSSM)
can have parameters, where there can be a first order EWPT 
\citep{GSW05,HKPS07}. 
A detailed text book discussion of the EWPT and the conditions
under which it can be a first order transition is given 
by \citet{Gor_Rub11}.

\subsubsection{Coherence scales and field strengths}

In such a first order phase transition, one may imagine that the
correlation scale $l_c$ of the generated magnetic field 
(and the corresponding comoving scale $L_c= l_c/a$), 
would be of order the largest bubble size before bubble collision.
This would in general be a fraction, say $f_c$, of the Hubble scale at the
epoch when the phase transition occurs. It is of interest
to estimate this scale. In the radiation dominated universe,
the Hubble radius is $d_H = H^{-1} = a/\dot{a} = 2 t$, where
from Einstein equations, 
the time temperature relation is given by \citep{Kolb_Turner},
\EQ
t = \frac{0.3}{g_*^{1/2}} \left(\frac{T}{{\rm MeV}}\right)^{-2} s.
\label{timetemp}
\EN
Here $g_*$ is the effective degrees of freedom which
contributes to the energy density of the relativistic plasma.
(Note $g_*$ could be slightly different from the $g$ in \Sec{postinf}
which contributes to the entropy).

For the EWPT, we can take $T\sim 100 GeV$, $g_* \sim 100$, which gives
$l_c \sim 1.4 f_c $ cm. Also using the constancy of entropy, we
have $a(t)T g^{1/3}$ constant. Adopting the present epoch 
$g_0 = 2.64$
(for two neutrino species being non-relativistic today),
$T_0 = 2.725$ K, $g\sim g_* \sim 100$, we have 
\EQ
L_c \sim 1.4 \times 10^{15} l_c \sim 2 \times 10^{15} f_c {\rm cm}.
\label{ewcor}
\EN
For the QCD phase transition, the corresponding numbers are 
$T\sim 150$ MeV, $g_* \sim 60$, which leads to 
$l_c \sim f_c \ 6.4 \times 10^5$ cm, and 
\EQ
L_c \sim 1.8 \times 10^{12} l_c
\sim \frac{f_c}{3} \  {\rm pc}.
\label{QCDcor}
\EN
Even for optimistic values of $f_c$, the above correlation 
scales are very small compared to the
coherence scales of magnetic fields, of order kpc, observed in 
galaxies and galaxy cluster plasma. An important question is
how much these scales can grow  
during the nonlinear evolution of the field, a feature to which
we return below.

It is also of interest to ask what would be the strength of the generated
fields. One simple constraint is that it cannot exceed a fraction, say $f_B$,
of the energy density of the universe at the time of the phase transition.
We already noted that a field of about 3 $\mu$G corresponds to
the CMB energy density today. With expansion, both the energy density
in radiation and the magnetic energy density, (ignoring 
nonlinear evolution) scale as $1/a^4$. Thus for getting nG strength
fields a fraction $\sim 10^{-7}$ of the radiation energy density
has to be converted to magnetic fields. Of course as photons are
only one component of the relativistic plasma in the early
universe, contributing a $g_* =2$ to the total $g_*$, an even smaller
fraction of the total energy density needs to be converted to 
magnetic energy, to result in nG strength fields today.
This of course ignores the decay of magnetic energy due to
generation of MHD turbulence and resistive effects, but gives
a rough idea of what is required.
The next question is to ask how the magnetic fields are generated.

One general idea which has been around from the early work of \cite{Hogan83},
is that, during a first order phase transition, 
a seed magnetic field is generated by a battery effect. And this
is subsequently amplified by a turbulent small scale (or fluctuation) 
dynamo (see \Sec{fd}).
The required turbulence could be generated in bubble collisions,
when some fraction of the free energy released during the transition from the 
false to the true vacuum goes into turbulent kinetic energy.
If the dynamo can saturate, then the magnetic energy can attain
a fraction of the turbulent kinetic energy, and optimally
the coherence scale could be a fraction of the bubble size
when they collide. There are uncertainties associated with
the saturation of the small-scale dynamo itself, as the dynamo is
likely to be a high magnetic Prandtl dynamo (cf. \Sec{fd}).
This general idea has been applied to both EWPT \citep{Baym_etal96}
and the QCD phase transition \citep{QLS89,Sigl_etal97}.

\subsubsection{Generation due to Higgs field gradients}

Another general possibility was suggested by \citet{Vachaspati91}
in the context of the EWPT, which has been subsequently extensively 
explored. The idea is that gradients in the Higgs field vacuum 
expectation value, which is the order parameter for the EWPT,
naturally arise during the phase transition, and directly induce
electromagnetic fields. The original work by \citet{Vachaspati91}
was applied to second-order phase transitions. From an analysis
of this picture, \citet{GR98} estimated the coherence scale of 
the resulting field to be of order the curvature scale of the 
Higgs effective potential at what is known as the Ginzburg temperature, 
$T_G \sim 100$ GeV. This is basically the critical temperature at which thermal 
fluctuations of the Higgs field inside a given domain of broken symmetry
can no longer restore the symmetry. This leads to an estimate of  
$l_c \sim 10/T_G$, for a range of Higgs masses, and also a 
physical
field strength $B \sim q^{-1}_{EW} T_G^2$
corresponding to a comoving field $B \sim 7\times 10^{-8}$ G.
Note however that the coherence scale can increase rapidly
due to nonlinear processing as discussed below.

The \citet{Vachaspati91} mechanism has also been numerically simulated
by several groups 
to estimate the nature of magnetic fields produced
\citep{DiazGil_etal08a,DiazGil_etal08b,CFVA08,SJ09,Stevens_etal12}.
\cite{Stevens_etal12} 
examined magnetic field generation 
in collision of bubbles of the true vacuum during the EWPT.
Numerically solving the equations of motion determined 
from an effective minimal supersymmetric standard model action, 
they find that bubble collisions result in $B \sim {\rm few} \ m_W^2$,
coherent on scales of $l_c \sim {\rm few} \ 10 m_W^{-1}$.
Here $m_W = 80.4$ GeV is the mass of the charged gauge bosons.
This field 
has a similar 
strength and 
coherence scale as what was estimated above for a second
order phase transition.

\subsubsection{Linking baryogenesis and magnetogenesis}

A remarkable connection between baryogenesis and
the helicity of generated magnetic fields during the EWPT, 
has been pointed out by several authors \citep{Cornwall97,Vachaspati01}. 
Note that Baryon number is classically conserved in the electroweak theory,
but this conservation law gets broken in quantum theory in the
presence of classical Gauge field configurations, by what is
referred to as an 'anamoly'. Specifically if $j^\mu_B$ is the
four-current density corresponding to Baryon number, 
then \citep{thooft76,Gor_Rub11},
\EQ
\partial_\mu j^\mu_B = N_F 
\frac{g^2}{16\pi^2} W^{\mu\nu a} \tilde{W}_{\mu\nu}^a.
\label{bnon}
\EN
Here $W^{\mu\nu a} = \partial_\mu W^a_\nu - \partial_\nu W^a_\mu 
+ g \epsilon^{abc} W^b_\mu W^c_\nu$ is the field strength 
corresponding to the $SU(2)_w$ gauge potential $W^a_\mu$, 
$\tilde{W}_{\mu\nu}^a = \epsilon_{\mu\nu\rho\lambda}  W^{\rho\lambda a}/2$ 
is dual tensor, $N_F=3$ the number of flavors 
and $g$ the gauge coupling. A corresponding
equation 
is also obtained
for each lepton number current,
without the factor $N_F$, and thus B-L (Baryon minus Lepton number)
is conserved although B and L separately are not.
Integrating \Eq{bnon} over the 4-d volume between two constant
time hypersurfaces, it turns out that baryon number changes
by $\Delta B = 3 \Delta N_{CS}$, where $N_{CS}$ is a topological index
called the Chern-Simons number \citep{Vachaspati01},
\EQA
N_{CS} = \frac{N_F}{32 \pi^2} \epsilon^{ijk} 
\int d^3x [g^2(W_{ij}^a W_k^a 
&-&\frac{g}{3} \epsilon_{abc} W_i^a W_j^b W_k^c) \nonumber \\
&-& g'^2 Y_{ij}Y_k ].
\label{NCS}
\ENA
Here we have included also the effect of the Hypercharge field $Y_\mu$.
For the electromagnetic field for example $N_{CS}$ is
just the usual magnetic helicity.

One can have pure gauge configurations with zero energy which have
different integer values of $N_{CS}$, and these different vacua 
are separated by configurations in field space with larger energy,
up to a maximum of $E_{sph} \sim m_W/g^2$. 
The field configuration with this `maximum' energy is in fact
a saddle point; energy decreases along one direction in configuration
space while it increases along all other directions, and is referred
to as a `sphaleron' (from the Greek, ready to fall).
Baryons are produced/annihilated as the sphaleron decays and $N_{CS}$ changes.
Low energy classical dynamics occurs
around one of these 'vacuua' conserving baryon number, while quantum dynamics
allows tunneling between vacua; albeit with such a small
probability, that baryon number violation is highly improbable. 
However at high enough temperatures, there
would be enough thermal energy to just go over the potential barrier
without tunneling and to transit from one of
these topologically distinct vacua to a neighbouring one,
through a sphaleron transition. If CP violation is also 
present such transitions could proceed more
efficiently to produce baryons than antibaryons.

\citet{Vachaspati01} argued that during the EWPT, 
such sphaleron type configurations can be produced
in the false vacuum phase, and their decay would change $N_{CS}$
resulting in electroweak baryogenesis. In addition, 
that this process also leaves behind magnetic fields 
with a net (left handed) helicity related to the baryon number, 
with helicity density $h \sim - 10^2 n_b$. ($n_b$ is the baryon
number density). A heuristic picture
suggested by \citet{Vachaspati01}, was to think of the sphaleron
configuration as two loops of linked electroweak strings carrying
the Z-magnetic flux. One channel of decay of this Z-string 
could be by nucleating a (electromagnetic) monopole-antimonopole pairs on the
string. A magnetic field then connects this monopole-antimonopole
pair. The pair then get pulled apart, the Z-string shrinks and disappears
leaving behind a linked loops of electromagnetic magnetic flux.
The coherence scale of the generated field is estimated
to be $l_c\sim 1/(\alpha_e T_{EW})$ initially. Here $\alpha_e$ is
the fine structure constant and $T_{EW}$ is the temperature
of the universe at the epoch of the EWPT. This can rapidly
evolve conserving helicity, and grow much larger in a Hubble
time. \citet{Vachaspati01} estimates that later evolution
conserving helicity (as described in \Sec{nonlin}) can lead to a field strength
of $10^{-13}$ G at recombination, coherent on comoving
scales of $0.1$ pc.

This heuristic picture has since been tested by numerically
solving the electroweak equations of motion on a lattice, 
starting initially from a sphaleron like configuration \citep{CFVA08}. 
These authors show that baryogenesis in this model does
generate helical magnetic fields, though $h$ created is somewhat smaller
than the above estimate. \citet{CFVA08} also argue that the magnetic energy
generated could be much larger than that associated with the helicity,
assuming that every baryon number violating reaction goes
via such a sphaleron. In this case, generating both baryons and
anti baryons will lead to magnetic field generation, but with
oppositely signed helicity. When baryons and anti baryons annihilate
to leave a small net baryon to photon ratio, not all the magnetic fields
with oppositely signed helicity need to annihilate. This could result
in a larger magnetic energy than that associated with just the net
helicity, 
with a comoving strength of even up to a nano Gauss. The comoving 
coherence scale $L_c$ will also be much larger in case the EWPT
goes through a first order phase transition of up to a $Mpc$;
but as \citet{CFVA08} themselves emphasize, the determination
of the precise strength, $L_c$ and spectrum remains an unsolved
problem.

The generation of helical fields during
a first order EWPT due to the inhomogeneities of the Higgs field, has
also been numerically analyzed by \citet{DiazGil_etal08a,DiazGil_etal08b}.
These authors set their problem in a model of electroweak
hybrid inflation, where initial fluctuations of the Higgs can
be naturally generated. They find that magnetic field get generated,
whose helical nature is linked to the winding of the Higgs field.
The nucleation and growth of Higgs bubbles squeeze the magnetic field
into string like configurations between the bubbles. 
The field energy at the end of the simulation is about a percent
of the total energy density and its correlation scale has grown 
to $l_c \sim 30 m^{-1}$, where $m= m_H/\sqrt{2}$ with $m_H$ the Higgs mass.

Note that the resistive dissipation time for fields generated 
during EWPT era is 
$\tau_{ohmic} \sim l_c^2 \sigma \sim 10^3 T^{-1}$, where
we have taken $l_c \sim 30 T^{-1}$, and the conductivity $\sigma \sim 10T$ 
(cf. \citet{DN13}), as appropriate for $T \sim 100$GeV. Then 
the ratio of the Ohmic dissipation time to Hubble time is, 
$\tau_{ohmic}/H^{-1} \sim 10^4 T^{-1}/((m_{pl}/T^2) \sim 10^4 (T/m_{pl}) 
\ll 1$. Thus the resulting fields could be strongly damped 
by resistivity. Of course the field itself will induce motions
at the Alfv\'en speed, and one has to consider its nonlinear
evolution more carefully (see \Sec{nonlin}).
One caveat is that, during the phase transition,
the charged particles which carry the current will not be in
thermal equilibrium, and the estimate of the conductivity
may not be relevant. Also if the phase transition is
of first order, the coherence scale at least along the field would 
be set by the size of the bubble of the new phase just before
bubbles collide, $l_c \sim f_c H^{-1}$. Also the collision of bubbles
would generate turbulence with typically large fluid and magnetic
Reynolds numbers \citep{Baym_etal96}. For example, if the
induced velocity is $v \sim 0.01 c$, and $f_c \sim 10^{-3}$, then
$\Rey \sim 10^{10}$ \citep{Baym_etal96} and $\Rm$ could be 
ten times larger.
Then one can have small scale dynamo action for some period of 
the time, and the strength of the fields and their coherence 
scale which survive nonlinear evolution could be greater. 
We will consider the nonlinear evolution
of the fields in more detail in \Sec{nonlin}.

\subsubsection{The chiral anomaly and magnetogenesis}

Another idea for magnetic field generation, 
which uses essentially chiral anomaly of weak interactions
in simple extensions of the standard model, is worth mentioning.
\citet{BRS12} show that large scale magnetic fields can arise spontaneously
in the ground state of the Standard model, due to the parity-breaking 
character of weak interactions and the chiral anomaly. 
The strength is at present not predicted, but the coherence scale
(wavenumber) is predicted to $k_{coh} \sim c_1 \alpha_e (G_F T^3) \eta_{L,B}$.
Here $c_1 \sim 2.5 \times 10^{-2}$ is a numerical co-efficient,
$\alpha_e$ the fine structure constant, $G_F \sim (300 {\rm Gev})^{-2}$ 
the Fermi coupling constant and $\eta_{L,B}$ the ratio of 
total lepton (baryon) number to the number of photons. For this
instability to operate at a rate faster then the Hubble rate, one
requires $\eta_L > {\rm few} \times 10^{-2}$, which is on the
threshold of being ruled out by neucleosynthesis constraints. 
\citet{BRS12} suggest that the required lepton
number can be realized in some models, just after EWPT but disappear
later. For such $\eta_L$, the
coherence scale $k^{-1}_{coh}$ is much larger than the
thermal wavelength $1/T$ and much smaller than the Hubble radius $1/H$.

However the coherence scale can increase further as shown by 
\citet{BFR12}. These authors argue that,
in the presence of strong magnetic fields, a left-right asymmetry
develops due the chiral anomaly. A net chemical potential for
the left-(right-) chiral electrons can persist. This results
in a modified Maxwell equation, and an "$\alpha$-effect", or
a source term to the current proportional to the magnetic
field itself, with the proportionality depending on the
chemical potential. This in turn can lead to further magnetic field
amplification and magnetic helicity transfer from
small scales to larger scales until the temperature of the universe
drops to $T\sim1$ Mev \citep{BFR12}. 
Such an evolution can then affect predictions
of the remnant field strengths and coherence scales arising
out of phase transitions. 
Similar ideas involving helicity generation due to
parity breaking effects of weak interactions, 
have also been explored by several other authors
\citep{JS97,FC00,Semi_Sok05,SSV12} (and references therein). 
It would be of interest to develop
such ideas in more detail, as it involves just the physics of
the standard model and its simplest extensions.

In summary, a number of ideas for magnetic field generation
during the QCD and electroweak phase transitions have been explored.
Especially interesting are the links between baryogenesis, leptogenesis
and magnetogenesis, the possibility that magnetic helicity could
be generated and the idea that parity breaking effects could lead
to a new form of the $\alpha$-effect and large scale dynamo action.
Which of these scenarios obtains in reality is uncertain at
present, as it depends on assumptions about the particle
physics model, and the nonlinear evolution of partially helical fields.
Thus
the exact predictions for the field strength and its coherence
scale are not yet fully developed. The 
energy going in to the magnetic field can be
a few percent of the radiation energy density, and the field
coherence scales can range from of few tens of the thermal
wavelength $1/T$ to a fraction $f_c$ of the Hubble scale.
For predictions of the present day field strength and coherence
scale, one has to examine how this initial field evolves
from the generation epoch to the present.

\section{Evolution of primordial magnetic fields: The linear regime}
\label{linearevol}

The primordial magnetogenesis scenarios 
discussed in the last section 
generally lead to fields which are Gaussian random.
For example, in case of inflationary generation, the vacuum fluctuations,
of the electromagnetic field that are amplified, are Gaussian random
and thus lead to classical, Gaussian random stochastic magnetic 
field fluctuations. For the EW and QCD 
phase transitions, the fields generated on the small sub-Hubble
scales could be non-Gaussian; but 
large astrophysical scales of relevance, may encompass 
a very large number of such domains. So the the central limit
theorem implies that the field averaged on such large scales
could be Gaussian random.
Thus to study the evolution of the field, one generically starts with
a Gaussian random initial field, 
characterised by a spectrum $M(k)$.
This spectrum is normalised by giving the field strength
$B_0$, at some fiducial scale, and as measured at the present epoch,
assuming it decreases with expansion as $B=B_0/a^2(t)$.
We will begin in this section by considering the evolution of
inhomogeneous magnetic fields in the radiation era.
This was first treated in detail by \citet{JKO98} in terms
of linear perturbations of the MHD modes, followed by a slightly
different approach by \citet{SB98a}, exploiting the conformal invariance
of the MHD equations in the radiation era (see also \citet{BEO96}), 
and some simple nonlinear solutions to the equations.
To get a feel for the possible evolution, we
first look at a simple nonlinear solution to the equations, following
closely the treatment in \citep{SB98a}.

\subsection{Alfv\'en waves in the early universe}
\label{alfven1}

Let us assume that the magnetic field can be written as ${\bf B}^{*}=%
{\bf B}_0^{*}+{\bf b}^{*}$, with a uniform ${\bf B}_0^{*}$. We assume ${\bf b%
}^{*}$ is perpendicular to ${\bf B}_0^{*}$, but do not put any restriction
on the strength of ${\bf b}^{*}$ so that it need {\it not be a small}
perturbation of ${\bf B}_0^{*}$.
Fix the co-ordinates such that ${\bf B}_0^{*}$ lies along the $z$%
-axis, that is ${\bf B}_0^{*}=B_0^{*}\hat {{\bf z}}$, where $\hat {{\bf z}}$
is the unit vector along $z$.
We also take the peculiar velocity ${\bf v}$
to lie perpendicular to ${\bf B}_0^{*}$ and assume that all the variables
depend only on $z$ and $\tau $. In this case, the velocity perturbation
automatically satisfies ${\bf \nabla }.{\bf v}=0$. 

Further, the ratio of the
magnetic energy density to the fluid energy density, $B^2/(8\pi \rho )\sim
10^{-7}B_{-9}^2 \ll 1$. So even when the magnetic field and 
induced velocities get damped by resistivity and viscosity respectively, 
$\rho $ will be perturbed negligibly. It is an excellent
approximation to neglect the viscous and resistive terms in 
Eq. (\ref{energ}). In this ideal limit we can assume, $\rho^*$ and $p^*$
are uniform, and from Eq. (\ref{energ}), also constant in time.

Moreover, the non-linear terms in the momentum equation \Eq{eulers} and
the induction equations \Eq{expind} are
individually zero because of the above properties (that $\vv$ and $\bb^*$ do
not have $z$-component, and all quantities vary only with $z$).
These equations then reduce to
\begin{equation}
(\rho^*+p^*){\frac{\partial {\bf v}}{\partial \tau }}=
-{\bf \nabla }p_T^*
+{\frac{B_0^{*}}{4\pi}} {\frac{\partial {\bf b}^{*}}{%
\partial z}}+
(\rho^*+p^*)\nu ^{*}\nabla ^2{\bf v}
\label{eulerred}
\end{equation}
\begin{equation}
{\frac{\partial {\bf b}^{*}}{\partial \tau }}=B_0^{*}{\frac{\partial {\bf v}%
}{\partial z}} -\eta^*\nabla^2b^* \;.
\label{inductred}
\end{equation}
Here $p_T^* = p^{*}+B^{*2}/8\pi$ is the sum of fluid and magnetic pressures.
The plasma in the early universe is highly conducting such that 
the resistive term in \Eq{inductred} can be neglected
(see \Sec{resist}).
Further, since $(\nab\cdot\vv) =0$, we have
$\nab^2p_T* =0$, which implies that $p_T*$ is uniform in space.
One can therefore drop the pressure gradient term in 
\Eq{eulerred}. Writing ${\bf b}^{*}=b_0(\tau ,z){\bf n}$
and ${\bf v}=v_0(\tau ,z){\bf n}$, eliminating $v_0$ from
Eqns. (\ref{eulerred}) and (\ref{inductred}), gives a damped wave equation
for $b_0(\tau ,z)$,
\begin{equation}
{\frac{\partial ^2b_0}{\partial \tau ^2}}-\nu^{*}(\tau )
{\frac \partial {\partial z^2}}\left( {\frac{\partial b_0}{%
\partial \tau }}\right) -V_A^2{\frac{\partial ^2b_0}{\partial z^2}}=0.
\label{almfin}
\end{equation}
where we have defined the Alfv\'en velocity, $V_A,$ as
\EQA
V_A&=&{\frac{B_0^{*}}{(4\pi (\rho ^{*}+p^{*}))^{1/2}}}={\frac B{(4\pi (\rho
+p))^{1/2}}} \nonumber \\ 
&\approx& 3.8\times 10^{-4}B_{-9}.  
\label{alfvel}
\ENA
For the numerical estimate, we have taken
$\rho =\rho _\gamma $, the photon energy density, as
would be appropriate in the later radiation-dominated era, 
after the epoch of $e^{+}e^{-}$ annihilation and neutrino decoupling 
(at much earlier epochs one has to take all relativistic degrees of
freedom in $\rho$ and $p$ in defining the Alfv\'en velocity).
This linear equation describes the nonlinear Alfv\'en mode in the viscous regime.
It can easily be solved by taking a spatial Fourier transform. For any mode $%
b_0(\tau ,z)=f(\tau )\e^{ikz}$, we have
\begin{equation}
{\ddot f}+ D {\dot f}+ \omega_0^2 f=0, \quad 
 \omega_0 = kV_A, \ D = \nu^* k^2.
\label{oscill}
\end{equation}

If $\omega _0\gg D$, one has damped oscillatory motion, while 
for $D \gg \omega _0$, the motion becomes overdamped. One
solution of the second-order differential equation, where the
oscillator starts with a large initial velocity, suffers strong damping.
However, the other independent solution is negligibly damped. 
This is because, under strong friction, any oscillator displaced from
equilibrium and released from rest has only to acquire a small `terminal'
velocity, before friction balances driving, freezing 
the motion
with energy 
decreasing negligibly. 

We focus primarily on damping by photon viscosity, which is the most
important source of viscosity, after $e^{+}e^{-}$ annihilation,
and has the potential to damp the largest scales.
The kinematic radiative viscosity coefficient is given by \Eq{earlynu}, 
where the photon mean-free-path is
\begin{equation}
l_\gamma (\tau )={\frac 1{\sigma _Tn_e(\tau )}}\approx 1.8
\left( {\frac T{0.25eV}}\right) ^{-3}f_b^{-1} 
x_e^{-1} {\rm kpc}.  \label{lgamnum}
\end{equation}
Here, $\sigma _T$ is the Thomson cross-section for electron-photon
scattering, $n_e$ the electron number density, $x_e$ the ionisation
fraction and $f_b = \Omega_bh^2/(0.022)$ with 
$\Omega _b$ the baryon density of the universe $\rho _b$,
in units of the closure density.
Using $\nu^* = \nu/a$, the damping-to-driving ratio is
\EQ
\frac{D}{\omega _0} ={\frac{\nu ^{*}k^2}{kV_A}}={\frac 1%
5}{\frac{k_p(\tau )l_\gamma (\tau )}{V_A}}  
\approx 526.3{\frac{k_p(\tau )l_\gamma (\tau )}{B_{-9}}},  
\label{bbyom}
\EN
where we have defined the proper wavenumber $k_p(\tau )=(k/a(\tau ))$.
For the diffusion
approximation to be valid, we require $k_pl_\gamma <1$; that is, we must
consider only wavelengths larger than the mean-free-path. Nevertheless, one
expects a large range of wavelengths for which modes will fall in the
overdamped regime. 
In this limit where $D \gg \omega_0$, $\dot f$ will adjust itself so that the
acceleration vanishes, so $\ddot f\approx 0$. In such a `terminal velocity'
approximation,  and $f$ satisfies the equation
\begin{equation}
\dot f=-\frac{\omega _0^2}{D}f; \
f(\tau )=f(\tau _T)\exp \left( -\int_{\tau _T}^\tau {\frac{\omega _0^2}{%
D(\tau ^{\prime })}}d\tau ^{\prime }\right) . 
 \label{termapp}
\end{equation}
Here, $\tau _T$ is the conformal time when the mode reaches the
terminal-velocity regime, or when the acceleration, $\ddot f,$ first
vanishes. 

Note that when $k$ is small enough for $D/\omega_0 < 1$, the actual phase
of the oscillation, given by
\[
\chi = kV_A\tau \sim 10^{-2} B_{-9} \left(\frac{k}{0.2h {\rm Mpc}^{-1}}\right)
\left(\frac{\tau}{\tau_*}\right)
\]
is very small, for galactic scales,
even by the conformal time $\tau_*$ corresponding to
the recombination epoch. 
Even for the largest $k$ where $D/\omega_0 < 1$, which from
\Eq{bbyom} is given by $k = 5 V_A a/l_{\gamma}$,
we have $\chi = 5 V_A^2 [\tau_*/L_{\gamma}(\tau_*)] \ll 1$.
Here we have adopted typical values of $\tau_* \sim 200$ Mpc and
$L_{\gamma}(\tau_*) \sim 2$ Mpc and $V_A^2 \sim 10^{-7} B_{-9}^2$. 
Thus even Alfv\'en modes whose wavelengths are large enough not to be
overdamped by diffusive photon damping, oscillate negligibly.

All in all Alfv\'en wave modes,
with scales larger than the photon mean free path,
do not get erased by radiative damping, during the radiation era.
Either their wavelength is so large that they oscillate negligibly, or
if the wavelength is small enough they are in the overdamped regime,
and so do not get damped.
This holds of course if the wavelength is large enough for the
diffusion approximation to hold.

In contrast to the Alfv\'en mode, compressible modes
have a phase velocity in the radiation era $c/\sqrt{3}$
which is much larger than $V_A$. They can then suffer strong
damping due to radiative viscosity, a process known as
`Silk' damping \cite{Silk68}. 
The damping of linear perturbations
in the expanding universe is further considered in detail
by \citet{JKO98} and \citet{SB98a}, illustrating the above features of
both the incompressible and compressible modes.
These authors show that the magnetically driven 
compressible modes get damped by a factor
\begin{equation}
\exp \left[ -{\frac{k^2}{%
k_D^2}}\right] ;\quad {\rm where}\quad k_D^{-2}={\frac 2{15}}\int {\frac{%
l_\gamma dt}{a^2(t)}}.  \label{sldamp}
\end{equation}
This agrees quite well with the Silk damping of sound waves in the radiation
era, derived in more detailed treatments \citep{Peebles80}, 
except for the small effects of baryon loading and polarization.
In the radiation-dominated epoch
one has $k_D^{-1}=(4/45)^{1/2}l_S(t)/a(t)\sim 0.3l_S(t)/a(t)$, where $%
l_S(t)=(l_\gamma t)^{1/2}$ is the Silk scale. The largest scales which
suffer appreciable damping are the compressible 
modes with wavelengths ($2\pi k_D^{-1})$, of order $L_S = l_S/a$.
What happens for the incompressible modes whose wavelength becomes smaller
than the photon mean free path? And what happens when the universe becomes
matter dominated.

\subsection{The free-streaming regime}

As the universe expands, the mean-free-path of the photon increases as $a^3$%
, while the proper length of any perturbed region increases as $a$. So the
photon mean-free-path can eventually become larger than the proper
wavelength of a given mode, even if it were initially smaller.
When this happens for any given mode, we will
say that the mode has entered the free-streaming regime. Modes with
progressively larger wavelengths enter the free-streaming regime up to a
proper wavelength $\sim l_\gamma (T_d)\sim 2$ kpc (see Eq. (\ref{lgamnum}%
) ), or a comoving wavelength of $L_\gamma(T_d)\sim 2$Mpc, 
at the epoch of decoupling.
After (re)combination of electrons and nuclei into atoms, $l_\gamma $
increases to a value larger than the present Hubble radius, and all modes
enter the free-streaming regime. (We will consider the pre- and
post-recombination epochs separately below.)

When photons start to free stream on a given scale of perturbation, the
tight-coupling diffusion approximation no longer provides a valid
description of the evolution of the perturbed photon-baryon fluid on that
scale. One has to integrate the Boltzmann equation for the photons together
with the MHD equations for the baryon-magnetic field system. A simpler
approximate method of examining the evolution of such modes in the linear
regime is to treat the radiation as isotropic and homogeneous, and only
consider its frictional damping force on the fluid. (The radiative flux
could have also contributed to the force on the baryons; however, for modes
with wavelengths smaller than $l_\gamma $, this flux is negligible since the
associated compressible motions have suffered strong Silk damping at earlier
epochs; when the wavelength was larger than $l_\gamma $). The drag force on
the baryon fluid per unit volume due to the radiation energy density $\rho
_\gamma $, is given by
\begin{equation}
{\bf F}_D=-{\frac 43}n_e\sigma _T\rho _\gamma {\bf v.}  \label{dragfor}
\end{equation}

Since, typically, less than one electron-photon scattering occurs within a
wavelength, the pressure and inertia contributed by the radiation can be
neglected when considering the evolution of such modes. The Euler equation
for the baryonic component then becomes
\EQA
{\frac{\partial {\bf v}}{\partial t}}+H(t){\bf v}+{\bf v}.{\bf \nabla }{\bf v%
}=&-&{\frac 1{a\rho _b}}{\bf \nabla }p_b+{\frac 1{\rho _b}}{\bf J}\times {\bf B%
}-{\frac 1a}{\bf \nabla }\phi \nonumber \\
&-&{\frac{4\rho _\gamma }{3\rho _b}}n_e\sigma _T%
{\bf v.}  
\label{eulerin}
\ENA
Here, $p_b$ the fluid pressure, and $%
H(t)=(da/dt)/a$ is the Hubble parameter. We have also included the
gravitational force, $(1/a){\bf \nabla }\phi $, due to any perturbation in
the density. Note that we have written this equation in the unstarred
conformal frame, (with the magnetic field defined in the `Lab' frame).
We have also transformed the time
co-ordinate, from conformal time, back to ''proper time'' $dt=ad\tau $.

It should be pointed out that the dramatic drop in the pressure, by a factor
of order the very small baryon to photon ratio $\sim 10^{-9}$, when a mode
enters the free-streaming regime, has important consequences. First, in the
absence of radiation pressure, the effect of magnetic pressure (if it
greatly exceeds the fluid pressure) is to convert what was initially an
incompressible Alfve\'n mode into a compressible mode (see below). Second,
the effective baryonic thermal 
Jeans mass decreases dramatically and compressible
modes can become gravitationally unstable. Thus, we have to retain the
gravitational force term in the above equation. The magnetic pressure will
also play a dominant role, providing pressure support against gravity on
sufficiently small scales.

The magnetic pressure $p_B$ and the fluid pressure $p_b$ are given by,
\begin{equation}
p_B={\frac{B^2}{8\pi }}(1+z)^4\approx 4\times 10^{-8}B_{-9}^2\left( {\frac{%
1+z}{10^3}}\right) ^4 \frac{{\rm dyn}}{{\rm cm}^2},  \label{magp}
\end{equation}
\begin{equation}
p_b=2n_ekT\approx 1.9\times 10^{-10}\left( {\frac{1+z}{10^3}}\right) ^4f_b%
\ \frac{{\rm dyn}}{{\rm cm}^2},  \label{fluidp}.
\end{equation}
Here we have assumed that the fluid temperature is locked to the radiation
temperature, and that the gas is still fully ionized electron-proton gas.
Thus magnetic pressure dominates the fluid
pressure, (\ie $p_B\gg p_b$ for $B\gg B_{crit}\sim 7\times 10^{-11}$ Gauss).

Consider first the case where the field $B$ is much smaller than $B_{crit}$.
In this case the motions can still be assumed incompressible. The Alfv\'en modes
which enter the free-streaming regime, remain Alfv\'enic. Following the
ideas of \Sec{alfven1}, we look again at non-linear 
Alfv\'en modes with ${\bf B}%
=({\bf B}_0+{\bf b})/a^2$, where ${\bf B}_0=B_0\hat {{\bf z}}$, with $B_0=%
{\rm constant}$, ${\bf b}={\bf n}\bar b_0(z,t)$ and ${\bf v}={\bf n}\bar v%
_0(z,t)$, with ${\bf n}$ perpendicular to $\hat {{\bf z}}$. Recall that $|%
{\bf b}|$ is {\it not necessarily small} compared to $|{\bf B}_0|$. We
assume $\rho _b$ to be uniform (but not independent of $t$), use the momentum
equation (\Eq{eulerin}) and the induction equation (\ref{expind}), change to
conformal time $\tau $, and look for solutions in the form $\bar b_0(z,\tau)={%
\bar f}(\tau)e^{ikz}$, following the same procedure as in \Sec{alfven1}. (For the
rotational Alfv\'en-type mode, the gradient terms in \Eq{eulerin} do not
contribute). We obtain 
\EQA
&&{\frac{d^2{\bar f}}{d\tau ^2}}+\left[ aH+\bar D\right] {\frac{d{\bar f}}{%
d\tau }}+\bar \omega _0^2{\bar f}=0,  
\nonumber \\
&&\bar \omega _0=kV_A({\frac{4\rho _\gamma }{3\rho _b}})^{1/2};\quad \bar D%
=n_e\sigma _Ta({\frac{4\rho _\gamma }{3\rho _b}}).
\label{barf}
\ENA
Note that $\bar\omega_0 = kV_{Ab}$ is the baryonic Alfv\'en frequency,
where $V_{Ab} = B/(4\pi \rho_b)^{1/2}$ is the baryonic
Alfv\'en velocity.

The evolution of this non-linear Alfv\'en mode depends once again on the
relative strengths of the damping and driving terms. First
before decoupling, viscous damping completely dominates expansion damping;
$\bar D/aH= (4\rho_\gamma/3\rho_b)(D_H/l_\gamma) \gg 1$,
since the Hubble radius $D_H\equiv H^{-1} \gg l_\gamma $.
Also, 
\EQA
{\frac{\bar D}{\bar \omega _0}}&=&{\frac{(4\rho _\gamma /3\rho _b)n_e\sigma _Ta%
}{kV_A(4\rho _\gamma /3\rho _b)^{1/2}}} \nonumber \\
&\approx& 3\times 10^3({\frac{\rho
_\gamma }{\rho _b}})^{1/2}{\frac 1{k_p(t)l_\gamma (t)B_{-9}}}.
\label{Dbyomegafree}
\ENA
When a given mode enters the free-streaming limit we will have $%
k_p(t)l_\gamma (t)\sim 1$. So, for the field strengths $%
B_{-9}<(B_{crit}/10^{-9}G) \ll 1$ that we are considering, all the Alfv\'en
modes are strongly overdamped. 
Then one can again apply the
terminal-velocity approximation, where one neglects $d^2{\bar f}/d\tau^2$,
assuming that  $d{\bar f}/d\tau $ has adjusts itself to 
the `zero acceleration' solution.
Then, $\bar f$ is given by
\begin{equation}
\bar f(\tau )=\bar f(\tau _f)\exp -\left[ \int_{\tau _f}^\tau {\frac{\bar
\omega _0^2}{\bar D}}dt\right] =
\bar f(\tau _f)e^{-k^2/k_{fs}^2}, 
\label{freedam}
\end{equation}
with the free-streaming damping scale $k_{fs}^{-1}$ given by
\begin{equation}
k_{fs}^{-2}=V_A^2\int_{t_f}^t{\frac{l_\gamma (t)dt}{a^2(t)}}.  \label{kfsdef}
\label{kfs}
\end{equation}
Modes with a scale for the magnetic field $k^{-1}<k_{fs}^{-1}$, get damped
significantly during the free streaming evolution. We see that the damping
in this regime is similar to Silk damping, except that the usual Silk
damping integral within the exponential (cf. Eq. (\ref{sldamp}) ) is
multiplied by an extra factor of $(15/2)V_A^2 \ll 1$.
After recombination, the viscous
damping is subdominant, compared to expansion damping (since $l_\gamma $
exceeds the Hubble radius), and so can be neglected. So the largest scale to
be damped is found by evaluating $k_{fs}^{-1}$ at the recombination
redshift, using \Eq{kfs}.  
Assuming that the universe is matter dominated at recombination,
we get $k_{fs}^{-1}\approx (3/5)^{1/2}V_AL_S^C(t_r)$. Hence, the damping
scale is of order the Alfv\'en velocity times the Silk scale. 
More specifically we have $k_{\rm max} = k_{fs}(t_r)$, where
\begin{equation}
k_{\rm max} \simeq  235 \, {\rm Mpc^{-1}} B_{-9}^{-1}
\left ( \frac{\Omega_b h^2}{0.02} \right )^{1/2}
\left ( \frac{h}{0.7} \right )^{1/4},
\label{eq:kmaxn}
\end{equation}
where we have used some typical cosmological parameters for the numerical
estimate. 
The largest wavelength mode to be damped, say 
$L_D^A\equiv 2\pi k_{\rm max}^{-1} \sim 30$ kpc for the 
parameters used above.

For $B>B_{crit}$, we noted that the evolution becomes
compressible, and gravitationally unstable for scales larger than the
magnetic Jeans length, $\lambda _J$ or wavenumber smaller than $k_J$.
On scales smaller than $\lambda_J$, we expect that fast compressible motions
dominated by magnetic pressure, 
will drive oscillations close to the baryonic Alfv\'en frequency $\bar\omega_0$
as in \Eq{barf} for incompressible modes. 
Also their damping by free streaming photons is the same.
Thus we expect such oscillations are also
initially overdamped, in the
pre-recombination era. The damping scale for such motions will then be
similar to $k_{fs}^{-1}$, as deduced above. 
This expectation is borne out by the linearised calculations of \cite{JKO98},
by perturbing around a homogeneous zero-order magnetic
field. 
Clearly, more detailed computations are
needed to get the exact damping scales, in this case.

In summary, we see that the Alfv\'en mode oscillates negligibly
on Mpc Scales by recombination.
Unlike the compressional mode, which gets
strongly damped below the Silk scale, $L_{Silk}$
due to radiative viscosity \citep{Silk68}, 
the Alfv\'en mode behaves like an overdamped oscillator.
Note that for an overdamped oscillator there is one mode which
is strongly damped and another where the velocity starts from
zero and freezes at the terminal velocity until the damping becomes
weak at a later epoch. The net result is that
the Alfv\'en mode survives Silk damping for scales
$L_A > (V_A/c) L_{Silk} \ll L_{Silk}$, much smaller than the
canonical Silk damping scale \citep{JKO98,SB98a}

\section{ Nonlinear evolution of primordial fields }
\label{nonlin}

Nonlinear evolution of magnetic inhomogeneities becomes 
important when the Alfv\'en crossing
time on any scale, $\tau_{NL} = (kV_A(k))^{-1}$ is smaller than the
comoving Hubble time. Here $k$ is the comoving
wavenumber as before and $V_A(k)$ is the Alfv\'en velocity at 
$k$, which we will define below. On small enough scales
(large enough $k > k_{NL}$) this condition is satisfied, and modes
with $k > k_{NL}$ undergo nonlinear processing.
Such small scale processing is especially important when considering
the evolution of primordial magnetic fields originating
in early universe phase transitions, like the electroweak or
QCD phase transitions. 
To study such nonlinear
evolution of primordial fields requires direct numerical
simulations, although considerable insight can also
be got through semi-analytic arguments. We now consider these
aspects below, following mainly the arguments due to \cite{BJ04},
and supplementing them where needed with more recent developments.

The dynamics of the magnetic field and the fluid component is 
governed by Maxwell and the fluid equations, \Eq{expind}, \Eqs{energ}{eulers},
conveniently expressed in conformally transformed variables, during 
the radiation era.
Suppose some early universe process were to generate magnetic fields,
which could be described as a statistically homogeneous and
isotropic, Gaussian random field.
The Fourier components $\Fo{B}_i(\kk,\tau)$ 
of the conformally transformed magnetic field  
$\BBB^*(\xx,\tau)$, satisfy
$\langle\Fo{B}_{i}({\bf k},\tau)\Fo{{B}}^\dagger_{j}({\bf q},\tau)\rangle
=(2\pi)^3 \delta(\kk -\qq)M_{ij} (\kk,\tau)$, where 
\EQ
M_{ij}(\kk,\tau) = \left[
P_{ij}(\kk) M(k,\tau) 
- \frac{i\epsilon_{ijk} k_k}{2k^2} H(k,\tau) \right],
\label{magspectra}
\EN
where $\Fo{B}^\dagger_i$ is the complex conjugate of $\Fo{B}_i$. This implies
\EQ
\frac{\bra{\BBB^{*2}}}{2} = 
\int \frac{d^3k}{(2\pi)^3} M(k,\tau) 
\equiv \int dk \ E_M(k,\tau)
\label{energy1d}
\EN
\EQ
\bra{\BBB^*\cdot(\nabla \times \BBB^*)}
= \int \frac{d^3k}{(2\pi)^3} H(k,\tau) 
\equiv \int dk \ H_M(k,\tau),
\label{hel1d}
\EN
where we have defined respectively, 
the 1-D energy and helicity spectra 
\EQ
E_M(k,\tau) = \frac{k^{2}M(k,\tau)}{2\pi ^{2}}, \quad  
H_M(k,\tau) = \frac{k^{2}H(k,\tau)}{2\pi ^{2}}. 
\label{1dsepc}
\EN
We define the power in the magnetic energy per unit logarithmic 
interval in $k$-space as $M_k(\tau) = kE_M(k,\tau)$, and a 
corresponding Alfv\'en velocity, $V_A(k,\tau) = \sqrt{M_k/(4\pi (\rho^*+p^*))}$.

The fluid could start initially from zero peculiar velocity $\vv$,
although it is also possible more generally, that the process which
led to the initial magnetization also induced peculiar velocities.
The Lorentz force acts on it to drive further motions. Note
that the standard inflationary scalar perturbations are also driving
compressional fluid motions at the same time. The Lorentz
force adds to this compressional driving,
but not by a large degree as we saw in the previous section
for nano Gauss fields as smoothed on any scale.
On the other hand, importantly, the Lorentz force also has
a rotational component, which can drive the vortical component
of the velocity. 

It is convenient to consider the evolution focusing on 
each particular scale, which we will sometimes refer to as 
a mode on that scale,
characterised by the comoving wavenumber $k$
and proper wavenumber $k_p = k/a$.
Such a scale enters the Hubble radius first when its proper wavelength
is equal to the Hubble radius, that is when 
$a/k = 1/H$. In the radiation era, $a(\tau) \propto \tau$ and so
$aH = da/dt = (da/d\tau)/a = 1/\tau$. So in terms of the conformal time, a given
scale enters the Hubble radius when $k\tau =1$ and is within the
Hubble radius when $k\tau > 1$.

It is useful to compare the relative importance of 
different forces on the fluid.
The magnitude of the Lorentz force can be estimated as 
$\vert\BBB^*\cdot\nabla\BBB^*\vert \sim kV_A^2(k) (\rho^*+p^*)$.
Assuming that initially the viscous force is negligible, 
the Lorentz force generates
a rotational component of velocity 
\EQ
v_R \sim k V_A^2(k) \tau = \chi(k,\tau) V_A(k,\tau),
\label{vr}
\EN
at any time $\tau$, where $\chi(k,\tau) = kV_A(k)\tau$ is the
phase factor defined earlier, except that now it is scale dependent.
As the velocity grows with time, the viscous force, which is
of order $\nu^* k^2 v_R (\rho^*+p^*)$, grows. 
The non-linear term in the momentum equation
which is of order $kv_R^2 (\rho^*+p^*)$, also becomes important. 
Its importance relative to the viscous force is given by the
fluid Reynolds number $\Rey$. For the diffusion damping regime, we have
\EQ
\Rey = \frac{kv_R^2 (\rho^*+p^*)}{\nu^* k^2 v_R (\rho^*+p^*)} = 
\frac{v_R}{k \nu^*} = \frac{5 v_R}{k L_d}
\label{Red}
\EN
where we have defined the comoving mean free path $L_d = l_d/a$.
We expect in general that the mean free path $L_d$ to be much smaller
than the comoving scales of importance, i.e. $kL_d \ll 1$.
Thus $\Rey \gg1$ and viscous damping can initially be neglected. 

Note that, as estimated above, 
the nonlinear term in the momentum equation becomes
comparable to the Lorentz force when $v_R\sim V_A(k)$, 
and so when $\chi(k,\tau) \sim 1$.
Thus we can define a timescale $\tau_{NL} \sim (kV_A(k))^{-1}$
when for any mode the nonlinear term becomes important.
At this time the fluid Reynolds number is given by
$\Rey = (5 V_A(k)/(kL_d)$.
Further evolution of a mode will be decided by whether
$\Rey > 1$ (case I) or $\Rey < 1$ (case II).
In general, modes become nonlinear before viscosity is important
and case I obtains. Let us first consider this case.

\subsection{Decaying MHD turbulence in early universe}

Note that for any spectra with say $M(k,t) \propto k^n$, 
$V_A(k) \propto k^{(n+3)/2}$, and thus is a monotonically 
increasing function of $k$, provided $n > -3$. Spectra with
$n=-3$ or $M(k) \sim k^{-3}$ is the marginal scale invariant
case, where $V_A(k)$ is independent of $k$. We will
always consider spectra with $n > -3$. Correspondingly, 
$\tau_{NL}(k) \sim k^{-(n+5)/2}$ decreases with $k$ and
so large $k$ modes go nonlinear first. 
We denote by $k_{NL}$ the wavenumber of the mode which is going
nonlinear at a time $\tau$, i.e. which satisfies 
$k_{NL}V_A(k_{NL})\tau =1$. When a given scale
goes nonlinear, the energy can be transferred to larger wavenumbers
and the energy decays. 
This decay depends on the form of the spectrum, or value of $n$,
and also whether magnetic helicity is present or not. 
This is because magnetic helicity is better conserved
than magnetic energy, and thus the decay of energy is constrained by this
conservation law \citep{Biskamp2003} (see \Sec{helicity_basic}).
Let us first consider non-helical magnetic fields with
a magnetic spectrum $M(k,t_i) = C k^n$. 

An inhomogeneous magnetic field by itself can drive MHD turbulence
due to the effect of the Lorentz forces. 
Due to nonlinear processing, the energy $M_k(\tau)$,
will decrease with $k$ for $k > k_{NL}$.
Thus $k_{NL}$ is at this stage also the approximately
the coherence scale of the field $k_{coh}$, provided $n > -3$.
The magnetic energy in a fixed comoving volume $\totalE_M$ 
for such a spectrum then scales as $\totalE_M \propto k_{NL}^{n+3}$.
Since velocities of order $V_A(k_{NL})$ are induced
by the Lorentz force at the nonlinear scale $k_{NL}$, 
the energy decay rate scales as 
\[
\frac{d\totalE_M}{d\tau} \propto -\frac{\totalE_M }{\tau_{NL}}
\propto -\totalE_M^{(3n+11)/2(n+3)},
\]  
where we have used the relation $\tau_{NL}(k) \sim k_{NL}^{-(n+5)/2}$
deduced above.
Integrating this equation then leads to a decay law for the magnetic field
similar to the decay law for hydrodynamic turbulence \citep{Davidson04},
$\totalE_M \propto \tau^{-2p}$, $L_{M} \propto k^{-1}_{NL}
\propto \tau^{q}$, with
\EQ
p = \frac{(n+3)}{(n+5)}, \quad 
q=\frac{2}{(n+5)}, \quad p+q=1.
\label{nonhel_decay}
\EN
The rate of energy decay and the growth of the magnetic correlation scale
depends on the spectral index $n$. For causally generated fields,
using the fact that magnetic fields satisfy $\nabla\cdot\BBB=0$,
\citet{DC03} have argued that 
the long wavelength tail of the magnetic spectrum must have
$n=2$ to maintain analyticity. In this case we have
$\totalE_M \propto \tau^{-10/7}$ and $L_M \propto \tau^{2/7}$.
However the velocity field could have
$n=0$ (a white noise 3-d power spectrum), as it is not strictly
divergence free. Then for magnetic fields generated by the presence of this
shallower tail of the velocity field, one may envisage a
slower decay law \citep{JS11}. For example, if the presence of a
turbulent velocity field, with $n=0$, maintains the same long wavelength 
spectrum for the magnetic field near the nonlinear scale,
then one would have a slower decay of the magnetic energy 
$\totalE_M \propto \tau^{-6/5}$ and a more rapid growth of its correlation
scale $L_M \propto \tau^{2/5}$.  
Direct numerical simulations of the decay of a nonhelical field
tend to find that the energy decays even more slowly,  
$\totalE_{M} \propto \tau^{-1}$ \citep{BM2000} to 
$\totalE_{M} \propto \tau^{-0.9}$ \citep{KTBN13,BKT15}.

\begin{figure}[t!]
\includegraphics[width=.48\textwidth]{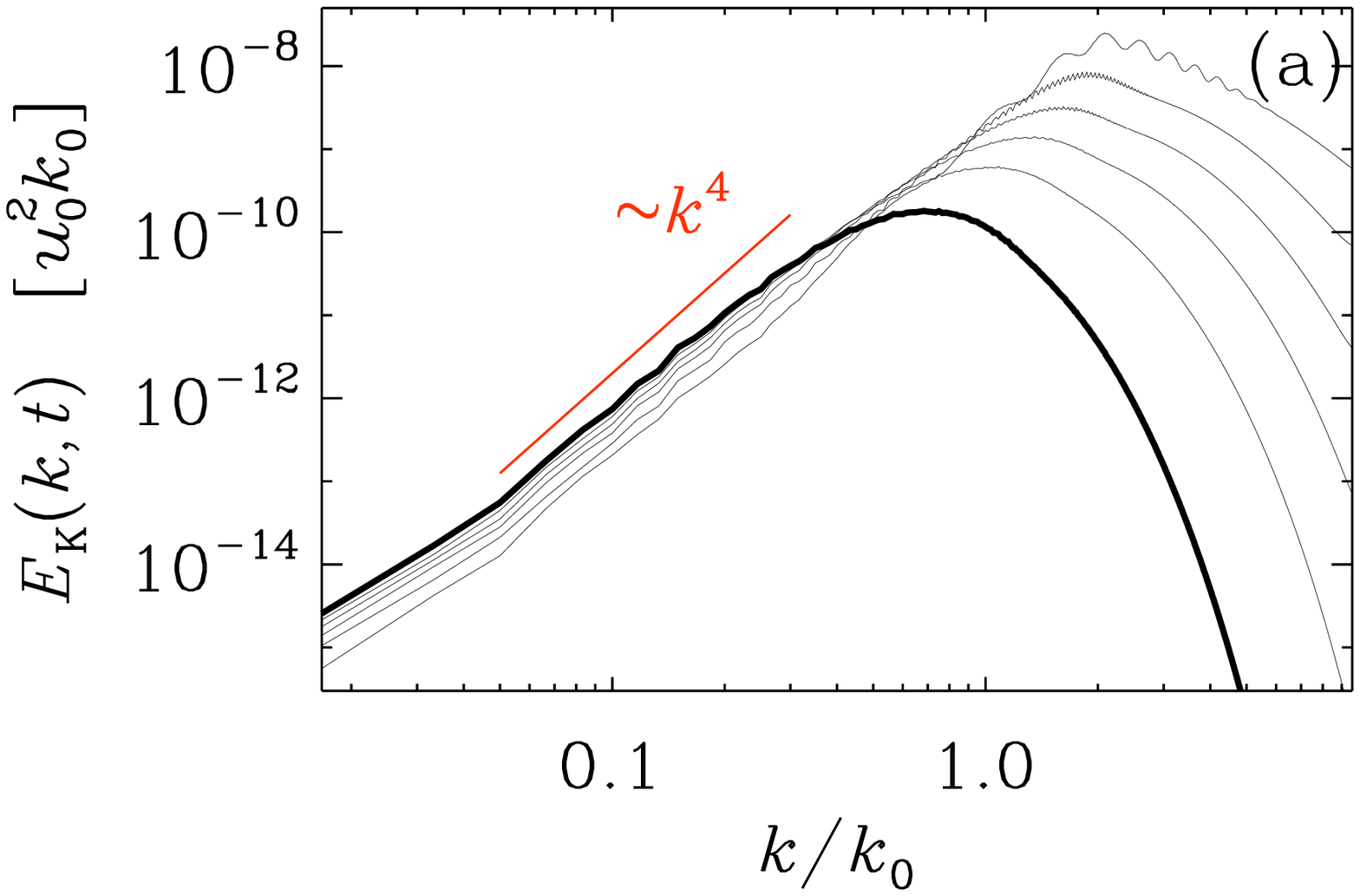}
\\
\includegraphics[width=.48\textwidth]{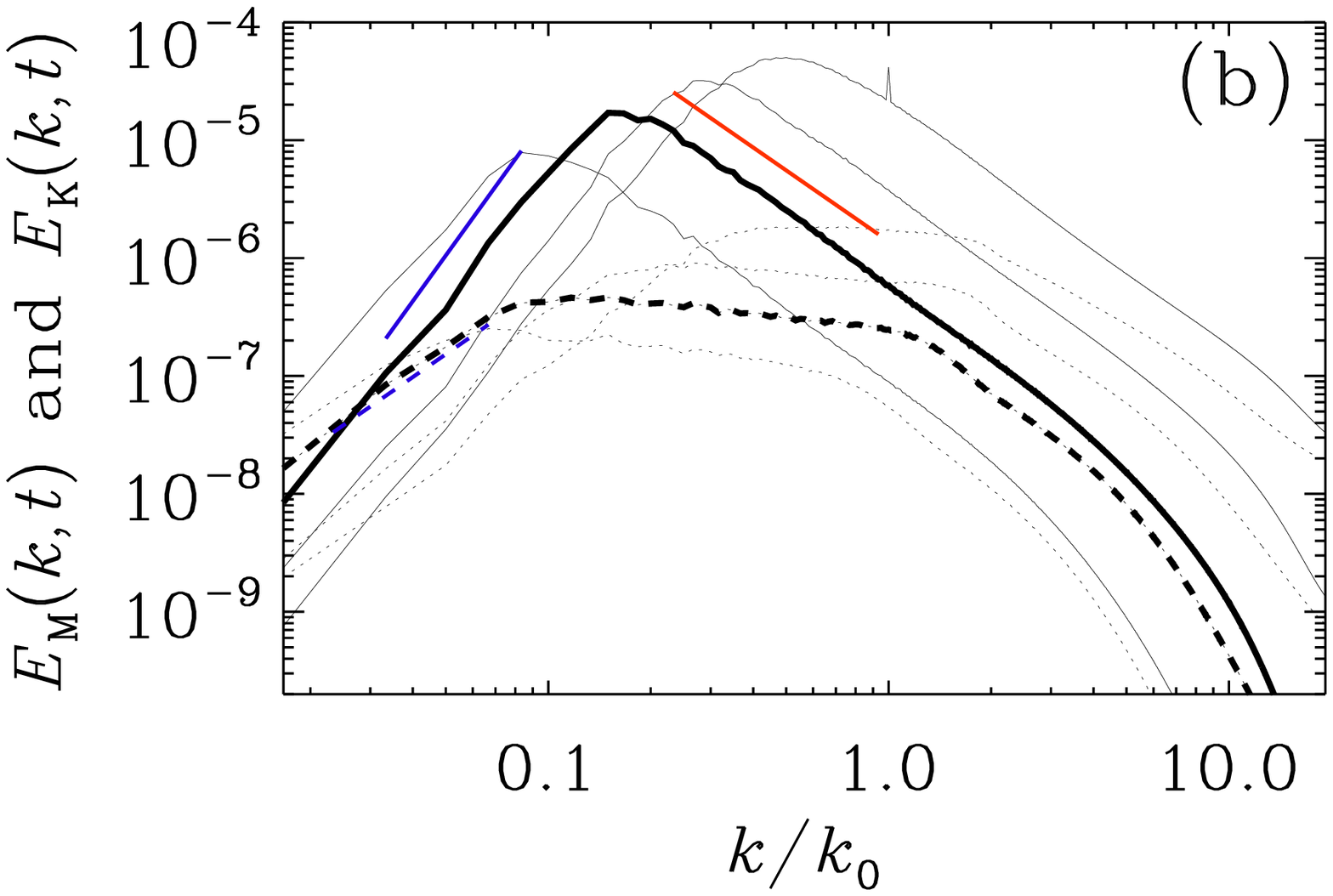}
\\
\includegraphics[width=.48\textwidth]{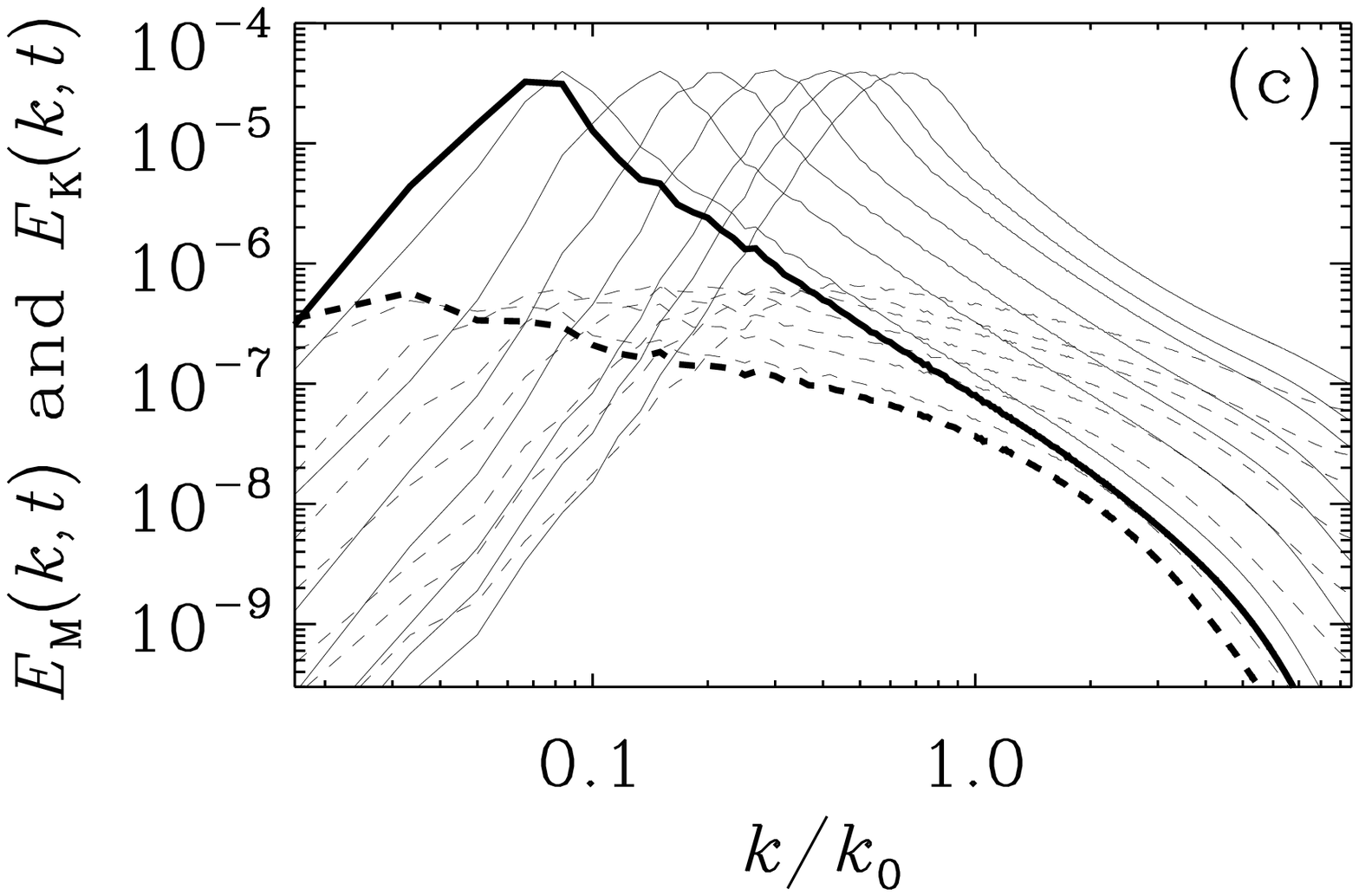}
\\
\caption[]{
Panel (a) shows the kinetic energy spectra in a 
simulation of decaying hydrodynamic turbulence, while
panels (b) and and (c) show respectively, 
the case of decaying MHD turbulence without and with helicity.
The evolving magnetic energy spectra in these latter panels are shown 
by solid lines while the kinetic energy spectra by dotted (panel b)
or dashed (panel c) lines. The solid and dashed straight lines
in panel (b) show $k^4$ and $k^2$ slopes. 
In the hydrodynamic case the coherence
scale increases by the progressive decay of smaller and smaller $k$ modes.
A clear sign of inverse cascade is seen for the MHD cases, which
is more pronounced when helicity is present. Adapted with permission 
from \citet{BKT15};
courtesy Axel Brandenburg.
}\label{axel15}
\end{figure}

\Fig{axel15} shows the results of direct numerical simulations (DNS) by 
\citet{BKT15} for the evolution of the magnetic and kinetic
spectra obtained in decaying hydrodynamical and MHD turbulence.
The hydrodynamical simulation (panel a) 
shows the expected behaviour of the energy spectrum preserving its
shape at small $k$, with a cut-off at 
progressively smaller and smaller $k$. The hydromagnetic simulation
without helicity (panel b), however seems to already show an inverse cascade of
energy to larger scales.
This could be the reason for the slower
decay obtained.
Such an inverse cascade behaviour for nonhelical MHD turbulence 
is surprising, as magnetic helicity conservation is usually regarded
as the key to produce an inverse cascade.
However it is also seen in simulations by \citet{Z14} of free decay
of nonhelical turbulence in a relativistic fluid.
An understanding of what exactly causes
this inverse cascade is still not completely clear
and is under active discussion; see the discussions
in \citet{BKT15,O15,C15}.
One possibility is the shallower slope of the kinetic
energy spectrum, as can be seen in panel (b) of \Fig{axel15}, 
and its effect on amplifying the magnetic fields at small $k$. 

\subsection{Helical field decay}

The most interesting difference between the decay of MHD turbulence 
and the purely hydrodynamic case occurs in case the
field is helical, due to
magnetic helicity conservation. 
Consider for example the decay of a fully helical magnetic field
with energy $\totalE_M$ and helicity $\totalH_M$ in a fixed co-moving volume $V$ 
and integral scale defined by
\EQ
L_M \totalE_M = \totalH_M.
\EN
Assume that there is rough equipartition between
the kinetic energy $\totalE_K$ and magnetic energy, and the two fields
have the same scale. Then on dimensional grounds, 
the total energy $E=\totalE_K+\totalE_M$ changes
at a rate $\dot E \sim - E/(L_M/E^{1/2})$. Using $L_M = \totalH_M/\totalE_M$ 
we get 
\EQ
\frac{d \totalE_M}{d\tau} = - \frac{\totalE_M^{5/2}}{\totalH_M}.
\EN
Noting that $H_M$ is approximately conserved during the decay
of the energy, this can be integrated to give
\EQ
\totalE_M \propto \tau^{-2/3}, \quad L_M \propto \tau^{2/3}.
\label{hel_scaling}
\EN
Thus the presence of helicity slows down the decay of the field even
further, and
more importantly, the coherence scale grows faster than in the nonhelical case.
Some earlier numerical simulations seem to show an even slower decay
with $E_M \propto \tau^{-1/2}$, $L_M \propto \tau^{1/2}$ 
\citep{BM1999,BM2000}i, while that of \cite{CHB01} showed
the expected decay of $E_M$ but $L_M \propto \tau^{1/2}$. 
However, more recent simulations \citep{BJ04,KTBN13,BKT15},
are consistent with
the scaling law given in \Eq{hel_scaling}.
Panel c of \Fig{axel15} gives the results from the DNS by
\citet{BKT15}, which shows clearly the inverse cascade
of the magnetic field to larger and larger scales.

The effect of having a partial helicity has also been discussed 
by \citet{BJ04}.
It turns out the field decays as if it were nonhelical conserving
helicity. As the energy deceases conserving helicity, the field is
eventually driven to the fully helical state and subsequently follows
the decay law for a helical field.
Suppose one starts with the helical fraction $h_g$ at generation, then
noting that the initial helicity $H_g$ 
is conserved while energy decays, fractional
helicity subsequently scales as
$h = H_g/(\totalE_M L_M) \propto h_g (\tau/\tau_0)^{2p-q} 
\propto h_g (\tau/\tau_0)^{3p-1}$.
Then the field will become fully helical at an epoch 
$\tau_h = \tau_0 (1/h_g)^{1/(3p-1)}$, and after this decay as if
it were fully helical.

\subsection{The effect of viscosity}

The importance of the viscous force increases
secularly with time as the mean free path of the least coupled
particle $l_d$, and hence $\nu$ increases. Recall that the fluid Reynolds
number $\Rey(k)$ at any scale $k$, 
is given by
\EQ
\Rey(k) = \frac{v_R(k)}{k\nu*} = \frac{(V_A(k) k \tau) 5 V_A}{k l_d^C}
= 5 V_A^2(k) \left(\frac{\tau}{L_d}\right).
\EN
Typically $L_d$ increases faster than $\tau$ as the universe expands;
for example for photons 
$L_d = (n_e \sigma_T a)^{-1} \propto a^2 \propto \tau^2$. Then the fluid 
Reynolds number $\Rey(k)$ decreases secularly with time. 
When $L_d$ increases to a value such that 
$k_{NL} L_d \sim 5 V_A(k_{NL})$, then $\Rey(k)$ drops 
to a value of order unity, and viscous damping 
damps the modes on the nonlinear scale $k_{NL}$ itself.
Both the velocity and magnetic field on this scale are expected
to be damped. However, very rapidly $L_d$ increases further so that
$\Rey(k) \ll 1$ on all larger scales (or $k<k_{NL}$), and 
the motions become overdamped. The Lorentz force induces 
then a fluid velocity got by balancing it against friction, that is
$\nu^* k^2 v_R \sim k V_A^2(k)$, which implies 
$v_R \sim \Rey(k) V_A(k) \ll V_A(k)$. This small induced velocity
does not distort the field significantly,
as it leads to a fractional displacement of order 
$k v_R \tau \sim \Rey(k) (k/k_{NL}) \ll 1$.  Therefore, the magnetic field
inhomogeneities are now frozen in this strong damping regime,  
a feature is very similar to what obtains for the linear
Alfv\'en waves discussed earlier.

As the comoving mean free path $L_d$ further increases, 
it becomes larger than the coherence scale $k_{coh}^{-1}$ 
of the magnetic field, i.e. $k_{coh} L_d > 1$. Then 
one transits from diffusive damping to the free stream damping
regime. In this regime 
the viscous force (in \Eq{eulers}), 
is given by $\FF_d = a(\tau)\FF_D = 
-(4/3) \rho_d (a/l_d)\vv = -(4/3) \rho_d\vv/L_d$,
where $\rho_d$ is the density of the particle species providing the drag,
and we have assumed it to be relativistic. This species could 
be neutrinos just before neutrino decoupling and photons 
just before recombination. We also define $\rho_R$ as 
the density of all the species coupled to the particle
providing the drag. 
As the damping will be large
when free stream damping first becomes important, 
the fluid velocity is given by the balance between the Lorentz force
and viscous force, that is
\EQ
k M_k \sim \frac{\rho_d v_R}{L_d} \quad \to \quad
v_R \sim k L_d V_{Ad}^2 .
\label{vrd}
\EN
Here $V_{Ad}$ is the Alfv\'en velocity defined
using the density of the particle providing the drag. For
the modes on the coherence scale, we have 
$k_{coh} L_d \sim 1$ intially and thus $v_R \ll 1$ as 
$V_{Ad} \ll1$ in general. The corresponding fluid Reynolds number 
during the free streaming regime is 
\EQ
\Rey(k) \sim \frac{\rho_R v_R^2}{(\rho_d v_R/L_d)}
= \frac{\rho_R}{\rho_d} (k L_d V_{Ad})^2. 
\label{reyk}
\EN
This is again much smaller than unity when free stream damping begins as
$k_{coh} L_d \sim 1$. Thus when free stream damping begins
for any mode, the fluid velocity which was already small during the
diffusive damping, will remain small. 
As the universe expands and $L_d$ increases, the free stream 
damping becomes weaker and the fluid velocity $v_R$ increases.
As $v_R$ increases the field will be advected due to flux freezing
and in a time $\tau$ all modes which satisfy $k v_R \tau \sim 1$ 
can oscillate significantly in the presence of free stream damping and
as a result be significantly damped. Thus the coherence scale
of the field now increases due to the damping and is given by
\EQ
k_{coh} v_R \tau \sim  k^2 L_d V_{Ad}^2 \tau \sim 1 \ \to \  
k_{coh}^{-1}  = V_{Ad} (L_d \tau)^{1/2}
\EN
where we have substituted $v_R$ from \Eq{vrd}. 
This is exactly the Silk damping scale of the Alfv\'en wave modes
discussed earlier.
 
Note that the velocity can at most increase to be of order the Alfv'en velocity
after which the Lorentz force due to the magnetic field acts to 
restore the fluid motion. This happens when $L_d$ has grown such that
$v_R \sim V_{A}$ or using \Eq{vrd}, 
$kL_d V_{Ad} \sim \sqrt{\rho_d/\rho_R}$. At this stage the mode 
which can suffer significant
damping satisfies, $k_{coh} v_R \tau = k V_{A} \tau \sim 1$, and so is
actually the mode at the nonlinear scale $k_{NL} \sim 1/(V_{A} \tau)$ itself. 
At the same time, the condition $kL_dV_{Ad} \sim \sqrt{\rho_d/\rho_R}$
also implies from \Eq{reyk}, that the fluid Reynolds number
grows to $\Rey(k) \sim 1$,
with the viscous evolution transiting again to turbulent decay.
Thus as pointed out by \citet{BJ04}, the magnetic coherence scale 
at the end of the viscous period,  
grows to the value it would have had, as if there had been no viscosity
dominated period of evolution at all. The viscous period in this sense
just delays the dissipation of the field. 
It is not of course obvious that $V_{A}$ or $k_{NL}$ will individually
tend to the value they would have had, if the viscous evolution interval
was not present, although their product $V_{A}k_{NL} \sim 1/\tau$
at the end of this phase. However, the simulations of 
\citet{BJ04} seem to suggest this is a good approximation.

The above picture holds
for example during the epochs when neutrino viscosity is important.
A similar evolution also holds during the epoch when photon viscosity
is important.
However, as the
scatterers, electrons and positrons become non-relativistic
below $T < m_e$, their number decreases rapidly with a corresponding
rapid increase of the photon mean free path. Also, after recombination,
the photon mean free path increases so rapidly $\sim$ Mpc to
the Hubble radius, that free stream damping
by photons is abruptly switched off. This leads to a smaller coherence
scale and a larger field strength at the end of the photon damping era,
compared to the case where the flow would have stayed turbulent all along.

\subsection{Summary}

The nonlinear evolution of primordial fields can be summarized
as follows. During turbulent evolution, the field
decays satisfying $B \propto \tau^{-p}$, $L_M \propto \tau^q$,
with $p=(5/7, 3/5, 1/3)$ and $q=1-p$, respectively  
for the $n=2$, $n=0$ nonhelical cases and for the fully helical case. 
And when viscosity dominates, the field first freezes during diffusive
damping and decays more rapidly during free-stream damping to a value as if
the viscous epoch was not present at all. Partially helical fields
first decay as if they are nonhelical conserving helicity till
the field becomes fully helical; then they decay as for the
fully helical case. 

These laws are applicable to the scaled
comoving magnetic fields, with $\tau$ the conformal time.
In the radiation dominated epoch, we have $a(t) \propto t^{1/2}$
and so $\tau \propto t^{1/2} \propto a(\tau)$, and we can
map the above power law scalings directly to that with the
expansion factor $a$ itself. On the other hand
in the matter dominated epoch, one needs a 
a different set of transformations to map MHD in the expanding
universe to flat space \citep{BJ04}, where the scaled 'conformal' time variable
$\tilde{t}$ satisfies, $d\tilde{t} = dt/a^{3/2} = dt/t$, 
with $a(t)\propto t^{2/3}$. Thus $\tilde{t} \propto \ln(t) \propto (3/2) \ln a$,
and any power law decay of $B$ or growth of $L_M$ in scaled 
$\tilde{t}$ time,  corresponds to only logarithmic decay/growth in terms
of physical time, or scale factor. Thus evolution of the field
virtually freezes at the end of the radiation domination epoch,
as also found in the detailed numerical calculations of \citet{BJ04}.   

These ideas have been put together by several authors to deduce the
present day strength and coherence scale of primordial magnetic fields 
causally generated at the QCD and electroweak
phase transitions, after it has gone through epochs of turbulent and
viscous decay.
An important general feature as discussed above is that
in turbulent epochs the field satisfies the condition $V_A k \tau \sim 1$.
\citet{BJ04} note that this relation is also applicable after reionization,
and so derive a general condition for the present day field strength
and coherence scale,
\EQ
B_0 \approx 5 \times 10^{-12} {\rm G} \left(\frac{L_c}{{\rm kpc}}\right).
\label{BJlimit}
\EN
A simple estimate for the field itself, 
can be got assuming the turbulent decay scaling
from generation era ($a=a_g$, $T=T_g$) to end of radiation era 
($a=a_{eq}$, $T \sim 1$ eV)
and subsequent freezing of the comoving field strength.
For nonhelical fields this gives,
$B_0 = (a_{eq}/a_g)^{-p}B_g$. Using the conservation
of $aTg^{1/3}$, adopting $g\sim 4$ at the equality epoch,
$g \sim 100$ at the generation epoch, and $p=-3/5$ 
gives 
\EQ
B_0\sim 10^{-7} B_g r_{-2}^{1/2} T_{100}^{-3/5}
\sim 3 \times 10^{-14} r_{-2}^{1/2} T_{100}^{-3/5} \ {\rm G},
\label{nonhelets}
\EN
where $r_{-2} = (r_B/0.01)$, $r_B$ is as in \Eq{Bestimate} 
and $T_{100} = T_g/(100)$ GeV.
For the partial helical case, with initial helical fraction $h_g$,
we have $B= B_g (a_{eq}/a_h)^{-1/3} (a_h/a_g)^{-p}
= B_g (a_{eq}/a_g)^{-1/3} h_g^{1/3}$, where $a_h$ is expansion
factor corresponding to $\tau_h$ when the field becomes
fully helical. Note that this result is independent of $n$, and
putting in numbers, 
\EQA
B_0 &\sim& 3.4 \times 10^{-4} B_g T_{100}^{-1/3} 
h_g^{1/3} \nonumber \\
&\sim& 0.1 r_{-2}T_{100}^{-1/3} h_g^{1/3} \ {\rm nG}.
\label{helest}
\ENA
The corresponding coherence scales can be got from using
$B_0$ in \Eq{BJlimit}.

These simple estimates agree reasonably with
the more detailed estimates by \citet{BJ04}. Taking $n=0$, they give
\EQA
B_0 &=&7.4\times 10^{-11} {\rm G} \ r_{-2}^{1/2} \ T_{100}^{-1/3} \ h_g^{1/3} 
\ {\rm (partially \ helical)}
\nonumber\\
B_0 &=&6\times 10^{-14} {\rm G} \ r_{-2}^{1/2} \ T_{100}^{-3/5} 
\ {\rm (nonhelical)}.
\label{Brobi}
\ENA
For $T_{100} =1$ 
as would be relevant to the EWPT 
and $h=1$, one gets from \Eq{Brobi},
$L_c \sim 15$ kpc
and $B_0 \sim 0.07$ nG. If fully helical fields could be
generated at the QCD phase transition, with 
generation at $T\sim 100 MeV$, the corresponding values 
for $L_c$ and $B_0$ are $(100 GeV/100 MeV)^{1/3} \sim 10$ times 
larger. For a nonhelical fields, 
assuming $n=0$ $B_0$ and $L_c$ would be 
$\sim 10^3$ smaller, compared to the fully helical case.
But predicting how much exactly, will depend 
on a better understanding of the 
possible inverse cascade seen for nonhelical fields. 
Nevertheless,  causally generated primordial fields surviving from
the early universe phase transitions, could have interesting 
strengths and coherence scales to influence physical processes
in the universe.

\section{CMB signals due to Primordial magnetic fields}
\label{cmb}

As the Universe expands it cools sufficiently such that 
below $z \la 1100$, the primeval ionized plasma recombines 
to form neutral atoms. The photon mean free path becomes larger
than the current Hubble radius and they can directly
free stream to the observer. These photons are seen as the CMB photons
today and they dominantly reflect the physical conditions
of the universe from the epoch when they last scattered,
known as the LSS or last scattering surface.
One of the most important ways of detecting or constraining the
existence of primordial magnetic fields is via the observations
of the CMB temperature and polarization
anisotropies (see \citet{S06,Durrer07} for reviews).
The signals that could be searched for include
excess temperature (T) anisotropies at both large angular scales and scales
below the Silk damping scale, E and B-mode polarization, non-Gaussian
statistics and Faraday rotation effects. 

CMB anisotropies in general arise in two ways. Firstly,
spatial inhomogeneities around the LSS lead to the
`primary' anisotropies in the CMB temperature as seen
at present epoch. Furthermore, variations in
intervening gravitational and scattering effects,
which influence the CMB photons as they come to us from
the last scattering surface, can lead
to additional secondary anisotropies
(\cf\ \citet{Padmanabhan02,Dodelson03,S05} for pedagogical reviews
of CMB anisotropies).

The CMB is described by its brightness (or intensity)
distribution. Since the spectrum of the CMB brightness, seen 
along any direction on the sky ${\bf n}$, is very close to 
thermal, it suffices in most cases to give the temperature $T({\bf n})$.
The temperature is very nearly uniform
with fluctuations $\Delta T(\nn)$ at the level of
$10^{-5} T$, after removing a dipole contribution.
It is convenient to expand the temperature anisotropies
$\Delta T(\nn)/T = \Theta(\nn)$ at the observer in 
spherical harmonics
(with the dipolar contribution, predominantly
produced by the Earth's motion in the CMB frame, subtracted)
\begin{equation}
\frac{\Delta T}{T}(\theta,\phi)
= \sum_{l=2}^{\infty}\sum_{m=-l}^{m=l} a_{lm} Y_{lm}(\theta,\phi), 
\end{equation}
with $a_{lm}^* = (-1)^m a_{l-m}$ as $\Delta T/T$ is real.
In the standard picture, the universe is assumed to have
evolved from density fluctuations initially described by a
random field, which is almost Gaussian. 
In this case $a_{lm}$'s are also random variables
with zero mean and a variance completely described by their power spectrum,
\EQ
\bra{a_{lm}a^*_{l'm'}} = C_l\delta_{ll'}\delta_{mm'}.
\label{cldef}
\EN
Here we have assumed also the statistical isotropy of $\Theta(\nn)$ field
because of which the power spectrum is independent of $m$.
Theoretical predictions of CMB anisotropy are then compared
with observations by computing the $C_l$'s or the
correlation function $C(\alpha) = \bra{\Theta(\nn)\Theta(\vec{m})}$, 
where if we have statistical isotropy, $C$ depends only on
$\cos\alpha = \nn\cdot\vec{m}$.
From \Eq{cldef} and the addition theorem for the spherical
harmonics, we have
\EQA
C(\alpha)&=&  \sum_{lm} \sum_{{l^\prime} {m^\prime}} \bra{a_{lm}
a^*_{{l^\prime} {m^\prime}}}Y_{lm} Y^*_{{l^\prime } {m^\prime }}
\nonumber \\ 
&=& \sum_l C_l \frac{2l +1}{4\pi} P_l(\cos\alpha).
\label{cobs}
\ENA
The mean-square temperature anisotropy, 
$\bra{(\Delta T)^2} = T^2 C(0)$ is  
\EQ
\frac{\bra{(\Delta T)^2}}{T^2} = \sum_l C_l {2l +1\over 4\pi} \approx 
\int \frac{l(l+1)C_l}{2\pi} \ d \ln l
\EN
with the last approximate equality valid for large $l$, 
and so $l(l+1)C_l/ 2\pi$ is a measure of the power
in the temperature anisotropies, per logarithmic
interval in $l$ space. 
This particular combination is used because
scale-invariant potential perturbations generate anisotropies,
which at large scales (small $l$) have a nearly constant
$l(l+1)C_l$. A convenient characterization
 of the scale-dependent temperature anisotropy
is $\Delta T(l) = T [l(l+1)C_l/2\pi]^{1/2}$. 
One can roughly set up a correspondence between 
angular scale at the observer $\alpha$, the corresponding  $l$ value
it refers to in the multipole expansion of $\bra{\Theta^2}$ and also 
the corresponding co-moving wavenumber $k$ of a perturbation
which subtends an angle $\alpha$ at the observer. One has 
$(\alpha/1^{o}) \approx (100/l)$ and $l \approx kR^*$ where
$R^*$ is the comoving angular diameter distance to
the LSS and is $\sim 10h^{-1}$ Gpc,
for a standard $\Lambda$CDM cosmology.
The predicted CMB temperature anisotropy $\Delta T^2(l)$ is shown
as topmost solid line, top left panel in \Fig{shaw} 
(and $\Delta T(l)$ in \Fig{vec} as a dashed-double-dotted line), 
for a standard $\Lambda$CDM model.

Primordial magnetic fields induce a variety of additional
signals on the CMB. 
First, a very large scale (effectively homogeneous) field
would select out a special direction, lead to
anisotropic expansion around this direction,
hence leading to a quadrupole anisotropy
in the cosmic microwave background (CMB).
\citep[see, for example,][]{Thorne67}.
The degree of isotropy of the CMB then implies a limit
of several nG on the strength of such a field redshifted to the
current epoch \citep{BFS97,ADFV11}.

Primordial magnetogenesis scenarios on the other hand, as we discussed above,
generally lead to tangled fields, plausibly Gaussian random,
characterized by say a spectrum $M(k)$.
The scalar, vector and tensor parts of the
perturbed stress tensor associated with such primordial magnetic fields
lead to corresponding metric perturbations, including gravitational
waves. Further the compressible part of the Lorentz force leads to
compressible (scalar) fluid velocity and associated density
perturbations, while its vortical part leads to
vortical (vector) fluid velocity perturbation.
These magnetically induced metric and velocity perturbations lead
to both large and small angular scale anisotropies in the CMB temperature
and polarization.
A helical field can also lead to odd parity, $T-B$ and $E-B$ correlations,
not expected for inflationary scalar perturbations.

In addition, the presence of tangled magnetic fields
in the intergalactic medium
can cause Faraday rotation of the polarized component
of the CMB, leading to the generation of new B-type
signals from the inflationary E-mode signal. Their damping
in the pre-recombination era can lead to spectral distortions
of the CMB \citep{JKO00},
while their damping in the post-recombination
era can change the ionization and thermal history of the Universe \citep{Sethi_KS05}.

\subsection{Scalar modes}

The scalar contribution has been the most subtle
to calculate, and has only begun to be understood by several groups
\citep{GK08,YIKM08,FPP08,SL10,BCD13}.
Three types of contributions to the 
curvature perturbation $\zeta$ are in principle possible.
(i) A mode known as the passive mode, 
which arises before neutrino decoupling, 
sourced by the magnetic anisotropic stress.
(ii) A compensated mode which remains after the growing neutrino
anisotropic stress has compensated the magnetic anisotropic
stress on large scales (cf. \cite{SL10} for detailed discussion).
(iii) In addition \cite{BCD13} have stressed the possibility of a
constant contribution to the curvature perturbation, which 
arises when magnetic fields are generated during
inflation. 

Specifically, the stress tensor (space-space part of the 
energy-momentum tensor) for magnetic fields in terms of the present 
day magnetic field value $ {\bf b_0}$ is
\EQ
T^i_j({\bf x}) = \frac{1}{4 \pi a^4} \left( \frac{1}{2} b_0^2({\bf x}) \delta^i_j - b_0^i({\bf x}) {b_0}_j({\bf x}) \right)
\label{Tij_b_nought}
\EN
In Fourier space, the product of magnetic fields becomes a convolution
\EQ
S^i_j({\bf k}) = \frac{1}{(2 \pi)^3} \int {b}^i({\bf q}) {b}_j({\bf k-q}) d^3 {\bf q}
\label{Tij_convolution}
\EN
\EQ
T^i_j({\bf k}) = \frac{1}{4 \pi a^4} \left( \frac{1}{2} 
S^{\alpha}_{\alpha}({\bf k}) \delta^i_j - S^i_j({\bf k}) \right),
\label{Tij_b_nought_fourier}
\EN
where $\bb(\qq)$ is the Fourier transform of $\bb_0(\xx)$.
This can be expressed in terms of the magnetic perturbations to the energy-momentum tensor as
\EQ
T^0_0 = -\rho_\gamma \Delta_B, \quad 
T^i_j({\bf k}) = p_{\gamma} \left( \Delta_B \delta^i_j + {\Pi_B}^i_j \right)
\label{Tij_perturb}
\EN
where $\Delta_B$ and ${\Pi_B}^i_j$ are the perturbations in the energy density and anisotropic stress, respectively, as 
defined in \cite{SL10}. Also $\rho_\gamma$ and $p_{\gamma}$ are
respectively the radiation energy density and pressure, and
as they are $\propto 1/a^4$, $\Delta_B$ and ${\Pi_B}^i_j$
are constant. The anisotropic stress can be decomposed to scalar, 
vector and tensor contributions.
The amplitude of the anisotropic stress for scalar perturbations 
is given by $\Pi_B(\kk)$, got by 
applying the relevant projection operator to $T^i_j({\bf k})$
\citep{BC05}.
\EQ
\Pi_B(\kk) = -\frac{3}{2} \left( \kkk _i \kkk_j - \frac{1}{3} \delta_{ij} \right) \Pi_B^{ij}.
\EN
Note that $\Pi_B(\kk)$ of \cite{SL10} is equal to 
$ -\tau^S(\kk)$ of \cite{BC05}.

The magnetic stresses are non-linear in the field but we assume that they are
always small compared to the total energy density and pressure of the photons,
baryons etc. Thus allowing a purely linear treatment of the perturbations. Hence
the scalar, vector and tensor perturbations decouple and evolve independently.
Prior to neutrino decoupling, the only source of anisotropic stress is 
the magnetic field. Once the neutrinos decouple, the anisotropic stress 
due to neutrinos also contributes but 
with an opposite sign to that of the magnetic field, thus compensating 
the contribution from the magnetic field on large scales \citep{L04}. 
The post inflationary evolution thus leads to two types of modes \citep{SL10}. 
The first one, the passive mode, is an adiabatic-like 
mode but has non-Gaussian statistics. It grows logarithmically 
in amplitude between the epochs of magnetic field generation and 
neutrino decoupling, driven by the magnetic anisotropic stress, 
but then evolves passively after neutrino decoupling. 
This behaviour has also been confirmed in \citet{BC10} 
in the context of deriving the magnetic Sachs-Wolfe effect for a 
causally generated primordial magnetic field. The second, more 
well-studied perturbation \citep{GK08,YIKM08,FPP08,PFP09}, 
is called the compensated mode, which is sourced by the
residual anisotropic stress and magnetic energy density. 
And as stated above, a third mode may 
arise due to the effects of magnetic stresses on the curvature during
inflation itself \citep{BCD13}  

The final curvature perturbation as derived using the conformal
Newton gauge is given by equation (32) 
of \citet{BCD13} (see also \citet{SL10}),
\EQ
\zeta=\zeta_{inf} +\zeta_{MI} + \frac{\Omega_B}{4} 
- \Omega_{\Pi} \left[ \ln \left( \frac{\tau_{\nu}}{\tau_B}\right) 
-\frac{1}{2}\right].
\label{curvature}
\EN
Here $\Omega_B$ and $\Omega_{\Pi}$ defined by \citet{BCD13}, 
are proportional to $\Delta_B$ and $\Pi_B$ defined in \citet{SL10}.
Specifically they are normalised by the energy density of the inflaton,
rather than radiation. (The power per unit logarithmic interval
defined in \citet{BCD13} is also a factor $2\pi^2$ larger than the standard
definition used by say \citet{SL10} and in CAMB code).
The $\zeta_{inf}$ term represents the standard inflationary contribution 
to the curvature perturbation, while $\zeta_{MI}$ is the result
of magnetic stresses during inflation, as estimated by \cite{BCD13}.
The passive mode contribution is the 
term, $\zeta_{pas}= -\Omega_{\Pi}[\ln (\tau_{\nu}/\tau_B) -1/2]$. 
It incorporates the logarithmic 
growth in the curvature driven by the uncompensated 
magnetic anisotropic stress, between the epochs of magnetic field 
generation $\tau_B$ and neutrino decoupling $ \tau_{\nu}$. 
In the radiation-dominated 
era the conformal time $\tau$ is inversely proportional to the 
temperature $T$ so that $\tau_{\nu}/\tau_B = T_B/T_{\nu}$. 
For magnetic field generation epoch such that 
$T_B \sim 10^{14}$ Gev and $T_\nu \sim 1$ Mev, $T_B/T_\nu \sim 10^{17}$
and $\ln(T_B/T_\nu) \sim 40$. The evolution of the curvature perturbation 
has also been discussed (in synchronous gauge) in \cite{KKM10} for the 
case of an extra source of anisotropic stress canceling the neutrino 
anisotropic stress. The role of anisotropic stresses on CMB has also 
been discussed by \cite{G10}.

It is useful to estimate the relative strengths of the various terms
in \Eq{curvature}. We focus on nearly scale invariant spectra,
which is perhaps the most interesting case, and also one where
the upper limits on the strength of the large scale field
from CMB observations is the weakest.
In this case, the amplitude of the magnetic inflationary mode is given by 
$\zeta_{MI}= -(2/\epsilon)\Omega_B \ln(k\tau_e)$, where
$\tau_e$ is the conformal time  at the end of inflation. And  
$\epsilon = [{\cal H}^2 - d{\cal H}/d\tau]/{\cal H}^2 = -(dH/dt)/H^2\ll 1$, 
is the standard inflationary slow roll parameter, with ${\cal H} = a H$.
It vanishes for purely exponential expansion with constant $H$.
The ratio of the magnetic inflationary mode contribution to the passive 
mode is given by $\zeta_*/\zeta_{pas} \approx (2/\epsilon) 
(\Omega_\Pi/\Omega_B) (\ln(k\tau_e)/\ln(T_B/T_\nu)$.
Note that $\ln(k\tau_e) \sim 50$ at $k=1$ Mpc$^{-1}$, and 
$H/M_{pl} \sim 10^{-5}$, and $\Omega_B \sim \Omega_\Pi$. Thus
in general $\zeta_{MI}/\zeta_{pas} \sim 1/\epsilon \gg 1$, and 
the magnetic inflationary mode can dominate the passive mode,
if both are induced by the field generated during inflation.
The passive mode itself dominates the compensated mode
due to the extra $\ln(T_B/T_\nu)$ factor \citep{SL10}.
(In case of fields generated in phase transitions, then
the magnetic inflationary mode will not be generated).
On the other hand, for a field produced during inflation,
$\zeta_{MI}/\zeta_{inf} \sim (H/M_{pl})/\sqrt{\epsilon} \ll1$ in general
and thus all magnetically induced scalar perturbations are expected
to be subdominant to the standard inflationary mode, if the
field is produced during inflation. Much more work on the
magnetic inflationary mode is warranted to firm up these
conclusions.

\begin{figure*}[t!]\centering
\includegraphics[width=0.9\textwidth,angle = 0]{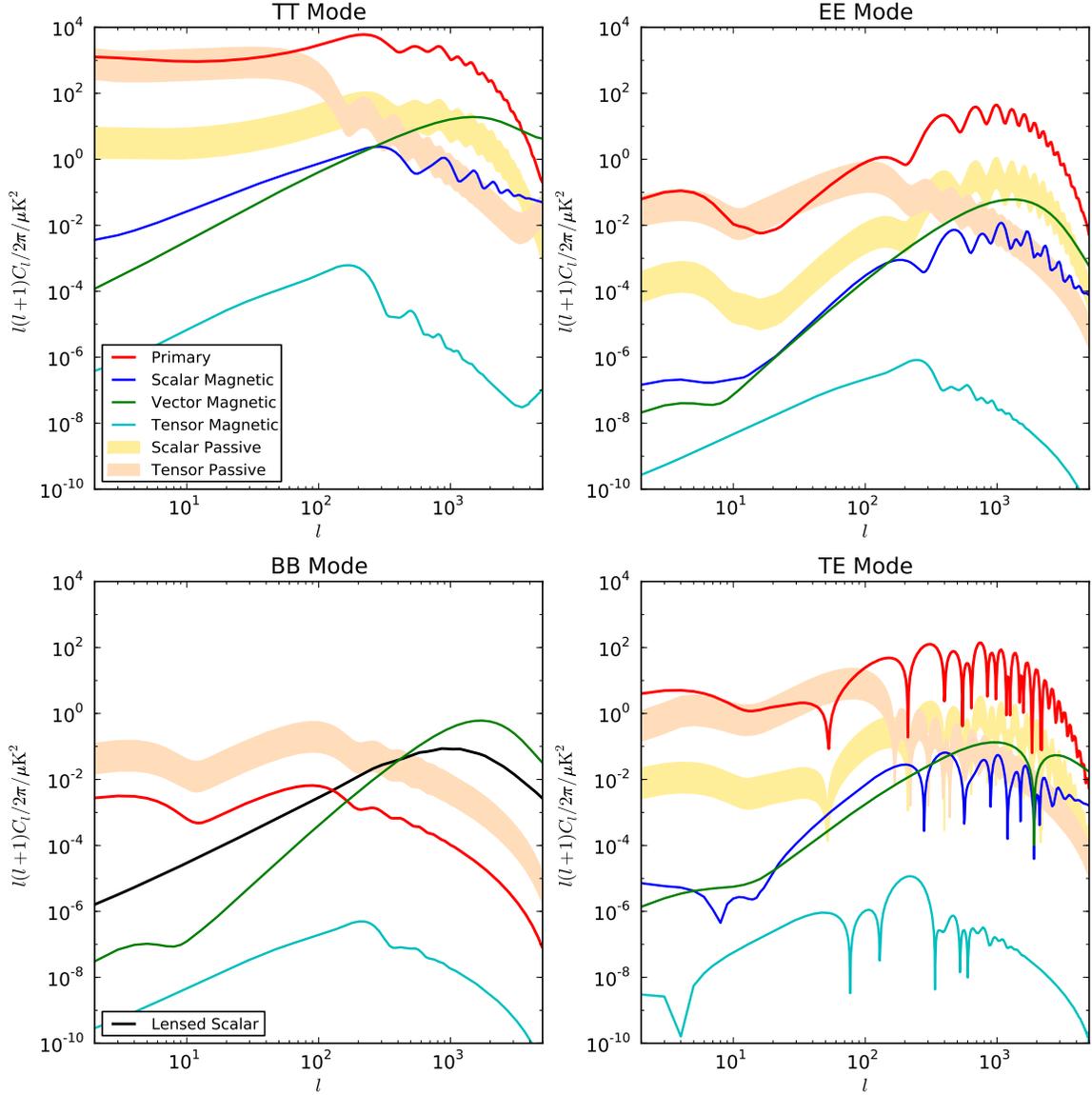}
\caption{
The four CMB power spectra versus $l$ giving the different signals
induced by a primordial magnetic field with 
$M(k)\propto k^{n}$, for $B_{-9}=4.7$ and $n=-2.9$. 
Also shown are the standard scalar primary contribution for 
the TT, EE and TE power spectra, and the tensor primary (with a tensor to
scalar ratio of 0.1) and for the BB power spectrum. 
The shaded regions give the expected range of signals for 
the passive modes, when their production epoch is
varied between the reheating and the electroweak transition.
The different modes shown are the passive scalar and tensor contribution,
the compensated scalar and tensor modes and the vector mode. Adapted 
with permission from
\citet{SL10} taking massless neutrinos; courtesy Richard Shaw.
}
\label{shaw}
\end{figure*}

Given the curvature perturbation at late times, and the 
evolution equation for the baryon-photon fluid, which includes the 
effect of the Lorentz force, one can calculate 
the CMB anisotropies due to scalar perturbations.
The magnetically induced compressible fluid perturbations,
also changes to the acoustic peak structure of the
angular anisotropy power spectrum \citep[see, for example,][]{ADGR96}.
We show in \Fig{shaw} the results of such a calculation done
using the CAMB code (http://camb.info/  by Lewis and Challinor), 
modified to incorporate the magnetic
effects (Richard Shaw; Private communication). 
Results are quoted for $B_0=4.7$ nG, and spectral indices $n_B=-2.9$.
In general, for nano Gauss fields, the CMB anisotropies 
due to the magnetized scalar mode are grossly subdominant to the 
anisotropies generated by scalar perturbations of the inflaton. 
They are also subdominant to
the anisotropies induced by the magnetically induced tensor modes
at large angular scales, and the those induced by vorticity 
perturbations (vector modes) at small angular scales.
They could dominate at intermediate angular scales with $l \sim 500$,
as can be seen in \Fig{shaw}.

\subsection{Vector modes}

A more important contribution to CMB anisotropies 
at large $l$, induced
by primordial magnetic fields, arises due to the 
Alfv\'en mode driven by the rotational component of
the Lorentz force \citep{SB98b,MKK02,SSB03,L04}.
Unlike the compressional mode, which gets
strongly damped below the Silk scale, $L_{S}$
due to radiative viscosity
\citep{Silk68}, we saw that the Alfv\'en mode behaves like an 
over damped oscillator, 
and survives Silk damping down to much smaller scales;
$L_A \sim (V_A/c) L_{S} \ll L_{S}$ \citep{JKO98,SB98a}.
The resulting baryon velocity can lead to CMB temperature
and polarization anisotropies, peaked below
the Silk damping scale (angular wavenumbers $l > 10^3$).
We estimate these more quantitatively below.

On galactic scales and above, note that the induced velocity due to the
Lorentz forces is generally so small
that it does not lead to any appreciable distortion of the initial field
\citep{JKO98,SB98a}. Hence,
the magnetic field simply redshifts away as 
$\BBB(\xx,t)=\BBB^*(\xx)/a^{2}$. The Lorentz force associated with
the tangled field ${\bf F}_{L}={\bf F}/(4\pi a^{5})$, with
${\bf F} = ({\bf \nabla }\times \BBB^*) \times \BBB^*$, 
pushes the fluid to create
rotational velocity perturbations. These can be estimated
by using the Navier-Stokes equation for the baryon-photon fluid
in the expanding Universe. In Fourier space the rotational component
of the fluid velocity satisfies,
\begin{equation}
\left( {\frac{4}{3}}\rho _{\gamma }+\rho _{\rm b}\right) {\frac{\partial
v_{i}}{\partial t}}+\left[ {\frac{\rho _{\rm b}}{a}}{\frac{{\rm d}a}{{\rm d}t}}+{\frac{
k^{2}\nu }{a^{2}}}\right] v_{i}={\frac{P_{ij}{\hat F}_{j}}{4\pi a^{5}}}.
\label{euler}
\end{equation}
Here, as before, 
$\rho _{\gamma }$ is the photon density, $\rho _{\rm b}$ the baryon
density, and $\nu =(4/15)\rho _{\gamma }l_{\gamma }$ the shear viscosity
coefficient associated with the damping due to photons, where
$l_\gamma$ is the photon mean free path.
The projection tensor, $P_{ij}({\bf k})=[\delta _{ij}-k_{i}k_{j}/k^{2}]$
projects ${\bf {\hat F} }$, the Fourier component of
${\bf F}$ onto its transverse (rotational) components
perpendicular to ${\bf k}$.

One can solve Eq.~(\ref{euler}) in two asymptotic limits.
For scales
larger than the Silk scale, $kL_{\rm S} < 1$, the radiative viscous damping
can be neglected to get
\EQ
v_i = \frac{3 P_{ij}{\hat F}_{j}}{16\pi\rho_0} D(\tau),
\label{rotvel}
\EN
where
$D(\tau) =\tau/(1 +S_*)$, with $S_*=3\rho_b/4\rho_gamma(\tau_*)$
and $\tau_*$ the conformal time at recombination. 
Since $\vert {\hat F}_{ij}\vert \sim k V_{\rm A}^2(k)$,
we get for large scales $v/c\sim \chi(k) V_{\rm A}/c$, as
estimated earlier.
For $kL_{\rm S} > 1$, a terminal velocity approximation, 
balancing viscous damping and the Lorentz force gives,
$D(\tau) \sim 5/ck^2L_\gamma$,
and $v/c \sim (5/kL_\gamma)V_{\rm A}^2(k)/c^2$, where
$L_\gamma = l_\gamma/a$.
Thus $v$ first increases with
$k$ and for $kL_S < 1$,
then decreases with $k$, with a maximum around the
Silk scale. 

This $v$ leads to CMB anisotropies $\Delta T/T \sim v/c$,
due to the doppler effect. For small $kL_S < 1$, we
have $\Delta T/T \sim V_A^2 k\tau_* \sim V_A^2 (\tau_*/R^*) l$.
Adopting $\tau_*/R^* \sim 10^{-2}$, we have $\Delta T/T \sim 10^{-6}
B_{-9}^2 (l/1000)$,  
indicating that significant CMB anisotropies 
at large $l$ can indeed result from the Alfv\'en mode.
 
The $C_l$ due to
rotational velocity perturbations can be calculated using \citet{HW96,HW97}.
\begin{eqnarray}
C_{l} &=& 4\pi \int_{0}^{\infty }{\frac{k^{2}{\rm d}k}{2\pi ^{2}}}\quad {\frac{%
l(l+1)}{2}} \nonumber \\
&\times& 
\bra{|\int_{0}^{\tau_{0}}{\rm d}\tau g(\tau_{0},\tau )v(k,\tau ){\frac{%
j_{l}(k(\tau_{0}-\tau ))}{k(\tau_{0}-\tau )}}|^{2}}
\label{deldef}
\end{eqnarray}
Here $v(k,\tau )$ is the magnitude of the rotational component
of the fluid velocity $v_{i}$ in Fourier space,
and $\tau_0$ the present value of $\tau$. 
There are also contributions from the vector metric
perturbation, and the polarization \citep{HW97}. 
But the vector metric perturbation
decays with expansion, even including magnetic sources \citep{MKK02}
and the polarization causes very small corrections to \Eq{deldef}.

The `visibility function' $g(\tau_{0},\tau )$ in \Eq{deldef} 
determines the probability that a photon reaches us at epoch $\tau_{0}$ if it
was last scattered at the epoch $\tau $.
We have shown as a solid line in \Fig{vis}
the visibility function for a standard $\Lambda$CDM model.
It is peaked about a small range of
conformal times, say $\sigma$, around $\tau_*$. 
The spherical Bessel function $j_{l}(z)$,
projects spatial variations, 
at the last scattering epoch, 
to angular (or $l$) anisotropies
at the present epoch.
It peaks around $k(\tau_0 -\tau) \approx l$, and for a fixed $l$,
probe a wavenumber $k \sim l/(\tau_0 -\tau)$ around last scattering.

We can obtain analytic estimates of $C_l$ in two limits. First, for
$k\sigma <<1$, $v(k,\tau$, $k(\tau _0-\tau ),$ and hence $%
j_l(k(\tau _0-\tau )),$ vary negligibly for $\tau $ where $g$
is significant. So they can be evaluated at $\tau =\tau _{*}$ and
taken out of the integral over $\tau $ in Eq. (\ref{deldef}). 
The remaining integral of $g$ over $\tau $ gives unity.
Also $v(k,\tau )$ 
does not vary rapidly with $k$, for $k\sim l/R_{*}$ where $j_l(kR_{*})$ 
is dominant ($R_{*}=\tau_0-\tau _{*})$. 
Thus, $v$ can also be evaluated at $k=l/R_{*}$ and pulled out of the $k$ 
integral. The remaining $k$-integral of over $j_l^2$ can be done analytically,
giving 
\begin{equation}
{\frac{l(l+1)C_l}{2\pi }}\approx {\frac \pi 4}\Delta _v^2(k=l{R}%
_{*}^{-1},\tau _{*}).  \label{powlar}
\end{equation}
where, $\Delta _v^2(k,\tau _{*})=k^3\bra{|v(k,\tau _{*})|^2}/(2\pi ^2)$ is the
power per unit logarithmic interval of $k$, residing in the rotational
velocity perturbation $v_i$. 

In the other limit, $k\sigma >>1$, 
$g$ is slowly-varying in $\tau$ compared to $j_l$.
There is a cancellation due to superposition of oscillating
contributions of $j_l$ over the thickness of the LSS. An approximate
evaluation of $C_l$ then gives 
\begin{equation}
{\frac{l(l+1)C_l}{2\pi }}\approx {\frac{\sqrt{\pi }}4}{\frac{\Delta
_v^2(k,\tau _{*})}{k\sigma }}|_{k=l/R_{*}}.  \label{powsmal}
\end{equation}
Note that when $(k\sigma >>1)$, $C_l$ is suppressed by a 
$1/k\sigma $ factor due to the finite
thickness of the LSS. 

The magnetic power spectrum $M(k)$ is normalized
using a top hat filter in $k$-space, and taken to
be of a power law form;
$k^{3}M(k)/(2\pi ^{2}) = (B_{0}^{2}/2)(n+3)(k/k_{G})^{3+n}$ with
$n > -3$. We generally adopt $k_G = 1$ h Mpc$^{-1}$,
and $B_0$ is the field smoothed over $k_G$.
$M(k)$ is cut-off at $k_{\rm c}$, determined
by dissipative processes.

We can now put together the above results.
Note that 
the power spectrum of the rotational velocity
involves not only $M(k)$, but also a mode
coupling integral $I(k)$ (see below).  
For $kL_S < 1$ we get \citep{SB98b,SB02},
\begin{eqnarray}
\Delta T_{B}(l) &=&T_{0}({\frac{\pi }{32}})^{1/2}I(k){\frac{kV_{A}^{2}\tau
_{\ast }}{(1+S_{\ast })}}  \nonumber \\
\  &\approx &5.8\mu K\left( {\frac{B_{-9}}{3}}\right) ^{2}\left( {\frac{l}{%
500}}\right) I({\frac{l}{R_{\ast }}}).  \label{an1}
\end{eqnarray}
Here, $l=kR_{\ast }$ and we have used cosmological parameters for the $%
\Lambda $-dominated model, with $\Omega _{\Lambda }=0.7$, $\Omega _{m}=0.3$
and $\Omega_{b}h^{2}=0.02$ 
We also use the fit given by Hu and White (1997b) to
calculate $\tau_{0}=6000h^{-1}((1+a_{eq})^{1/2}-a_{eq}^{1/2})(1-0.0841\ln(\Omega
_{m}))/\Omega _{m}^{1/2}$, valid for flat universe.

On scales smaller than the Silk scale, where $kL_{S}>1$ and $k\sigma >1$, 
but $kL_{\gamma }(\tau _{\ast})<1$, we get 
\begin{eqnarray}
&&\Delta T_{B}(l) =T_{0}{\frac{\pi ^{1/4}}{\sqrt{32}}}I(k){\frac{5V_{A}^{2}}{%
kL_{\gamma }(\tau _{\ast })(k\sigma )^{1/2}}}  \nonumber \\
&\approx& 13.0\mu K\left( {\frac{B_{-9}}{3}}\right) ^{2}\left( {\frac{l}{%
2000}}\right) ^{-3/2}f_{b}h_{70}^{-1}I({\frac{l}{R_{\ast }}}),  \label{an2}
\end{eqnarray}
where $h_{70}\equiv (h/0.7)$.
Here $I(k)$ is a mode coupling integral
\begin{eqnarray}
I^{2}(k)&=& {\frac{8}{3}}(n+3)({\frac{k}{k_{G}}})^{6+2n}; \quad n < -3/2
\nonumber \\
&=& {\frac{28}{15}}{\frac{(n+3)^{2}}{(3+2n)}}({\frac{k}{k_{G}}})^{3}({%
\frac{k_{c}}{k_{G}}})^{3+2n}; \quad n>-3/2.
\nonumber
\end{eqnarray}
Note that for $n< -3/2$, $I$ is independent of $k_{\rm c}$.
For a nearly scale-invariant spectrum, say with $n=-2.9$,
and $B_{-9} =3$, we get
$\Delta T(l)\sim 4.7\mu {\rm K}\,(l/1000)^{1.1}$ for scales larger than the Silk
scale, and $\Delta T(l)\sim 5.6\mu {\rm K}\,(l/2000)^{-1.4}$ for scales smaller
than $L_{\rm S}$ but larger than $L_{\gamma }$. Larger signals will be expected
for steeper spectra, $n>-2.9$ at the higher $l$ end.

One can also do a similar calculation for the expected CMB
polarization anisotropy \citep{SS01,SSB03}.
Note that polarization of the CMB arises
due to Thomson scattering of radiation
from free electrons and is sourced by the quadrupole component of
the CMB anisotropy.
For vector perturbations, what is referred to as 
the B-type contribution dominates
the polarization anisotropy \citep{HW97}, unlike for
inflationary scalar modes. 

\begin{figure}[t!]\centering
\includegraphics[width=0.5\textwidth,angle = 0]{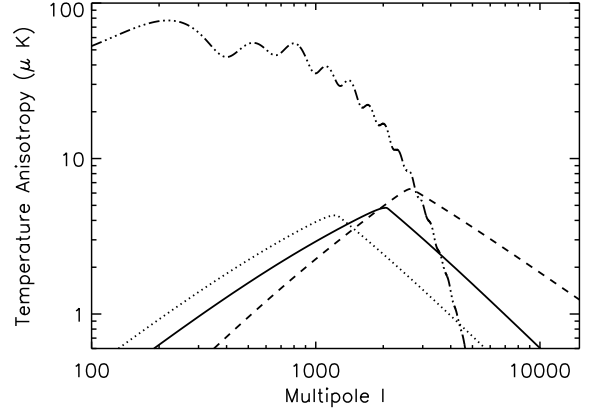}
\caption{$\Delta T$ versus $l$ predictions for the vector
mode assuming different cosmological models
and $M(k)\propto k^{n}$, for $B_{-9}=3$. The bold
solid line is for a canonical flat,  $\Lambda $-dominated
model, with $\Omega _{\Lambda }=0.7$, $\Omega _{\rm m}=0.3$,
$\Omega_{\rm b}h^{2}=0.02$, $h=0.7$ and almost scale-invariant spectrum $n=-2.9$. The
dotted curve (....) obtains when one changes to $\Omega _{\rm m}=1$ and $\Omega
_{\Lambda }=0$ model. The dashed line is
for the  $\Lambda $-dominated model with a larger baryon density $\Omega
_{\rm b}h^{2}=0.03$, and a larger $n=-2.5$. We also show for qualitative
comparison (dashed-triple dotted curve),
the temperature anisotropy in a
'standard'  $\Lambda$CDM model, computed using CMBFAST
(Seljak \& Zaldarriaga 1996) with cosmological parameters as for the
first model described above. Adapted with permission 
from \citet{SB02,SSB03}.}
\label{vec}
\end{figure}

\begin{figure}[t!]\centering
\includegraphics[width=0.5\textwidth,angle = 0]{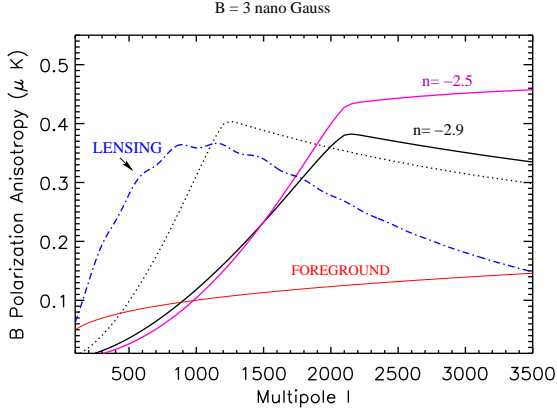}
\caption{
Predictions for the B-type polarization anisotropy. 
$\Delta T_P^{BB}$ versus $l$ for different cosmological
models and magnetic power spectrum $M(k)\propto k^{n}$, for $B_{-9}=3$.
The bold solid line is for a standard flat,
$\Lambda $-dominated model, with $\Omega _{\Lambda }=0.73$,
$\Omega _{\rm m}=0.27$, $\Omega_{\rm b}h^{2}=0.0224$, $h=0.71$ and almost scale
invariant spectrum $n=-2.9$. The dashed curve obtains
when one changes to $n=-2.5$.
The dotted curve gives results for a $\Omega _{\rm m}=1$ and
$\Omega_{\Lambda }=0$ model, with $n=-2.9$.
We also show for qualitative comparison
(dashed-dotted curve), the B-type polarization anisotropy due to
gravitational lensing, in the canonical $\Lambda$CDM model, computed
using CMBFAST \citep{SZ96}.
The signal due to magnetic
tangles dominate for $l$ larger than about $1000$. Finally, the thin
solid line gives the expected galactic foreground contribution
estimated by \citet{Prunet98}, which is also smaller
than the predicted signals. Adapted with permission from \citet{SSB03}.}
\label{vecpol}
\end{figure}

We show in Figs.~\ref{vec} and \ref{vecpol} the temperature
and polarization anisotropy for the magnetic field induced
vector modes obtained by evaluating the $\tau $ and $k$ integrals in
Eq. (\ref{deldef}) numerically. We retain the analytic
approximations to $I(k)$
and $v(k,\tau)$. These curves show the build up of power in temperature and
B-type polarization due to vortical perturbations
from tangled magnetic fields which survive Silk damping at high
$l \sim 1000$--$3000$. The eventual slow decline is due to the
damping by photon viscosity,
which is only a mild decline as the magnetically sourced vortical
mode is over damped.
By contrast, in the absence of magnetic tangles there is a
sharp cut-off due to Silk damping.
Our numerical results are consistent analytic estimates
given in Eqs.~(\ref{an1}) and (\ref{an2}).
They also qualitatively agree with the more detailed
calculations presented in \Fig{shaw} as thin, solid green lines.

A scale-invariant spectrum of tangled fields with
$B_{0}=3\times 10^{-9}$ Gauss, produces
temperature anisotropies at the $5\mu$K level
and B-type polarization anisotropies
$\Delta T_B \sim 0.3$--$0.4~ \mu$K between
$l\sim 1000$--$3000$. Larger signals result for steeper spectra
with $ n > -3$.  Note that the anisotropies in hot or cold
spots could be several times larger, because the non-linear dependence of $%
C_{l}$ on $M(k)$ will imply non-Gaussian statistics for the anisotropies.

\subsection{Tensor modes}

The magnetic anisotropic stress due to a stochastic magnetic field 
also induces tensor or gravitational wave perturbations. These can 
lead to CMB temperature and polarization anisotropies, but now peaked on large
angular scales of a degree or more or small $l$ \citep{DFK00,MKK02,CD02}.
The perturbed FRW metric describing tensor metric perturbation $h_{ij}$
is given by $ds^2 = a^2(\tau)[-d\tau^2 + (\delta_{ij} + 2 h_{ij}) dx^i dx^j]$.
Here $h_{ij}$ is transverse ($h^{ij},j =0$) and traceless ($h^i_i =0$) and
obeys the equation
\[
h_{ij}'' +2 {\cal H} h_{ij}' -\nabla^2h_{ij}
= 8\pi G a^2\delta T_{ij}^{TT},
\]
where ${\cal H}=a'/a$,
a prime denotes derivative with respect to the conformal
time, and $\delta T_{ij}^{TT}$ is the transverse, traceless component
of the energy momentum tensor (due to the magnetic field).
The gravitational wave affects the photon trajectory and its frequency,
and results in change in the CMB temperature.
The resulting CMB anisotropy is then computed using
\[
(\Delta T/T) = \int_{\tau_i}^{\tau_0}
h_{ij}' n^i n^j {\rm d}\tau,
\]
where $n^i$ is a unit vector along the line of sight, and
prime denotes a conformal time derivative.
The integration is from an epoch $\tau_i \approx \tau_*$, the 
epoch of last scattering to the present epoch $\tau_0$. 
To make progress one expands $h_{ij}$ and $\delta T_{ij}^{TT}$, 
in a Fourier and polarization basis. The polarization tensor
for any Fourier mode $\kk$ is defined as 
$ \hat\ee_{ij}(\kk,\pm) 
= (1/2) (\ee_1 \mp i\ee_2)_i \otimes (\ee_1 \mp i\ee_2)_j$,
where $(\ee_1,\ee_2,\kk)$ form mutually perpendicular unit vectors.
We define 
\[
h_{ij}(\xx,\tau) = \sum_{k,\lambda} h(k,\tau,\lambda) 
\hat\ee_{ij}(\kk,\lambda) e^{-i \kk\cdot\xx},
\]
\[
\delta T_{ij}^{TT} = \sum_{k,\lambda} (\Pi^T(k,\tau,\lambda)/a^4)  
\hat\ee_{ij}(\kk,\lambda) e^{-i \kk\cdot\xx},
\]
\[
\Pi^T \ee_{ij} = \frac{1}{2}\left[ P_{mi}P_{nj} + P_{mj}P_{ni} 
- P_{ij}P_{mn}\right](a^4p_\gamma){\Pi_{B}}^{mn}.
\]
The amplitudes $h(k,\tau,\lambda)$ then satisfy the damped harmonic
oscillator equation,
\[
h'' +2 {\cal H} h' + k^2h = 8\pi G \Pi^T/a^2,
\]
whose particular solution sourced by the magnetic anisotropic stress 
is given by \citep{DFK00,MKK02},
\EQ
h(\kk,\tau,\lambda) \approx 4 \pi G \tau_0^2 z_{eq} 
\ln\left(\frac{\tau_\nu}{\tau_B}\right) k \Pi^T(\kk,\lambda) 
\frac{j_2(k\tau)}{k\tau}.
\label{tensol}
\EN
Here we have modified the argument of the log factor, to take
account of the fact that the magnetic anisotropic stress gets
compensated on large scales after neutrino decoupling \citep{L04,SL10}.
This mode is therefore referred to as the `passive' tensor mode.

We see that the amplitude of a given gravitational wave mode remains constant
for small $k\tau$, when it remains outside the Hubble radius.
Once it enters the Hubble radius, and $k\tau >1$, it undergoes
damped oscillations, as can be seen from the $j_2(k\tau)/(k\tau)$
term in \Eq{tensol}. As smaller scales (larger $k$ or $l$) modes
enter the Hubble radius at earlier epoch, they are damped more
by the epoch of last scattering. Thus for a nearly scale invariant
spectrum of magnetic fields, their tensor mode contribution 
to CMB anisotropies will be largest at large scales (small $l$)
and will oscillate and decay at smaller and smaller scales.
Such a behaviour can indeed be seen in \Fig{shaw}, where the Tensor passive
mode signals are seen as top most magenta band at low $l$, 
in all the panels. The band represents the uncertainty in the value
of $\tau_B$ which is chosen to range from $10^{-6}-10^{-12}\tau_\nu$. 

Using the formalism described in \citet{DFK00,MKK02}, 
an analytical estimate of the tensor contribution at small $l<100$
is $\Delta T \sim 7 (B_{-9}/3)^2 (l/100)^{0.1}~\mu{\rm K}$, for
$n=-2.9$.
The tensor mode also contributes to the B-type polarization
anisotropy at large angular scales ($l <100$ or so), with
$\Delta T_B < 0.1~\mu $K for $B_{-9} < 3$.
The production of gravitational waves has been used in an indirect
manner by \citet{CD02} to set strong upper limits on
$B_0$ for spectra with $n > -2.5$ or so.

\subsection{Faraday rotation due to primordial fields}

Another interesting effect of primordial fields is the
the Faraday rotation it induces on the polarization of
the CMB \citep{KL96,KKLR05,CDGV04}
The rotation angle is
\EQ
\Delta\Phi 
\approx 1.6^\circ B_{-9} (\nu_0/30\, {\rm GHz})^{-2},
\EN
where $\nu_0$ is the frequency of observation.
So this effect is important only at low frequencies,
and here it can lead to the
generation of B-mode polarization from the Faraday
rotation of the inflationary
E-mode. From the work of Kosowsky et al. (2005) one
can estimate a B-mode signal
$\Delta T_B \sim 0.4 (B_{-9}/3)\, (\nu/30\, {\rm GHz})^{-2} \mu {\rm K}$,
for $n=-2$, at $l \sim 10^4$.
The signals are smaller at smaller $n$.
The Faraday rotation signal can be distinguished
from the B-mode polarization generated by say
vector modes, or gravitational lensing, because of
their frequency dependence ($\nu^{-2}$).

\subsection{CMB non Gaussianity}

A crucial difference between the magnetically
induced CMB anisotropy signals compared to those
induced by inflationary scalar and tensor perturbations,
concerns the statistics associated with the signals.
Primordial magnetic fields lead to non-Gaussian statistics
of the CMB anisotropies even at the lowest order,
as magnetic stresses and the temperature anisotropy they induce
depend quadratically on the magnetic field.
In contrast, CMB non-Gaussianity due to inflationary scalar
perturbations arises only as a higher order effect.
A computation of the non-Gaussianity of the magnetically induced
signal has begun \citep{SS09,CFPR09,CHZ10},
based on earlier calculations of non-Gaussianity in the
magnetic stress energy \citep{BC05}.
Anisotropy and non-Gaussianity can also result
during inflationary magnetogenesis in particular models where conformal
invariance is broken by coupling to the inflaton 
\citep{BNP12,BMPR13,FY13}.
This new direction of research 
on CMB non-Gaussianity
currently leads to
constraints on the field at sub nG level \citep{TSS14}, and
promises to lead to tighter constraints or a detection of
strong enough primordial magnetic fields 
\citep{SNYIT10,SNYIT11,TSS10,TSS12,TSS14}.

\subsection{A summary of CMB constraints}

We have outlined some of the possible ways one could
detect/constrain primordial magnetic fields using the CMB
anisotropies and polarization. 
For a field of $B \sim 3$~nG and a nearly scale-invariant spectrum
one predicts CMB temperature anisotropies with a $\Delta T \sim 5 \mu$K,
at $l< 100$ (induced by tensor modes) and $l> 1000$ (due to the Alfv\'en mode)
and polarization anisotropies about 10 times smaller.
Especially interesting is that the vector modes induced
by primordial fields 
can contribute significantly below the Silk
scale, where the conventional scalar modes are exponentially damped.
Further, the magnetically induced polarization signal 
will be dominated by B-mode polarization.
A unique signature of primordial fields on the CMB 
is that magnetically induced anisotropies
are predicted to be strongly non-Gaussian.
Also if fields have helicity, this would induce further parity
odd cross correlations, which would not otherwise arise
\citep{CDK04,KR05,KMLK14,BFP15}.

The most systematic analysis of CMB constraints has been carried out
using the Planck data \citep{PlanckB15}. 
This analysis assumes a spatially flat universe, with the CMB temperature
$T_0 = 2.7255$ K, a primordial Helium fraction of $y_p =0.24$, 
three massless neutrinos and lensing effect only for the primary CMB spectrum.
It also marginalises over astrophysical residuals and secondary
anisotropy contamination, varies a set of 6 standard cosmological
parameters, the amplitude $B_0$ of the magnetic field smoothed
over scales of 1 Mpc and the magnetic spectral index $n> -3$.
Overall this analysis constrains the primordial magnetic field strengths
to be less than a few nG. More specifically, using the angular power spectra
and the Planck likelihood, the 95$\%$ confidence level 
constraints are $B_0 < 4.4$ nG on Mpc scale
with correspondingly, $n < -0.008$
for non-helical fields and $B_0< 5.6$ nG for helical fields.
These tighten to $B_0 < 2.1$ nG for nearly scale invariant fields
with $n=-2.9$, $B_0 < 0.011$ nG for causally generated fields where one
expects $n=2$ and $B_0 < 0.55$ nG for any $n > 0$. For the nearly 
scale invariant case, the limits
further tighten to $B_0 < 0.7$ nG if their effects on heating and
ionization \citep{Sethi_KS05,KK15,Chluba15} 
(see \Sec{postrecomb}) are included.
The limits from the magnetically induced tensor and compensated 
scalar bispectrum are at the 3 nG level. 
It has been pointed out by \citet{TSS12,TSS14}
that stronger sub nG limits are possible from analysis of
the CMB trispectrum. Moreover stronger limits would also be potentially 
possible from the magnetic inflationary mode \citep{BCD13}.
A comparison of the estimated trispectrum from the scalar passive
mode and Planck 2013 limits on CMB non-Gaussianity, gave 
$B_0 \la 0.6$ nG, and that from the magnetic inflationary
mode suggests $B_0 \la 0.05$ nG \citep{TSS14}.

Primordial magnetic fields have also been constrained by the
POLARization of the Background Radiation (POLARBEAR) experiment
from the detection of the B-mode polarization at high $l \ga 500$ 
\citep{POLARBEAR}.
Comparing the theoretically expected B-mode polarization 
signal from vector mode gives a limit $B_0 < 3.9$ nG at the 95\% 
confidence level. There has also been a claimed detection of the
B-mode polarization on degree scales 
(low $l\sim 80$) by the BICEP2 experiment \citep{BICEP2}, which
could be due to tensor modes from the inflationary
era, or potentially also seeded by primordial magnetic fields.
This signal is however now thought to due to contamination by
dust emission \citep{bicep2keck,bicep2keckplanck}. Even were it present, the
amplitude of the observed B-mode polarization is difficult to explain 
as due to primordial magnetic fields \citep{BDM14}, 
given the constraints from the CMB trispectrum \citep{TSS14}.
However the B mode induced by such fields, combined with the dust
contribution, could ease the required level of tensor modes 
from standard inflationary models \citep{BDM14}.
Clearly, the detection of B-modes at both large and small
angular scales, and also the limits (or detection) of non-Gaussianity
in both temperature and polarization anisotropies, are likely to
provide the strongest future CMB probes of primordial magnetic fields.

The current limits on primordial magnetic fields from CMB observations 
are summarized in Table~(\ref{B0limits}),
where we have also given constraints coming from other physical probes
discussed below.

\begin{table*}\centering
\caption{Limits on primordial magnetic fields from magnetic 
mode contributions to the 
CMB power spectra, bispectra, trispectra, reionization, weak lensing, 
Lyman-$\alpha$ forest and Faraday rotation of background quasars. 
We quote limits derived for close to scale-invariant magnetic fields
(except where we say general) 
and an early generation epoch ($10^{14}$ GeV) for magnetic passive modes.
The value $B_0$ refers to the magnetic field smoothed at a scale of 1 Mpc.
The last row gives the approximate lower limit from $\gamma$-ray
observations of TeV blazars. 
\\}
\begin{tabular}{cccc} 
\hline \hline \\
\multicolumn{1}{c}{\,\,\,\,\, Probe \,\,\,\,\,} &
\multicolumn{1}{c}{\,\,\,Magnetic modes\,\,\,} &
\multicolumn{1}{c}{Upper limit $B_0$ (nG)} &
\multicolumn{1}{c}{\,\,\,Reference\,\,\,}
 \\ [3 pt]
\hline \hline
\\
CMB Power Spectra &
scalar, vector \&  tensor &
4.4 (non helical, general)&
Planck-2015 \\ [5 pt] 
 &
scalar, vector \&  tensor &
5.6 (Helical, general)&
Planck-2015 \\ [5 pt] 
 &
scalar, vector \&  tensor &
2.1 (Scale invariant) &
Planck-2015 \\ [5 pt]
 &
Ionization History &
0.7 &
Planck-2015 \\ [5 pt]
CMB Polarization &
Vector, B Mode &
3.9 &
POLARBEAR \\ [5 pt] 
\hline  
\\
CMB Bispectrum &
energy density&
\,\,22 
& \cite{SS09} 
\\ 
& (Compensated scalar) & & 
\\ [5 pt]
 &
Passive-scalar &
2.4 &
\cite{TSS10} \\ [5 pt]
 &
Compensated-scalar &
3 &
Planck-2015 \\ [5 pt]
 &
vector &
10 &
\cite{SNYIT10} \\ [5 pt]
 &
Passive-tensor &
3.2 &
\cite{ShSe14} \\ [5 pt]
 &
Passive-tensor &
2.8 &
Planck-2015 \\ [5 pt]
\hline
\\
CMB Trispectrum &
energy density &
19 &
\citet{TSS14} \\
& (Compensated scalar) & & \\ [5 pt]
 &
Passive-scalar &
0.6 &
\citet{TSS14} \\ [5 pt]
 &
magnetic inflationary mode &
0.05 &
\citet{TSS14} \\ 
 & \citep{BCD13} & & \\ [5 pt]
\hline \hline
\\
Reionization & $n=-2.85 \ {\rm to} -2.95$ & 
0.059-0.358 & 
\citet{PCSF15}\\ [5 pt]
\hline
\\
Weak Lensing & & $\sim 1-3$ & \citet{PS12} \\ [5 pt]
\hline
\\
Lyman-$\alpha$ forest & $n\approx -3$ & $0.3-0.6$ & \citet{PS13} \\ [5 pt]
\hline
\\
Faraday Rotation & uniform to 50 Mpc & 1-6 & \citet{BBO99} \\ [5 pt]
& uniform to 1 Mpc & 0.5-1.2 (2 $\sigma$) & \citet{PTU15} \\ [5 pt]
\hline \hline
\\
\multicolumn{2}{c}{Absence of GeV halo from Tev Blazars} & $B_0 \ga 10^{-16}$ G & \citet{NV10} 
\\ [5 pt]
\multicolumn{2}{c}{(Lower Limit)} & ($l_B \gg l_{ic}$) & \citet{TGBF11} \\ [5 pt]
\hline \hline
\end{tabular}
\label{B0limits}
\end{table*}


\section{Primordial magnetic fields post recombination}
\label{postrecomb}

After recombination, the ionized fraction  decreases by several orders of 
magnitude by $z \simeq 1000$, finally reaching a value of $\simeq 10^{-4}$
for $z \la 100$ (for details see \citet{Peeb93}). 
However, the residual free electrons are still sufficient in density
to carry the currents required to sustain primordial magnetic fields. 
In the standard picture, the matter temperature
continues to follow the CMB temperature, both falling as $\propto 1/a$ for
$z \ga 100$. At smaller redshifts, matter 'thermally' decouples from the 
radiation and the matter temperature falls as $\propto 1/a^2$. 
This  thermal and ionization history holds up to $z \simeq
10\hbox{--}20$, when the formation of first structure can lead to 
reionization and reheating. 
Primordial magnetic fields can alter this picture in several
interesting ways \citep{Sethi_KS05}.

First, the magnetic fields exert forces on the electron-ion fluid,
but not on the neutral atoms, causing a relative drift velocity and hence
a friction between these components. This can lead to the dissipation
of the magnetic energy, called ambipolar diffusion, into the 
intergalactic medium. Also the increased photon mean free path due to
recombination, leads to a reduced viscosity and increased
fluid Reynolds number, leading to the possibility of generating
decaying fluid turbulence. These processes will affect the
thermal and ionization history of the universe. In addition the Lorentz
force due to tangled primordial magnetic fields, can also induce
density perturbations, which grow due to gravitational instability
and cause early formation of structures in the universe. Such
enhanced density perturbations can also lead to weak lensing
signatures, and together with the altered ionization history
lead to new signals in the 21 cm and the CMB. We consider
some of these effects in more detail below. 

\subsection{Magnetic field dissipation post recombination}

A rough estimate of the field strength $B_0$ which will
result in significant changes to the ionization and thermal history of the 
universe can be obtained as follows. If, at a certain epoch, 
a fraction $f$ of magnetic field energy is dissipated into the 
IGM then it will typically raise it to a temperature: 
$T = f B_0^2/(8 \pi)/n_b k$; with $n_b = n_b(t_0) (1+z)^3$ and 
$B_0 = B_0(t_0)(1+z)^2$. Taking 
$f = 0.1$, this give $T \simeq 10^4 \, \rm K [(1+z)/100]$ for 
$B_0 \simeq 10^{-9} \, \rm G$. For $z \ga 100$, this
could be an overestimate because, owing to inverse
Compton scattering off CMB photons, matter temperature cannot increase 
much above CMB temperature. The fraction of the energy dissipated $f$ 
will depend on the magnetic field power spectrum and also on the rate 
of energy dissipation compared to the expansion rate. However it does 
give a rough estimate of the magnitude of $B_0$ that are of interest. 

The volume rate of energy dissipation 
due to ambipolar diffusion is (\citet{C56} Eq.~(27)).
\begin{equation}
\Gamma_{\rm in} = {\rho_n \over 16 \pi^2 \gamma \rho_b^2 \rho_i} 
|({\bf \nabla x B) x B}|^2
\label{eq:amdif}
\end{equation} 
Here $\rho_n$, $\rho_i$, and $\rho_b$ are the densities of 
neutral particles mostly hydrogen, ionized hydrogen, 
and total baryon density,  respectively. 
Also $\gamma = \langle w \sigma_{\rm in} \rangle /(m_n + m_i)$ \citep{Shu92}; 
where $w$ is the ion-neutral relative velocity and $\sigma_{\rm in}$ is the 
cross section for the collision between ions and neutrals. 
For $w \la 10 \, \rm km \, sec^{-1}$, 
$\langle w \sigma_{\rm in} \rangle \simeq 3 \times 10^{-9}$ independent 
of the relative velocity of ions and neutrals \citep{Shu92}. 
This energy is deposited into the neutral
component of the medium. However collisions between electrons,
protons, and neutrals lead to rapid thermalization \citep{MMR97}.
The volume rate of energy deposition in electrons, 
is $\Gamma_e = x_e \Gamma_{\rm in}$, where $x_e$ is the ionization
fraction. It can be seen from \Eq{eq:amdif} that the rate of 
dissipation is dominated by the smallest
scale (largest $k$) for the scale-free
magnetic field power spectrum. This will correspond to the 
large-k cut-off, $k_{\rm max}$ in \Eq{eq:kmaxn}.

\begin{figure}
\includegraphics[width=0.35\textwidth,angle = 270]{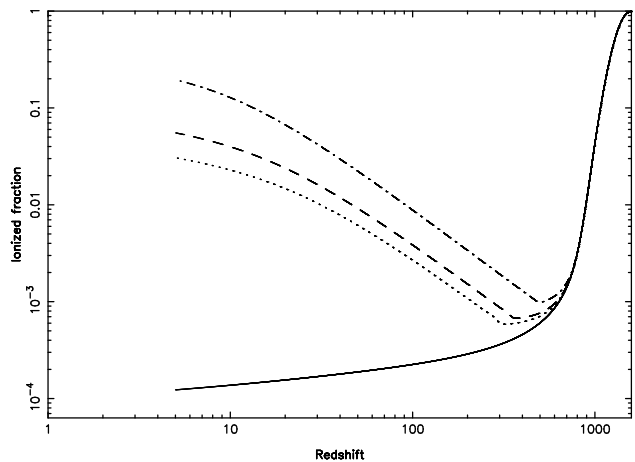}
\caption{Evolution of the ionization state of the universe is
shown for ambipolar dissipation. Different curves are:
standard recombination (solid curve); the dotted and  dashed curves
correspond to nearly scale free magnetic field power spectra with 
$n = -2.9$ and $n = -2.8$ with $B_0 = 3 \times 10^{-9} \, \rm G$;   the
dot-dashed curves correspond to the delta function magnetic field power spectrum
with  $B_0 = 3 \times 10^{-9} \, \rm G$ and $k_\star = k_{\rm max}$. 
Reproduced with permission from \citet{Sethi_KS05}. 
}
\label{fig:f1}
\end{figure}

\begin{figure}
\includegraphics[width=0.35\textwidth,angle = 270]{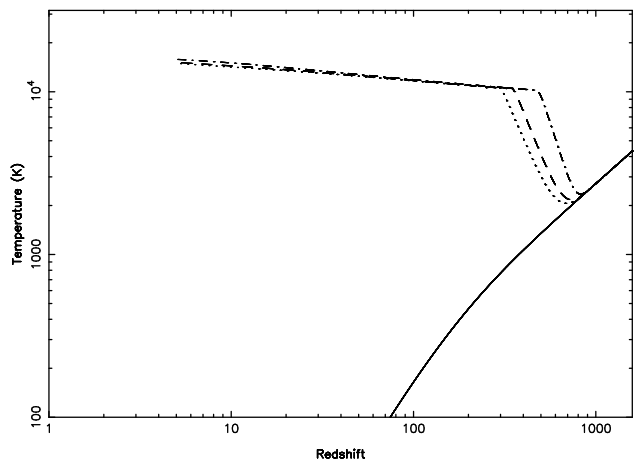}
\caption{Evolution of the thermal  state of the universe is
shown for ambipolar dissipation. Curves are for the same 
 parameters as in \Fig{fig:f1}. 
Reproduced with permission from \citet{Sethi_KS05}. 
}
\label{fig:f2}
\end{figure}

Figure~\ref{fig:f1} and~\ref{fig:f2} show the ionization and 
thermal history of the universe
for some interesting values of $B_0$ for both delta function and power
law power spectra, as computed by \citet{Sethi_KS05}.  
For delta function power spectrum, the power is assumed to be 
at $k_{\rm max}$ and for the power law 
power spectrum $B_0$ is 
defined as RMS value smoothed at $k= k_{\rm max}$. 
We see that the IGM can indeed be significantly heated and ionized
for $B_0 \sim 3$ nG.
Similar results have also been obtained in subsequent calculations
of heating and ionization by primordial magnetic fields 
\citep{SBK08,SBK09,KK15,CPFR15}.

\begin{figure}[t!]\centering
\includegraphics[width=0.5\textwidth,angle = 0]{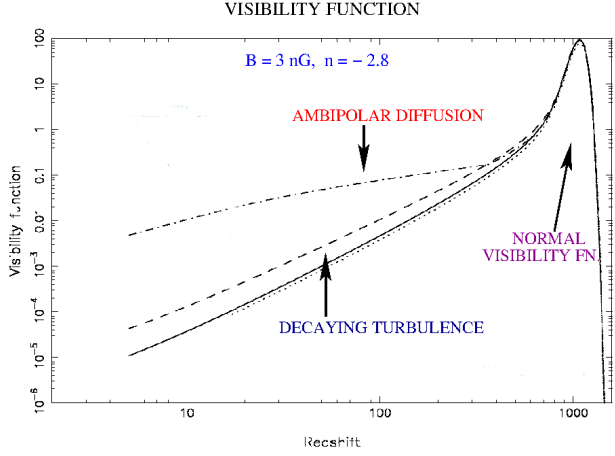}
\caption{Visibility function, is plotted for different models. The 
solid curve is for the standard recombination. 
Dashed curve corresponds  to a  decaying turbulence model with   
$B_0 = 3 \times 10^{-9} \, \rm G$. Dot-dashed  curve
corresponds to the ambipolar diffusion case with 
$B_0 = 3 \times 10^{-9} \, \rm G$
and $n = -2.8$.
Adapted with permission from \citet{Sethi_KS05}.
}
\label{vis}
\end{figure}

There are several consequences of such heating and ionization:
(i) The thermal Jeans mass is raised by the increase in the IGM
temperature which affects  
subsequent structure formation. 
(ii) The number density of free electrons which can catalyze 
formation of molecular hydrogen is increased. This 
results in a larger molecule abundance at the onset of
in the first galaxies \citep{SNS08,SGG09}, 
and its subsequent influence on the formation of first stars.
(iii) The extra ionization also modifies the visibility function 
$g(\tau_0,\tau)$ for the CMB photons.
Recall that the visibility functions measures the normalized probability 
that a photon last scattered between  $\tau$ and $\tau + d\tau$,
or a redshift interval $(z,z+dz)$. 
We show in \Fig{vis} 
visibility functions for the some representative models in 
comparison with the standard visibility function.
One can see that the effect of ambipolar diffusion, with $B_0 = 3$ nG, 
can cause significant changes to the visibility function, and this
will also lead to distortions in the CMB anisotropies.
Such changes are only beginning to be explored \citep{CPFR15},
and appear to provide strong nG level constraints on the primordial
fields 
\citep{PlanckB15}.

\subsection{Primordial magnetic fields and structure formation}

In the radiation dominated era, we saw that the pressure associated with
a primordial magnetic field is generically much smaller than the pressure of 
the baryon-photon fluid.
Therefore any non-uniform magnetic field would only generate motions which
are nearly incompressible.
However, as we mentioned earlier, once
any particular scale becomes smaller than the photon mean free path, 
the baryons and photons begin to decouple for perturbations on this scale. 
Then this `mode' enters the free-streaming regime, where the radiative
damping transits from being diffusive to free-stream damping.
More importantly, the baryons no longer feel the pressure of the photons, and 
there is a dramatic fall in the
fluid pressure, by a factor of order the very small baryon to photon ratio 
$n_b/n_\gamma \sim 10^{-9}$. Of course this happens for all scales 
after the recombination era, when atoms form, and the electron density drops
so drastically that the photon mean free path becomes of order the Hubble
radius. As a result, the pressure of the non-uniform magnetic field,
associated with what might well have been an Alfv\'en-type mode, can no
longer be ignored. This 
will now also induce gravitationally
unstable, compressible motions. The resulting growth in density perturbations
in the baryons lead, via their gravitational influence, to
density perturbations in any dark matter component. 
Such perturbations then grow due to their self gravity and when 
they become comparable to the background density,
they become non linear, and cause the collapse of these regions,
forming structures. We study this magnetically induced structure
formation in this section.

Let us assume that the perturbations in density and velocity are small enough so
that non-linear terms in the perturbed density and velocity can be
neglected. In the Euler equation (\Eq{eulerin}), we neglect the
non-linear term, ${\bf v}.{\bf \nabla }{\bf v}$ and take the 
density of the baryonic component to be 
$\rho_b = \overline\rho_b (1+ \delta_b)$, where   
$\overline\rho_b$ is the  
unperturbed FRW background density of baryons, 
and $\delta_b$, its fractional perturbation. 
This equation has
to be supplemented by the continuity equation for 
$\delta_b$,
the induction equation (\ref{expind}), with $\BBB^*= a^2 \BBB$, and
the Poisson equation for the gravitational potential, $\phi$. 
We have \citep{W78,KOR96,SB98a,Sethi_KS05},
\begin{eqnarray}
\overline\rho_b\left[
{\frac{\partial {\bf v}}{\partial t}}+H(t){\bf v}\right]
=-{\frac 1{a}}{\bf \nabla }p_b &+&\frac{{\bf J}\times {\bf B}}{c}
-{\frac 1a}\overline\rho_b {\bf \nabla }\phi \nonumber \\ 
&-&{\frac{4\rho _\gamma }{3}}n_e\sigma _T%
{\bf v},  \label{eulerinp}
\end{eqnarray}
\begin{equation}
{\frac{\partial \delta_b}{\partial t }}+{\frac{1 }{a}}{\bf \nabla }.{\bf v}
=0,  \label{conta}
\end{equation}
\EQ
\frac{\partial (a^2\BBB)}{\partial t} 
= {\frac 1a} \nab \times \left [ \vv \times (a^2\BBB) - \eta
{\frac 1a} \nab \times (a^2\BBB) \right],
\label{expind2}
\EN
\EQ
{\bf \nabla }^2\phi =4\pi Ga^2\delta \rho _T=
4\pi Ga^2\left[ \overline\rho_b\delta
_b+\overline\rho_{DM}\delta_{DM}\right] .  \label{poison}
\EN
We recall that all spatial derivatives are with respect to co-moving
spatial co-ordinates.
In the Poisson equation, we have taken account of the possibility 
that there may be other
forms of matter, like collisionless dark matter (DM)
whose background density is given by $\overline\rho_{DM}$ and its fractional
density perturbation
by $\delta_{DM} = \delta\rho_{DM}/\overline{\rho}_{DM}$. 
Thus $\delta \rho _T$ is the sum of the perturbed density 
due to both the baryonic plus dark matter. 
We shall adopt the equation of state $%
p_b=\rho _b c_b^2$, where $c_b^2 = (kT/\mu)$ is the sound speed and $\mu$ the
mean molecular weight (for fully ionized hydrogen $\mu=0.5 m_p$, with $m_p$ 
the proton mass).

In treating the resulting evolution, it is usual to assume firstly 
ideal MHD ($\eta \to 0$), and further that the perturbed velocity does not 
significantly distort the initial magnetic field \citep{W78,Peebles80}).
So one takes ${\bf B}={\bf B}_0({\bf x})/a^2$, which solves the induction 
equation (\ref{expind2}), if ${\bf v}$ and $\eta$ are neglected.
Then the perturbations to the Lorentz
force, due to the perturbed magnetic field, is subdominant with respect to
the zeroth-order contribution of the Lorentz force itself. 
Of course, this approximation will break
down once significant peculiar velocities have been developed, as will
always happen on sufficiently small scales, or at sufficiently late times,
for any given magnetic field strength. 
For galactic scales, it turns out that the
distortions to the magnetic field will become important only at late times,
even for $B_{-9}\sim 1$. 
Taking the divergence of \Eq{eulerinp}, substituting for $\nab\cdot\vv$ from 
\Eq{conta}, using \Eq{poison} and the equation of state,
leads to the evolution equation 
for $\delta _b$ (see \citet{W78} for the case without dark matter),
\EQA
{\frac{\partial ^2\delta _b}{\partial t^2}}&+&\left[ 2H+{\frac{4\rho _\gamma }{%
3\overline\rho_b}}n_e\sigma _Ta\right] 
{\frac{\partial \delta _b}{\partial t}}-c_b^2{%
\nabla }^2\delta _b \nonumber \\
&=& 4\pi Ga^2\left[ \overline\rho_b\delta _b
+\overline\rho_{DM}\delta_{DM}\right] 
+\frac{1}{a^3}S_0({\bf x})
\label{delb}
\ENA
where the source term $S_0$ is given by
\begin{equation}
S= {\frac{{\bf \nabla }.\left[ {\bf B}_0\times ({\bf \nabla }\times {\bf B}%
_0)\right] }{4\pi \overline\rho_b(t_0)}}.  \label{sorce}
\end{equation}
Here, $\overline\rho_b(t_0)$ is the baryon density at the present time, 
$t_0$. We see from \Eq{delb} that perturbations in the baryonic fluid
are generated by any inhomogeneous magnetic field even if they
were initially zero. 
These perturbations can grow due to gravity (the first two terms on
the right hand side of \Eq{delb}), are damped by the expansion of the
universe and radiative viscosity (respectively 
the 2nd and 3rd terms on the LHS of \Eq{delb}). Further baryonic pressure
can provide support against collapse on small scales (the 4th term
on the LHS of \Eq{delb}).

If we assume dark matter to be cold, one can also derive a 
similar equation for its fractional perturbed density $\delta_{DM}$ 
\citep{Peebles80,Padmanabhan02}, 
\begin{equation}
{\frac{\partial ^2\delta_{DM}}{\partial t^2}}+2H{\frac{\partial \delta _{DM}}{%
\partial t}}=4\pi Ga^2\left[ \overline\rho_b\delta _b
+\overline\rho_{DM}\delta_{DM}\right] .
\label{delc}
\end{equation}
The dark matter perturbations
are not directly affected by the magnetic field, but are coupled to the field
via the baryonic perturbations by gravity.

After recombination, the mean-free-path of the
photon increases rapidly to a value exceeding the Hubble radius,
viscous damping becomes subdominant compared to expansion damping,
and can be neglected. Also, for large enough scales,
larger than the thermal Jeans mass we can
neglect the fluid pressure term.
The perturbation equations can then be solved by first defining,
$\delta_{\rm m} = (\overline\rho_{DM} \delta_{DM} 
+ \overline\rho_b \delta_b)/\rho_{\rm m}$
with $\rho_{\rm m} = (\overline\rho_{DM} + 
\overline\rho_b)$. Then \Eqs{delb}{delc} become
\EQA
{\partial^2\delta_b \over \partial t^2} &=&
-2{\dot a \over a} {\partial \delta_{b} \over \partial t} 
+ 4 \pi G   \rho_{\rm m}\delta_ {\rm m}  +  \frac{S_0(x)}{a^3} \nonumber \\
{\partial^2\delta_{\rm m} \over \partial t^2} &=&
-2{\dot a \over a} {\partial \delta_{\rm m} \over \partial t} 
+ 4 \pi G   \rho_{\rm m}\delta_ {\rm m} 
+ {\rho_{b} \over \rho_{\rm m}}\frac{S_0(x)}{a^3}
\label{eq:denevol1}
\ENA
The homogeneous solutions correspond to perturbations generated by 
sources before recombination, \eg during inflationary epoch. 
We focus here on the
particular solution, whereby density perturbations are induced
by inhomogeneous magnetic field. Then the solution to
\Eq{eq:denevol1} for $z \gg 1$, in a spatially flat universe is
\begin{eqnarray}
\delta_{\rm m}({\bf x},t)
&\simeq& {3 \Omega_b \over 5 \Omega_m}\left ({3 \over 2}\left
({t \over t_i} \right )^{2/3} + \left ({t_i \over t} \right ) -
{5 \over 2} \right )  \frac{t_i^2 S_0({\bf x})}{a^3(t_i)} 
\nonumber \\
&\sim& \frac{3}{5} \frac{\Omega_b}{\Omega_m} 
\left(\frac{2}{3H_0\Omega_m^{1/2}}\right)^2 S_0(x)  
\left({t \over t_i} \right )^{2/3} 
\label{eq:n5pp}
\end{eqnarray}
The initial time $t_i$ corresponds to the epoch of recombination 
as the perturbations cannot grow before this epoch.  
In the second line in \Eq{eq:n5pp}, 
we have used the Einstein equation 
$H^2 = (8\pi G /3) \rho_m$ valid at early times
to estimate $t_i^2/a^3(t_i)$ and also retained
only the growing mode solution.
Using this solution, one can show that the fastest growing
mode also has $\delta_{b} \propto t^{2/3}$ and so 
both the baryonic and the dark matter perturbations
grow at the same rate.

We can make an estimate of the typical scales which can grow
for nano Gauss field strength by using \Eq{eq:n5pp}.
We adopt $\Omega_m=0.3$, $\Omega_b=0.05$, $h=0.7$ and approximate
$S_0 \sim k V_{Ab}(k,t_0)$. Here $V_{Ab}(k,t)$ is the Alfv\'en velocity 
relevant in the post-recombination epochs using the baryon density, and
is given by
\EQA
\frac{V_A(k,t)}{c} &=& {\frac{B(k,t)}{(4\pi \overline\rho _{b}(t))^{1/2}}}
\nonumber \\
&\approx& 1.5\times 10^{-5}\left({ B(k,t) a^2 \over 10^{-9}{\rm G}}\right)
\ a^{-1/2}.
\label{alfven}
\ENA
where $B(k,t)$ is the magnetic field smoothed on
a scale $l=a/k$ at time $t$. 
We get 
\EQ
\delta_m(z) \sim  0.7 \left(\frac{k}{1 {\rm Mpc}^{-1}}\right)^2 
\frac{B_{-9}^2}{1+z}
\label{delm}
\EN
We see that galactic scales can become non-linear with $\delta_m \sim1$
by the present, if all the field is concentrated on this scale.
Subgalactic scales can become non-linear even earlier for nano-Gauss
strength fields. We give a more precise calculation of this effect below.

\begin{figure}[t!]\centering
\includegraphics[width=0.5\textwidth,angle = 0]{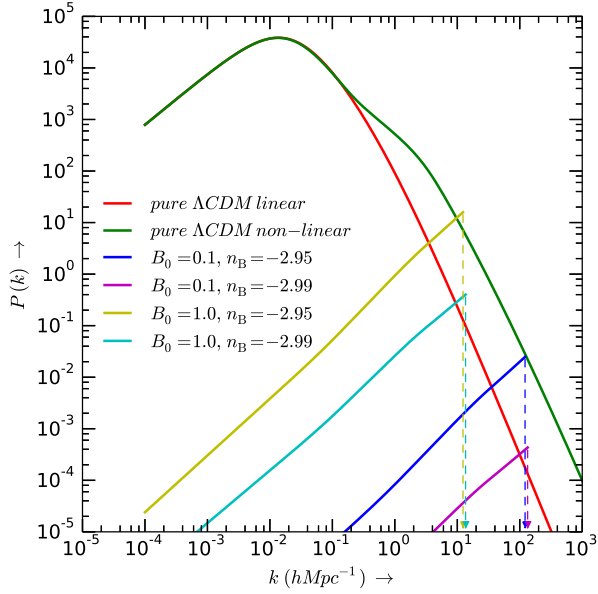}
\caption{Matter power spectrum induced by nearly scale invariant
magnetic fields, compared to that produced during inflation 
in a standard $\lambda$CDM model. One can see the steep, nearly $k^4$, 
rise of the magnetically induced contribution. The magnetically induced 
spectra have been cut-off at the magnetic Jeans scale. $B_0$ in the figure
corresponds to the rms field when the spectrum is cut off
at a wavenumber $k_c= 1$ Mpc$^{-1}$, in units of nG.
Reproduced with permission from 
\citet{PCSF15}. 
}
\label{matpow}
\end{figure}

The power spectrum of matter  perturbations can written as,
\begin{equation}
P(k,t) = \langle \delta_{\rm m}(k,t) \delta_{\rm m}^*(k,t) 
\end{equation}
and can be computed given the magnetic power spectrum
$M(k)$ \citep{GopalSethi03}.
In particular, for a nearly scale invariant magnetic spectrum, 
we find that  
$P(k) \propto k^4$ (since $S(x)$ involves two spatial derivatives).
This steep increase in $P(k)$ will lead to the magnetically induced
density perturbations to dominate those induced by inflation
at small enough scales for $B_{-9} \sim 1$, as can be seen in 
\Fig{matpow} from \cite{PCSF15}. 
This feature also provides the potential
to constrain the strength of the primordial field
on subgalactic scales by using observations of weak gravitational lensing
\citep{PS12} and the Lyman-$\alpha$ forest
clouds \citep{PS13}.

Note that the field gets distorted by the induced motion for
all scales smaller than a nonlinear scale say $l_{NL}$ where,
$v(l_{NL})/l_{NL} \sim H(t) \sim 1/t$.
This scale is also approximately equal to the
magnetic Jeans length, below which
the distortion of the field can lead to
magnetic pressure gradients which counteract the gravitational
collapse \citep{KOR96,SB98a}.
In a linear analysis 
the proper magnetic Jeans' wave number, say $K_J$, can be got 
from equating the two terms: $4 \pi G \rho_m =
K_J^2 B^2/(8 \pi\rho_b)$, giving
\begin{equation}
K_J = {4\pi \sqrt{2\rho_m \rho_b G} \over B}.
\end{equation}
The first term incorporates gravitational instability, while
the second one the restoring effects due to magnetic pressure.
Noting that $H^2(t) = 8\pi G\rho_m/3$,
the above condition is equivalent to
the condition 
\[
K_JV_A(k_J,t) = \sqrt{3} H(t).
\]
which is explicitly very similar to 
the condition determining the nonlinear scale.
Interestingly, the comoving Jeans scale
$k_J = aK_J$, and the comoving Jeans' length
$ \lambda_J= 2\pi/k_J$, do not depend on time,
at early epochs where the universe is matter dominated; and
assuming that the field even at the scale $k_J$
just redshifts as $\propto 1/a^2$,
without significant distortion \citep{SB98a}. This is because
in this case $V_A \propto a^{-1/2}$ and $H(t) \propto t^{-1} \propto a^{-3/2}$
and hence $k_J \propto a(t)H(t)/V_A$ is constant with time.
So any scale which is linear/nonlinear just after recombination,
is approximately linear/nonlinear at all epochs
(until the vacuum energy starts dominating).
Putting in numerical values we get
\begin{equation}
k_J \simeq 14.8\, {\rm Mpc^{-1}}
\left ({\Omega_m \over 0.3 } \right )^{1/2} \left ({h \over 0.7}
\right )  \left ({B_J \over 10^{-9} {\rm G}} \right )^{-1}
\label{jeansscale}
\end{equation}
where $B_J = B(k_J,t) a^2(t)$ is the redshifted
value of the field smoothed on the scale $k_J$.
Perturbation growth is suppressed below the magnetic Jeans scale.
This is shown schematically 
as a sharp cut-off in the power spectrum in \Fig{matpow}.
The first structures to collapse will have scales
close to the magnetic Jeans' scale $\lambda_J$.

A standard measure of the stochastic density perturbation $\delta_m$ 
is the mass dispersion $\sigma(R,t)$ smoothed over a given radius $R$.
In the real space representation one first defines a smoothed out
density perturbation field as a convolution of $\delta_m$ with 
a window function of radius $R$. Then looks at the dispersion
of this quantity. 
This can then be calculated more conveniently in Fourier space 
\begin{equation}
\sigma^2(R,t) =  4 \pi \int_0^{k_J}  dk k^2 P(k,t)  W^2(kR)
\label{eq:massdis}
\end{equation}
Here $W(kR)$ is the window function; 
for example one can use a Gaussian
window function with $W(kR) = \exp(-k^2R^2/2)$. 
The usefulness of $\sigma(R,t)$ lies in the fact that its value
roughly determines when a structures on a given scale $R$ can collapse.
For spherically symmetric perturbations, collapse of a structure
corresponds to $\sigma(R,t) = 1.68$ \citep{Peebles80}. Of course
for a power spectrum with a cut-off the first collapses will be more
pancake like. And there is the added complication of
taking account of magnetic pressure effects. Nevertheless it seems reasonable
to demand that $\sigma(R,z) \sim 1$ for the formation of
structures. 
For $n \le -1.5$, the matter power spectrum is
$P(k) \propto B_0^4 k^{2n+7}/k_c^{2n+6}$, 
where $B_0$ now is the rms value filtered at scale $k_c$ \citep{GopalSethi03}. 
This gives $\sigma^2(R) \propto B_0^4/ (k_c^{2n+6} R^{2n+10})$.
For $k_c \simeq k_J$, $\sigma^2(R) \propto B_0^4k_J^4$, and
so from \Eq{jeansscale} does not depend on the value of $B_0$.
Therefore the redshift at which the first structures collapse
becomes nearly independent of the value of $B_0$, although the mass
contained in these structures, which depends on the scale
$k_J$, does depend on $B_0$, through the $k_J$ dependence.
For $n \simeq -3$, $\sigma(R) \propto 1/R^2$, and therefore, 
even though first structures
might form early, the formation of larger structures is suppressed.

\begin{figure}
\includegraphics[width=0.35\textwidth,angle = 270]{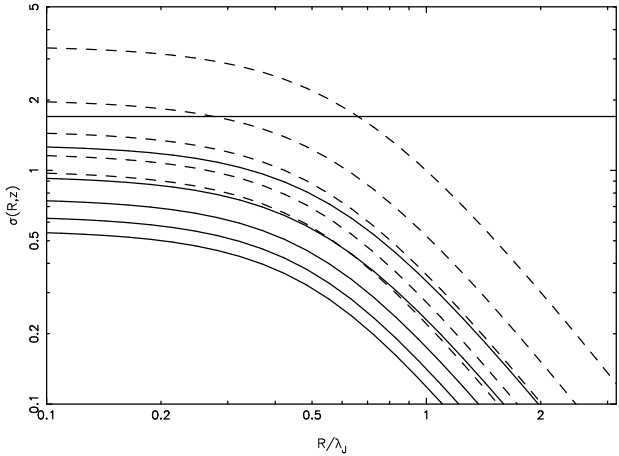}
\caption{The mass dispersion $\sigma(R,z)$ is shown for two
models with nearly scale free  magnetic field power spectra. The solid
and dashed curves correspond to $n = -2.9$ and $n = -2.8$, respectively.
   Different curves, from top to
bottom,  correspond to redshifts $z = \{10,15,20,25,30\}$, respectively.
 The horizontal line corresponds to $\sigma = 1.68$.
Reproduced with permission from \citet{Sethi_KS05}. 
}
\label{fig:f7}
\end{figure}

In Figure~\ref{fig:f7}  we
show the  evolution of $\sigma(R,t)$
for the power law magnetic spectra with $n \simeq -3$, for $\Omega_m = 0.3$.
We can adapt the standard 
\citet{PS74} type prescription
to compute the abundance of objects \citep{Peebles80,PS92,Padmanabhan02}.
For the power law models
with nearly scale invariant spectrum, with say $n \sim -2.8$,
we see from Figure~\ref{fig:f7},
that the first structures can collapse at high redshift $z \sim 10-20$.
This can significantly influence the reionization of the universe. 
Also, as $\sigma(R) \propto 1/R^2$ beyond the Jeans scale 
collapse of larger structures occur much later. 
This means that even though magnetic fields can induce  the
formation of first structures, it would have less impact on the
formation of galactic and larger scale structures at the present
epoch.
For $B_0 \sim 3 \times 10^{-9} \, \rm G$ the total mass 
of a collapsed halo $M_c \simeq 1-3 \times 10^{10} \, \rm M_\odot$,
which is much smaller than a typical $L_\star$ galaxy.
For a field $B_0 \sim 10^{-9} \, \rm G$, the mass $M_f$ of the first
collapsed objects will be smaller, by a factor $ \sim 30$.
Therefore the first structures to collapse
would be sub-galactic. 

From our discussion  above we can conclude that
(i) collapse of first structures
could have commenced for  $z \simeq 10\hbox{--}20$, (ii) only
a small fraction of mass range close to the magnetic Jeans' scale collapse
(iii) the collapse redshift is nearly
independent of the strength of the magnetic field, if
the magnetic field is specified as rms filtered at the Jeans scale and
(iv) the mass of the first collapsed objects will be sub-galactic their
exact value depending on $B_0$. These conclusions hold for magnetic
field strengths for which the magnetic
Jeans' length exceeds the thermal Jeans' length.

\subsubsection{Reionization signals}

The early formation of subgalactic structures in the 
presence of primordial magnetic fields, can significantly 
add to the ionizing photon budget and lead to an early 
onset of reionization 
\citep{Sethi_KS05,TS06a}.
Note that the universe, which was predominantly neutral after recombination,
transited to an ionized state after the first stars formed
and emitted ionizing radiation. This period is known as the 
epoch of reionization (EOR), and its understanding
is one of the outstanding challenges of 
modern cosmology. Observations of the spectra of high redshift quasars
have revealed the existence of neutral hydrogen (HI) at redshifts $z \ga 6$ 
\citep{fan}. Meanwhile the detection
of temperature-polarization cross-correlation and the 
polarization-polarization correlation
at large angular scales ($\ell \la 10$) in the WMAP and Planck 
data have given firm evidence that the universe reionized by
$z \simeq 10$ \citep{wmap,Planck_Cosmo15}. 
The presence of primordial magnetic fields could well
strongly impact on EOR predictions, and one can in turn
constrain such fields from observations related to the EOR.
 
\begin{figure}
\includegraphics[width=0.35\textwidth,angle = 270]{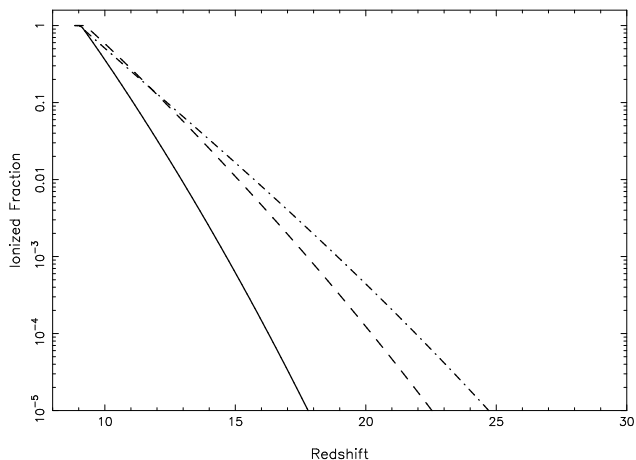}
\caption{
The ionization history of the universe is shown for two magnetic field
models along with a standard $\Lambda$CDM model without magnetic field. 
The dashed and the dot-dashed curves correspond to  $B_0 =1$ nG and
$B_0=3$ nG, respectively. The magnetic spectrum is nearly
scale invariant with a spectral index $n=-2.9$. The solid curve
correspond to the case without magnetic field.
Reproduced with permission from \citet{Sethi_Sub09}. 
}
\label{reion}
\end{figure}

In \Fig{reion} we show the reionization history for two
values of magnetic field strength and a standard $\Lambda$CDM
model with zero magnetic fields. The magnetic spectrum is fixed
to be nearly scale invariant with a spectral index $n=-2.9$. 
The reionization of the universe is modeled as described
in \citet{sethi05,Sethi_Sub09}. In this
model, an HII sphere is carved around collapsed  haloes. 
Reionization
proceeds as more sources are born and as the radius of the HII region
around each source increases. 
It is completed when the HII regions coalesce.
The formation rate of dark matter halos is computed as the time
derivative of halo abundance which in turn is derived
using the Press-Schechter formalism \citep{PS74}.  
The radius of the HII region $R$ is computed by following 
its  evolution around a source with a 
given photon luminosity $\dot N_\gamma$ (in $sec^{-1}$) which is assumed to 
grow  linearly with the halo mass; we fixed the fiducial value of the photon
luminosity to be $\dot N_\gamma(0)$ 
at the mass scale $M = 5 \times 10^7 \, \rm M_\odot$ 
(see e.g. \citep{sethi05}). The baryons in the IGM are assumed to be
clumped with a clumping factor of $2$. 
The figure shows the evolution of the ionized fraction of 
the IGM, $f_{ion}(z)$, adopting 
$\dot N_\gamma(0) = \{4 \times 10^{48}, 10^{49} \} \, \rm s^{-1}$ 
for $B_0 = 3$ nG and $B_0 = 1$ nG respectively, and 
$\dot N_\gamma(0) = 2.5 \times 10^{50} \, \rm s^{-1}$, for the $B_0=0$
$\Lambda$CDM model. All the reionization histories are
normalized to obtain an optical depth of $\tau_{\rm reion} = 0.1$
to scattering off the ionized electrons between $30 \ge z \ge 0$.

This photon luminosity can be cast in terms of the star formation 
efficiency, $f_{\rm eff}$ and escape fraction of hydrogen-ionizing 
photons $f_{\rm esc}$ from star-forming halo. The adopted value
for the zero field standard model is such that both the star formation 
efficiency and escape fraction are roughly $10\%$, assuming a Scalo
IMF for the forming stars. We see from \Fig{reion} that one needs
a much lower efficiency of star formation (about 25 times lower for
$B_0 = 1$ nG) for models with primordial
magnetic fields, to achieve the same degree of ionization as
a standard $\Lambda$CDM model \citep{Sethi_Sub09}.
Note that due to the uncertainties associated with both star formation 
and galaxy formation in presence of strong fields, it is not
easy to derive very precise constraints on primordial fields.
For some recent efforts in this direction see \citet{PCSF15},
who infer sub nG limits on $B_0$ for nearly scale invariant spectra.

\subsubsection{Redshifted 21 cm signatures}

The most direct approach to study the EOR is through observations 
of the redshifted 21-cm line of the neutral hydrogen \citep{Fur15}. There is 
substantial on-going effort in that direction and 
this also remains one of the primary goals of upcoming and 
future radio interferometers like the SKA 
\footnote{\tt www.skatelescope.org/pages/page\_sciencegen.htm} 
\citep{KoopSKA,PritSKA},
and its percursors 
(e.g. LOFAR \footnote{\tt www.lofar.org}). Since primordial magnetic
fields cause changes in the matter power spectrum and also
in the thermal and ionization history of the universe, 
they would leave characteristic imprints on the redshifted 21 cm 
emission. These signals have been worked out by several
authors \citep{TS06,SBK09,Sethi_Sub09}.

There are two types of signals which are usually considered;
the global HI signal and the fluctuating component. 
The global HI signal is observable at the redshifted HI hyperfine transition
frequency ($\nu_0 = 1420/(1+z)$), 
against the CMB, the only radio source at 
high redshifts. As the matter temperature falls off faster with
expansion
than the CMB temperature below a $z \sim 100$, the global HI signal
is generally expected to be seen in absorption against the CMB. 
However in the presence of sufficiently strong primordial fields,
there are two effects which can alter this expectation. First
the dissipation of tangled magnetic fields can 
significantly alter
the thermal history of the universe \citep{Sethi_KS05}. Moreover, 
the build-up of the Lyman-$\alpha$ radiation during the 
reionization era, which determines the spin temperature $T_s$
of the HI line, depends on the mass function of 
collapsed haloes. As we noted above 
primordial fields can enhance the abundance of such
small mass halos, and also lead to changes in the $T_s$.

\begin{figure}
\includegraphics[width=0.35\textwidth,angle = 270]{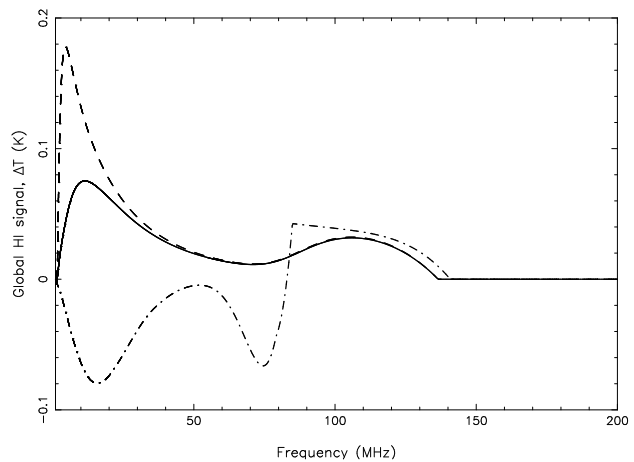}
\caption{
The Global HI signal is shown for two values of magnetic field strengths.
The solid and dashed curves correspond to the magnetic field strength 
$B_0 = \{5 \times 10^{-10}, 10^{-9}\} \, \rm G$, respectively. The dot-dashed
curve corresponds to HI signal for one possible  scenario in the 
zero magnetic field case.
Reproduced with permission from \citet{Sethi_Sub09}.
}
\label{globalHI}
\end{figure}

The global HI signal in the post-recombination era predicted
by the calculations of \citet{Sethi_Sub09} is shown in \Fig{globalHI},
for some magnetic field models and for a standard 
$\Lambda$CDM model with zero field \citep{sethi05}.
The figure shows that unlike 
the usual case with zero magnetic fields,
the HI signal  in the magnetised universe is only observable in emission
throughout the post-recombination era. This is mainly because  the matter
temperature does not fall  below the CMB temperature 
for the magnetic field models considered (see also \cite{SBK09}). 
The lack of detection of the HI in absorption could therefore 
be one signal for the presence of primordial magnetic fields.

The fluctuating component of the HI signals also gets
altered in the presence of primordial fields \citep{Sethi_Sub09}. 
Note that this
component is determined by density inhomogeneities in HI
resulting both from density fluctuations in matter and also
the result of inhomogeneous ionization. We saw that the matter
spectrum induced by nearly scale invariant primordial fields
has a sharp rise which is then sharply cut-off at the magnetic Jeans
scale (cf. \Fig{matpow}). Such a sharp feature leads
to corresponding oscillations in the HI angular correlation
function, with a scale length which increases  
with increasing value of the magnetic field strength.
Detecting such oscillations could point to 
the influence of primordial magnetic fields
and also help in determining the field strength. 
Another major difference between the $\Lambda$CDM model 
and the magnetic field scenario, is in the scale
of the signal. Both the density coherence scale and 
the scale of ionization inhomogeneities turn out to be
typically larger in the $\Lambda$CDM case \citep{Sethi_Sub09}. 

\subsubsection{Weak lensing signatures}

Another possibility 
to directly probe the excess power in the matter 
power spectrum induced by primordial fields is via weak 
gravitational lensing \citep{BaS01,Munshi08}. 
In such lensing the light rays of distant
galaxies are deflected by the intervening inhomogeneities resulting
in the shearing of their shapes. Galaxies are also intrinsically elliptical.
Thus the shearing distortion caused by weak lensing has to be measured
statistically by looking at the shapes of a large sample of background galaxies,
and calculating the shear correlation function. 

The shear field due to scalar density perturbations is to leading order
curl-free and is referred to as an E-type field. Thus typically one decomposes
the observed shear signal into an E (non-rotational) and B (rotational)
components. The level of B-modes can also used to estimate systematic 
errors in the data. The shear correlations due to  
the matter power spectrum induced by primordial fields has been calculated
by \citet{PS12}. They find that, for
nearly scale invariant spectra
and nG field strengths on $k=1$ Mpc$^{-1}$ scales, 
primordial fields lead to an excess in the
shear correlation function on few arc min angular scales, compared to
a standard $\Lambda$CDM model. A comparison to the observed data from  
\citet{Fu_wl08} was used by \citet{PS12} to set 
upper limits on primordial fields
at the nG level. 

\subsubsection{Influence on Lyman-$\alpha$ clouds}

The matter power spectrum at small spatial scales can also be 
sensitively probed from observations of the Lyman-$\alpha$ forest
absorption lines in the spectra of high redshift quasars 
\citep{Croft02,Mcdonald05}.
These lines are thought to trace the mildly over dense IGM and so
reflect the matter power spectrum. There are of course uncertainties
related to the mapping between dark matter and baryons, and the
background radiation which keeps the IGM ionized. Typically numerical
simulations including the effect of baryonic physics are needed to infer
the Lyman-$\alpha$ flux power spectrum \citep{VHS04}, but one can
also use semi-analytic models \citep{BiD97,CSP01}.

\citet{PS13} have use semi-analytical methods to simulate density
fluctuations along the line of sight, including the contribution 
of matter perturbations due to primordial magnetic fields.
They compare the effective Lyman-$\alpha$ opacity with the 
data given by \citet{Fauch08}, and derive an upper bound on the 
magnetic field strength of about $0.3-0.6$ nG for a range of
nearly scale invariant magnetic field power spectrum
 
\subsection{Constraints from Faraday rotation observations}

The polarization plane of a linearly polarized EM wave with wavelength
$\lambda$, when propagating through a magnetized plasma, with electron
density $n_e$ and line of sight magnetic field $B_{\parallel}(x)$,
gets rotated by an angle
\EQ
\psi=\frac{e^3}{2\pi m_e^2c^4}\,\lambda^2 \int_0^D n_e(x) B_{\parallel}(x)
\dd x+\psi_0 
\label{FRM}
\EN 
Here $\psi_0$ is the initial polarization angle within the source, 
$e$ the electron charge, $m_e$ electron mass,
and the integral is over the
line of sight to the source which is at a distance $D$.  
We can separate out the $\lambda^2$ dependence and write
$\psi= \RM \ \lambda^2+\psi_0$, where 
$\RM$ is called the Faraday rotation measure. Its determination
for a distant polarized source, like a high redshift radio source, 
probes the magnetic field, including any primordial component, 
along the line of sight to the source.

In the cosmological context, the observed wavelength will be related
to that in the medium at redshift $z$, by $\lambda_0= (1+z) \lambda$
Then $\psi = \RM \lambda_0^2$, where now
\EQ
\RM = \frac{e^3}{2\pi m_e^2c^4}\,\int_0^z 
\frac{n_e(z) B_{\parallel}(z)}{(1+z)^2} \frac{dx}{dz} dz,
\label{rmz}
\EN
with now $dx/dz = -cdt/dz = -(c/H)(d\ln a/dz)$. From Einstein equation, 
$H= H_0 \sqrt{\Omega_m(1+z)^3 +\Omega_\Lambda}$ and $a \propto (1+z)^{-1}$ and
so this gives 
\EQ
\frac{dx}{dz} = \frac{c}{H_0(1+z)\sqrt{\Omega_m(1+z)^3 +\Omega_\Lambda}},
\label{dxdz}
\EN
where $\Omega_m$ and $\Omega_\Lambda$ are present day matter and dark energy
densities in unit of the critical density.
For an estimate of the resulting $\RM$, assume an uniform field
which dilutes with as $B(z) = B_0/a^2 =B_0 (1+z)^2$, and that
the IGM is ionized with $n_e(z) = n_{e0} (1+z)^3$. Then one can 
integrate \Eq{rmz} exactly to get
\EQA
\RM(z) = 11.5 \left[\frac{B_0}{ 1 {\rm nG}}\right]
&&\left[\frac{\sqrt{\Omega_m(1+z)^3 +\Omega_\Lambda} - 1}{11.5 \Omega_m}\right]
\nonumber \\ 
&&\left[\frac{n_{e0}}{10^{-7} {\rm cm}^{-3}}\right]
\frac{rad}{m^2},
\label{rmzval}
\ENA
where the numerical value is got by normalizing to a redshift $z=3$.
We have also normalised $n_{e0}$ by its expected value
for a fully ionized universe with $\Omega_B =0.02h^{-2}$.
On the other hand, if the field is random, with an integral scale
$l_B$, then the mean value of $\RM(z)$ would be zero but its
rms value, would be smaller than that given in \Eq{rmzval} by
a factor of order $(l_B/D(z))^{1/2}$. 
Thus, assuming a homogeneous universe, an upper limit to the IGM contribution
to $\RM \sim 10$ rad m$^{-2}$ by $z=3$, results in upper limit
of a nG for the homogeneous field, and an upper limit $\sim 50$ nG for
a field coherent on a Mpc scale. 

It was however pointed out by \citet{BBO99} that the universe for redshifts
of interest is far from homogeneous, as indicated by the Lyman-$\alpha$
forest absorption lines, thought to arise in mildly over dense
regions of the IGM. The resulting RM would then be larger in such
an inhomogeneous universe, both because the density in such regions
would be larger and the magnetic field would be larger if it is flux frozen.
Assuming $B \propto n_e^{2/3}$ and a log normal density distribution for 
$n_e$ \citep{BiD97}, 
these authors simulated the RM resulting from
a large number of random lines of sight to high redshift quasars,
until $z=2.5$. Including the IGM inhomogeneity implied
by the Lyman-$\alpha$ forest, results in more than an order of magnitude
increase in the predicted RM, a large scatter and also a much less 
sensitivity to the coherence scale.
Comparing with the then existing data of \citet{WPK84}, \citet{BBO99} 
set upper limits of a nG for horizon scale fields, 6 nG for
fields coherent on scales of 50 Mpc. 
Using the RM data from the NVSS survey, \citet{PTU15} 
set stronger constraints of $0.5$ nG on horizon scale fields and
$1.2$ nG at the 2$\sigma$ level on fields with Mpc coherence lengths.
These limits are especially valuable as they use
a completely independent technique. They are also
likely to improve considerably with the advent of new 
generation of radio telescopes like the SKA \citep{TaylorSKA,JSKA15}.
There have been interesting work on RM due to magnetic fields
in the IGM using cosmological simulations \citep{AR11,Ryu_etal12} 
and techniques for their detection \cite{AGR14}, and it would be of interest
to extend such studies to include a primordial component.

\subsection{Constraints from Gamma ray observations}

Recent developments in gamma ray astronomy has provided 
an intriguing possibility to detect and constrain
very weak magnetic fields in the IGM. The basic idea reviewed
in detail by \citet{DN13} is the following:
Suppose we have a source of very high energy gamma rays in the
Tev range, like a blazar. These TeV photons can interact with
the low energy (eV) photons in the IGM, the 
extragalactic background light (EBL), and produce
a beam of electron-positron pairs. 
The resulting $e^\pm$ pairs are highly
relativistic and in turn inverse scatter of the much more
abundant CMB photons leading to GeV gamma rays. These Gev gamma rays
should be detectable by telescopes like Fermi \citep{Atwood09}, as a gamma
ray source associated with the Tev source. However in several
blazars where the TeV emission is detected, the corresponding 
secondary GeV cascade emission is not. 

\citet{NV10} suggested that
this could be explained by the presence of weak intergalactic magnetic fields.
In the presence of such fields, $e^\pm$ pairs get deflected
by the magnetic field. As one sees the GeV gamma ray only when the 
highly relativistic electrons and positrons move towards the observer,
this means that we see the GeV photons from a direction
displaced from the direction of the primary emission.
Thus in the presence of a magnetic field, the secondary 
GeV source becomes more extended, and its surface brightness
could drop below the detectability threshold of Fermi.
This was then used to put a lower limit on the IGM field.

We now give more quantitative estimates following \citet{NV10,DN13}.
First, the necessary condition for a TeV photon with energy $E_\gamma$ 
to interact with an EBL photon of energy $E_{2}$ and 
create electron-positron pairs is that the geometric mean of the two
photon energies has to be larger than the electron rest mass. That
is $\sqrt{E_\gamma E_{2}} > m_e c^2$, which implies one needs  
a primary photon energy $E_\gamma > 250 (E_2/1 {\rm eV})^{-1}$ GeV. 
The mean free path for gamma rays above this threshold energy is 
\citep{NV10}, 
$l_\gamma \sim 80\kappa (E_\gamma/10 {\rm Tev})^{-1}$ Mpc, where 
$\kappa \sim 1$ accounts for uncertainties in the EBL intensity.
Thus the Tev photons propagate far away from their source, in general
into the void regions of the IGM, before pair creation
take place, and the gamma ray constraints therefore probe the magnetic field
in the general IGM.

The created pairs have an energy $E_e \sim E_\gamma/2$ and so 
a relativistic gamma factor 
$\gamma_e \sim E_\gamma/(2 m_ec^2) \sim 10^6 (E_\gamma/1 {\rm TeV})$.
The inverse Compton scattering of the CMB photons off these
electron then produce secondary gamma rays with energy 
$E_{\gamma 2} \sim (4/3)\gamma_e^2 \epsilon_{CMB} \sim 0.8 
(E_\gamma/ 1 {\rm TeV})$ GeV, where we have used the fact that
typical energy of the CMB photon $\epsilon_{CMB} \sim 6 \times 10^{-4}$ eV.
The mean free path for such inverse Compton scattering is
$l_{ic} \sim 0.3 (E_e/1 {\rm TeV})^{-1}$ Mpc, much smaller than 
$l_\gamma$ the mean free path for pair creation. Thus inverse compton scattering
of the CMB drains the energy from the pairs, a short distance after 
they are created. 

Now the Larmour radius of a relativistic electron
is given by $R_L = \gamma_e m_e c^2/(e B) = E_e/(e B)
\sim 100 (E_e/ 1 {\rm TeV}) (B/ 10^{-17} {\rm G})^{-1}$ Mpc.
Thus while the electron traverses a distance $l_{ic}$ before it loses energy
to inverse compton scattering, it will be deflected by an angle
\EQ
\delta \sim \frac{l_{ic}}{R_L} \sim 3 \times 10^{-4} 
\left(\frac{E_e}{ 10 {\rm TeV}}\right)^{-2}
\left(\frac{B}{10^{-16} {\rm G}}\right) 
\label{delta1}
\EN
assuming the field is coherent on scales
much larger than $l_{ic}$.
If the coherence scale of the field $l_B$ is much smaller than
than $l_{ic}$, then the deflection of the charged particle
performs a random walk in angle. Then the deflection angle
only grows as $\sqrt{N}$ of the number of steps $N=l_{ic}/l_B$, and
$\delta \sim (l_B/R_L) \sqrt{N} \sim \sqrt{l_{ic}l_B}/R_L$.
This gives 
\EQ
\delta \sim 5 \times 10^{-5} \left(\frac{E_e}{ 10 {\rm TeV}}\right)^{-3/2}
\left(\frac{B}{10^{-16} {\rm G}}\right) \left(\frac{l_B}{1 {\rm kpc}}\right)
\label{delta2}
\EN
The size of the extended GeV cascade source $\Theta_\gamma$, 
can be estimated from the Geometry of the problem, where 
the blazar (B), the observer (O) and the source of cascade gamma 
rays (S) form a triangle, with angle $\angle BSO = (\pi-\delta)$ and 
$\angle SOB = \Theta_\gamma$. We have $\sin(\pi -\delta)/D_B =
\sin\Theta_\gamma/l_\gamma$, where $D_B$ is the angular diameter
distance to the blazar. Thus $\Theta_s \sim \delta (l_\gamma/D_B)
\sim \delta/\tau_\gamma$, where $\tau_\gamma = D_B/l_\gamma$ is
the optical depth for TeV gamma rays from a source at distance $D_B$,
due to absorption by the EBL.
Note that $\delta$ and the source size $\Theta_\gamma$ is larger 
for lower energy $e^\pm$ particles and in higher magnetic fields. 
If this deflection $\delta$ becomes greater than the PSF $\Theta_{PSF}$ 
of the telescope, then the surface brightness of the secondary gamma
ray source will decrease.

From the non detection of the secondary gamma ray emission by 
Fermi telescope, \citet{NV10,TGBF11} deduce a lower limit on the
magnetic field in the IGM of $B \ga 10^{-16}$ G, provided $l_B \gg l_{ic}$.
In the opposite limit $l_B \ll l_{ic}$ the limit becomes more stringent
and the lower limit increases as $l_B^{-1/2}$ with decreasing $l_B$.
These limits are derived assuming that the cascade emission by the
source becoming suppressed due to source extension.
Another possibility is that there is large time delay between the
direct primary emission and the secondary cascade emission.
Such a time delay is expected because for seeing the secondary 
emission, the TeV gamma rays have to first propagate from B to S and
then the secondary gamma rays propagate from S to O, where as the
primary emission is seen directly along the path B-O.
This lowers the limits on the magnetic field by an order of magnitude
or so \citep{Dermer11,TVN11}.

The basic idea that the relativistic beam of $e^\pm$ 
plasma loses its energy primarily to
inverse Compton emission against the CMB, has been questioned by 
\citet{BCP12}, who argue that plasma instabilities can drain
the energy from the $e^\pm$ beam at a faster rate. 
Whether this indeed obtains also depends on the efficiency of the
nonlinear Landau damping, which can suppress the growth of the
instability and stabilize the beam. There is however
disagreement about the range of parameters for which 
nonlinear Landau damping is important; while
\citet{ME13} favor the plasma instabilities being stabilized,
\citet{SIS12,Chang14} do find that plasma instability to be
important. As pointed out by \citet{DN13}, the development
of plasma instabilities is highly sensitive to the angular
and energy distribution of the particles in the beam.
A recent particle-in cell simulations also shows that, while
plasma instabilities broaden the $e^{\pm}$
beam distribution, they do not provide enough energy loss
to account for the missing GeV photons, and so magnetic field
deflections would then still be important \citep{KKS15}.
Thus much more work is perhaps needed to settle this issue.

Of relevance in this context is some intriguing work by \citet{TCFV14,CCFTV15}, 
who examine a parity-odd triple correlator $Q$ of $\gamma$-ray arrival 
directions in Fermi-LAT data, from possibly unseen blazars.
They find a non-zero $Q$
for certain energy bins, which seems to indicate the existence
of a helical (left-handed) IGM field $B\sim 10^{-14}$ G on 10 Mpc scales.
As these $\gamma$-rays are actually detected, the issue of whether
plasma instabilities drained some energy from the $e^{\pm}$ beam
becomes irrelevant.
Of course a positive detection of an extended halo of GeV emission 
around TeV blazars or a time delay effect, would provide more
conclusive evidence for the intergalactic magnetic field.
Nevertheless, 
$\gamma$-ray observations 
at present appears to be the only
manner in which such weak fields can be potentially
detected, and is therefore very exciting!

\section{ Astrophysical batteries and dynamos}
\label{batdyn}

Primordial magnetic fields are 
the main focus of this review.
However, it is also interesting to mention astrophysical batteries
which can generate coherent seed magnetic
fields in the
late universe. In addition magnetic fields are currently only
detected in systems like galaxies, galaxy clusters, stars and planets.
In all these systems, even if the field was originally
seeded by an early universe mechanism, they would decay
if not maintained by a dynamo mechanism. 
Thus it is of interest to
also briefly discuss dynamos.

\subsection{Cosmic batteries}
\label{seed}

Cosmic battery mechanisms are generally based on the fact that 
positively and negatively charged particles in a charge-neutral 
universe, do not have identical properties. For example, in an ionized
plasma, the electrons have a much smaller
mass compared to ions. Then for a given pressure gradient
of the gas electrons would be accelerated much more than the ions.
However this rapidly leads to an electric field, which couples back
positive and negative charges. If such a thermally generated electric
field has a curl, then from Faraday's law of induction a magnetic
field can grow. 
The resulting battery effect, is known as the Biermann battery and
was first proposed as a mechanism for the
thermal generation of stellar magnetic fields \citep{Bier50}.

More quantitatively, this thermally generated electric field is given by 
$\EE_{bier} = -{\bf \nabla} p_e /e n_e $, got by
balancing the pressure gradient force on the electrons, 
with the electric force and assuming the ions move negligibly as
they are much more massive. 
The curl of $\EE_{bier}$ leads to an extra source term
in the 
induction equation, which if we adopt 
$p_e = n_e k_{\rm B}T$, where $k_{\rm B}$
is the Boltzmann constant, 
gets modified to,
\EQ
{\partial \BBB \over \partial t} =
\nabla\times\left({\bf v}\times{\bf B} -\eta\nabla\times{\bf B} \right)
-{c k_{\rm B} \over e} {\nab n_e\over n_e}\times\nab T.
\label{modb}
\EN
We have assumed that the expansion factor hardly varies on the time-scale
of interest for the battery to generate magnetic fields.
We see from \Eq{modb} that over and above the usual
induction and diffusion terms we have a {\it source }
which is nonzero if density and
temperature gradients, $\nab n_e$ and $\nab T$, are not
parallel to each other.

Such non-parallel density and
temperature gradients can arise for example, in cosmic ionization fronts 
produced when the first ultraviolet photon sources,
like star bursting galaxies and quasars, turn on to reionize the universe 
\citep{SNC94}.
The temperature gradient in this case is 
normal to the front.  However, density gradients associated with
fluctuations which will later collapse to form galaxies and clusters,
can be uncorrelated with the ionizing sources
and point in a different direction.
This leads to a thermally generated
electric field which has a curl, and magnetic fields
correlated on galactic scales can grow. After compression during
galaxy formation, they turn out to have a strength
$B \sim 3 \times 10^{-20}$G \citep{SNC94}. 
This scenario has in fact been confirmed in detailed numerical
simulations of IGM reionization by \citet{GFZ00}.
Note that the large coherence is set here by the coherence
scale of the density gradients, even though this  battery
operates purely by a plasma physics process.
The Biermann 
type baroclynic term 
could also generate both
vorticity and magnetic fields 
just after recombination \citep{NN13} or 
in oblique, curved, cosmological shocks which arise
during cosmological structure formation
\citep{KCOR97}.

The asymmetry in the mass of the positive and negative charges
also implies that radiation interacts more strongly with electrons
than ions. In the presence of rotation or vortical motions this can 
also leads to battery effect \citep{Harrison69}. The resulting
magnetic fields, for example generated from 2nd order
perturbations during the radiation era, 
are very small $B \sim 10^{-30}$G on Mpc scales
up to $B \sim 10^{-21}$G at parsec scales 
\citep{GS05,MMNR05,TIOH05,KMST07}.

Although these astrophysically generated magnetic fields
cannot directly explain the observed fields, they are coherent 
on large scales and can provide a seed for dynamo action to which
we turn.

\subsection{Cosmic dynamos}

Dynamos convert the kinetic energy associated with motions of the 
fluid to magnetic energy by electromagnetic induction. 
An important point to clarify is the following:
Even in the
presence of strong initial fields, and long ohmic decay times
due to having a high $\Rm$, dynamos could be needed to maintain fields.
This is because given a strong tangled field the
Lorentz forces transfer magnetic energy to motions in the fluid.
This can in turn be dissipated due to viscous forces for small $\Rey$. 
Or for large $\Rey$ drive decaying MHD turbulence, 
with a cascade of energy to smaller and smaller scales and
eventual dissipation on the dynamical timescales associated with the motions. 
This timescale is $\sim 10^7$ yr in galaxies and only 10 times 
larger in clusters, and so much smaller than their ages.
Similarly, if the fluid is already turbulent, the associated larger turbulent
resistivity will lead to the decay of large-scale fields.
One exception to these arguments would be if the field is
strong and helical, in which case it can resist such turbulent
decay, due to a more subtle helical dynamo action driven by the
field itself (see below and \citet{Blackman_Sub13,BBS14}).
Therefore, generically one needs dynamos even if strong
primordial fields are generated.

All the astrophysical systems are strongly turbulent. 
This turbulence is either driven (like in galaxies and galaxy clusters)
or due to instabilities (like in stars and accretion disks).
Thus all astrophysical dynamos are turbulent dynamos.
Turbulent dynamos are conveniently divided into
fluctuation (or small-scale) and mean-field (or large-scale) dynamos.

\subsubsection{Fluctuation dynamos}
\label{fd}

The importance of fluctuation dynamos in cosmic objects obtains
because they are generic to any random flow
of a sufficiently conducting plasma, and operate whenever 
$\Rm$ exceeds a modest critical value $\Rmc \sim 100$ \citep{Kaz68}.
Fluid particles in such a flow randomly walk away from each other. A
magnetic field line frozen into such a fluid is extended by the
random stretching and exponentially amplified. 
The amplification is also typically 
rapid compared to the age of the system. For example, in the galactic
interstellar medium it could occur on the eddy turn over time scale
of about $10^7$ years, while in clusters the corresponding
time scale is about $10^8$ years. These time scales are much smaller
than the ages of these systems. Thus the fluctuation dynamo is a good
candidate for explaining observed magnetic field strengths 
in galaxy clusters and young galaxies.

However, this amplification comes at a cost, as the field is 
squeezed into smaller and smaller volumes as rapidly as it is amplified. 
The generated field then gets highly intermittent in the kinematic stage 
\citep{ssd} and concentrated on resistive scales $l_\eta \sim l/\Rm^{1/2}$, 
where $l$ is the eddy scale. 
A critical issue for all applications is how
coherent are the fields when the fluctuation dynamo 
saturates. Note that some of the early universe mechanisms which involve 
the amplification of field due to turbulence generated in a 
phase transition, assumed that fluctuation
dynamos generate fields coherent on eddy scales (see \Sec{genpt}).

A simple model of \citet{S99} suggests that the dynamo can saturate
by the Lorentz force driving the dynamo to its marginal state. In 
such a case the magnetic field in the saturated state concentrates
on scales $l_c \sim l/\Rmc^{1/2}$. As $\Rmc \ll \Rm$ typically, this implies
a much more coherent field in the saturated state of the dynamo
than during the kinematic stage.
Using direct numerical simulations (DNS) with large magnetic Prandtl numbers 
($\Pm=\Rm/\Rey\gg 1$), but small fluid Reynolds numbers ($\Rey$),
\citet{Schek04} argued that the fluctuation dynamo saturates 
with the magnetic field still concentrated on resistive scales.
On other hand simulations of \citet{HBD04,Eyink_etal13} 
with $\Pm=1$ and a large $\Rm=\Rey \approx 10^3$,
found that the magnetic integral scale, $l_B$, is just a modest 
fraction of $l$, and much larger $l_\eta$.
The generated field could then have a sufficient degree of coherence
to explain cluster and young galaxy fields
\citep{SSH06,EV06,CR09,BS13}, and also the level of coherence usually
assumed when applying the fluctuation dynamo to the early universe
context.
The case when both $\Rey$ and $\Pm$ are large, 
obtains
in galactic, cluster and early universe plasmas, 
while dense plasmas as in stars has $\Rey$ large but $\Pm\ll1$. 
At large $\Pm$, the fluctuation dynamo is easier to excite, 
than when $\Pm$ is small. But the case when $Pm$ is very different
from unity
is of course not easy to simulate,
as one has to resolve widely separated resistive and viscous scales.
The saturation of the fluctuation dynamo,
in these cases
could be quite different \citep{Eyink11}, and needs more work.
Understanding fluctuation dynamos and their saturation
is also important to understanding the magnetization of the
first galaxies \citep{Sch10,SSK13} and the first stars \citep{Sur10}.

\subsubsection{Turbulent mean field dynamos}

In systems like disk galaxies, one observes the field
to have a larger scale (say of order several kpc) 
than the coherence scale of the turbulent
motions (which are of order 100pc). 
Mean field dynamos (MFD) which grow or maintain 
large-scale fields correlated on scales larger
than the turbulent eddy scales, would be relevant.
They typically require more special conditions
(like turbulence to be helical) and amplify fields 
on a much longer time scale $\sim {\rm few} \times 10^8$ yr
rotation timescale of the disk galaxy.

To understand MFDs more quantitatively, 
consider a velocity field which is 
the sum of a mean velocity $\mean{\VV}$
and a turbulent, stochastic velocity $\vv$.
The induction equation becomes a stochastic
partial differential equation.
Split also the magnetic field as $\BBB = \mean{\BBB} + \bb$, 
into a mean field $\mean{\BBB}$ and a
fluctuating component $\bb$. We again assume the expansion
timescale is much larger than relevant dynamo time-scale
and take \Eq{expind} as valid for $\BBB$ itself.
Taking the average of \Eq{expind}, one gets
the MFD equation for the mean field $\mean{\BBB}$,
\EQ
\frac{\partial \mean{\BBB}}{\partial t} =
{\bf \nabla } \times \left( \mean{\VV} \times \mean{\BBB} + \vec{\emf}
- \eta \nab \times \mean{\BBB} \right) .
\label{MFDeqn}
\EN
This averaged equation now has a new term,
the mean electromotive force (emf)
$\vec{\emf}=\overline{\vv\times\bb}$, 
which crucially depends on the statistical properties of 
the small-scale velocity and magnetic fields.
A central closure problem in MFD theories is
to compute $\vec{\emf}$
and express it in terms $\mean{\BBB}$ itself.
Under the simplest assumptions of isotropy and sufficient scale separation in
space and time, we have just $\vec{\emf}=\alpha\mean{\BBB}-
\eta_t(\nab \times \mean{\BBB})$, where \citep{BF02b,BS05a}
\EQ
\alpha=-\sfrac13\tau\left(\overline{\oo\cdot\vv}
-\frac{\overline{(\nab\times\bb)\cdot\bb}}{4\pi\rho}\right),
\
\eta_t=\sfrac13\tau\overline{\vv^2}.
\label{alptilde+etatilde}
\EN
Here $\alpha$ is the dynamo $\alpha$-effect, where
$\alpha_K = -(\tau/3)\overline{\oo\cdot\vv}$ proportional to the
kinetic helicity obtains when Lorentz forces are not important,
and $\eta_t= (\tau/3)\overline{\vv^2}/3$ 
is the turbulent magnetic diffusivity proportional to the kinetic energy
of the turbulence \citep{Mof78}. 
The extra term  $\alpha_M = 
-(\overline{(\nab\times\bb)\cdot\bb})/(4\pi\rho)$, depending 
on the current helicity, is a magnetic contribution
to the $\alpha$-effect and incorporates the  
back reaction due to the Lorentz forces on the dynamo \citep{PFL76}.
The turbulent diffusion can in principle lead to the decay
of large scale fields, while the
 alpha effect allows its generation.
The dynamo in disk galaxies works with differential rotation
shearing radial fields to generate toroidal fields,
while the $\alpha$-effect is crucial for regeneration
of poloidal from toroidal fields.
This leads to exponential growth of the mean field, 
typically on time-scales a few times the rotation time scales,
of order $3-10 \times 10^8$ yr, during the kinematic phase. 

This picture of the galactic dynamo faces two potential problems.
First, while the mean field dynamo operates to generate
the large-scale field, the fluctuation dynamo is also operating.
This could again lead to concentration of power
on resistive scales \citep{SB14}.
However, recent work by \citet{BSB15} find that
Lorentz forces
help to move power to larger scales
and the mean field can grow to a significant fraction of the total field
even in the presence of the fluctuation dynamo. Moreover, both small
and large-scale fields grow in a unified manner, initially at the fast
rate corresponding to the fluctuation dynamo.

Second, the very nature of the MFD, which involves 
toroidal-poloidal-toroidal cycle, implies the growth of links 
or magnetic helicity in the mean field. Due to magnetic helicity 
conservation, this implies the growth of oppositely signed 
small scale helicity, which then contributes to $\alpha_M$ and
goes to suppress the dynamo (see \citet{BS05a,BL14} for detailed discussion
and references).

This can be made more precise by defining a gauge invariant
small scale magnetic helicity density $h$, 
using the Gauss linking formula for helicity \citep{SB06}.
In physical terms $h$ of a random small scale field 
is the density of {\it correlated} links of the field.
\citet{SB06} then derived a {\it local} conservation
law for $h$,
\EQ
\frac{\partial h}{\partial t} + \nab\cdot\mean{\FF}
= -2\vec{\emf}\cdot\mean{\BBB}-2\eta(4\pi/ c)\overline{\jj\cdot\bb}.
\label{finhel}
\EN
where $\mean{\FF}$ gives a flux density of helicity.
(For a weakly inhomogeneous system, $h$ is approximately
$\overline{\aaa\cdot\bb}$ in the Coulomb gauge.)
We have
in the stationary limit, 
$\vec{\emf}\cdot\mean{\BBB} = -(1/2)\nab\cdot\mean{\FF}
-\eta(4\pi/c)\overline{\jj\cdot\bb}$,
and therefore $\vec{\emf}\cdot\mean{\BBB}$ will
go to zero as $\eta \to 0$ in the absence of helicity fluxes.
However this need not be the case for a non-zero $\FF$.
Large scale dynamos then seem to need helicity fluxes
which transfer the twists in the small scale field 
out of the dynamo active region,
to work efficiently \citep{BF00,SSSB06,SSS07}.
They could even be completely driven by such fluxes (Ethan Vishniac,
Private communication).

What do these ideas on dynamos imply for the primordial magnetic field?
If the primordial field is small, it can provide a seed for the dynamo;
however as we saw in \Sec{seed}, astrophysical batteries could also
equally provide coherent seed fields. Moreover, whatever is the
source of the seed field, the fluctuation dynamo can amplify it
rapidly to near equipartition with the turbulence, and this
small scale field itself could be the seed for the mean field
dynamo. In fact, when helicity is present both the small-scale
and large-scale can grow in a unified manner \citep{BSB15}. 

\begin{figure}
\includegraphics[width=0.49\textwidth,height=0.4\textheight]{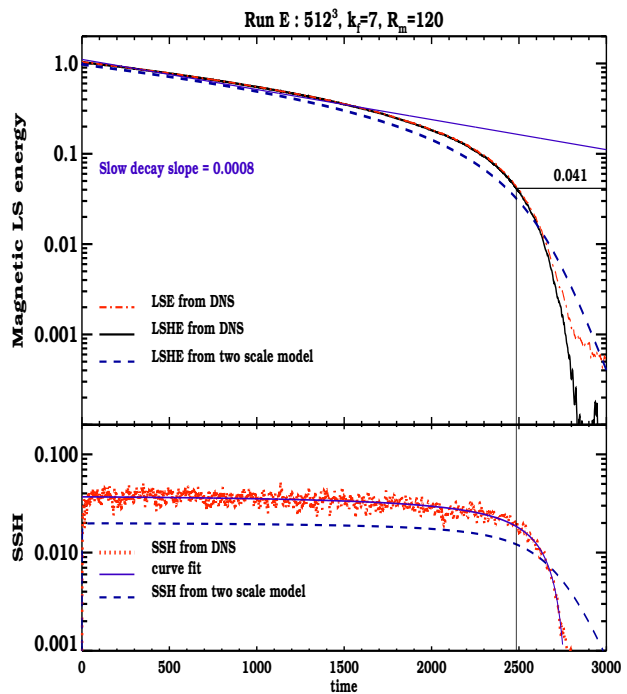}
\caption{
The evolution of the large scale field energy (LSE), 
its helical counterpart (LSHE) and the small scale helicity (SSH) 
for a DNS of helical field decay in the presence of
forced nonhelical turbulence (forced at a wavenumber $k_f=7$)..
All the quantities are normalised by $M_{eq}$.
The thin vertical line marks the time by when the SSH decreases 
by $50\%$ of its initial 
steady state value, and intersects the LSHE
curve at the transition energy indicated by the horizontal thin line. 
The thin blue line in the top panel shows the fit 
to the slow resistive decay phase.
Adapted with permission from \citet{BBS14}.}
\label{figdecay512}
\end{figure}

The situation gets more interesting if primordial fields
are strong enough, when they are captured and amplified in the
formation of a galaxy or cluster, that their energy density 
is comparble to the turbulent energy density.
In this case, naively turbulent diffusion would still lead
to its decay. However, $\eta_t$ itself could be affected and
moreover, the nature of the dynamo could change qualitatively
in the presence of strong initial fields.

Even more interesting is when the primordial
field is helical. It turns out that a large scale helical field
is resilient to turbulent diffusion, and only decays on the
large resistive time scale \citep{KBJ11,Blackman_Sub13,BBS14}.
The reason has again to do with helicity conservation.
Suppose we have an initially large scale helical field in the
presence of small scale nonhelical turbulence. Then $\alpha_K=0$, but
$\eta_t$ is not and contributes to $\vec{\emf}$ in \Eq{finhel}.
This turbulent diffusion then leads to the transfer of helicity to smaller
scales, and this buids up $\alpha_M$ until $h$ reaches a steady state 
(bottom panel of \Fig{figdecay512}).
In the presence of $\alpha_M$ the large scale field has a generation
term and decays only on the slow resistive timescale. This argument
made by \citet{Blackman_Sub13} on the basis of a two scale
model, was checked by \cite{BBS14} through DNS. The result
of one such run, shown in \Fig{figdecay512}, confirms the
picture of initially slow decay of the helical field until it reaches
a few percent of the equipartition value, after which it undergoes
a rapid decay. Thus if the primordial field were helical, it would
seem to be preserved for a resistive times (which is generally longer
than the age of the universe) in turbulent galaxies and clusters. 

\section{Final thoughts}

Primordial magnetic fields are important not only because
they may explain to some extent the observed fields in the universe,
but also because they could provide a window to probe the
physics of the early universe. A vast range of mechanisms for their
origin have been suggested, but none are as yet natural and or 
compelling. 
Theoretical predictions being highly parameter dependent, we have 
for most part taken a more pragmatic approach, assumed 
that such a field could be generated in the early Universe
and asked how it would evolve and what signals it could leave behind.

The generation during inflation is attractive as it naturally
provides the needed coherence; however apart from having to fine
tune parameters, there is also the problem of strong
coupling. Perhaps the solution to this issue is 
having theories where some field appears outside the full 4-d action,
and then finally goes to renormalize the 4-d distances, rather than
the coupling constants. This idea, which was mentioned as
a consequence of an evolving higher dimensional scale factor, 
needs to be better developed.
To the author it is also appealing that origin of 
primordial fields could reflect naturally the influence of evolving extra 
dimensions.
It would also be intriguing if such fields are just a residue
left after a proper regularization of the electromagnetic 
energy momentum tensor during the inflationary era. More work
on this idea, which avoids adhoc breaking of conformal invariance, 
would be of interest.

The idea that the origin of primordial fields is intimately linked to 
baryogenesis in the simplest extensions of the standard model
seems also attractive. 
One needs of course fairly optimistic assumptions
about the amount of energy going into the field at say the EWPT,
and it possibly being helical, to get interesting field strengths 
and coherence scales. However, even very weak fields if they
are sufficiently coherent can provide fields in voids,
being suggested by high energy gamma ray astronomy.

The linear evolution of large-scale primordial fields is reasonably 
well understood. The nonlinear evolution, of particularly nonhelical
fields has thrown up a surprise; the possibility of inverse cascade
of such fields. This unexpected result seems to either reflect
an additional conservation law, or the fact that the joint
evolution of the velocity and magnetic fields is more complicated
than hitherto realized.
 
We have discussed several signals of the primordial field
on the CMB. Currently the Planck data puts upper limits
at the few nG level. As the CMB power spectrum goes
as the fourth power of the field strength, it may not be
possible to easily improve these limits by more than a factor 10
or so. As the magnetic field induced signal is inherently
non Gaussian, this would seem to provide a good way to
isolate its effects. Clever ideas are also required, combining all the
different magnetic field signals to make more progress.

Regarding the effect of primordial fields on structure formation,
the work so far has been quite preliminary. The magnetic field is
quite a complicated "entity", which influences baryonic collapse
in an anisotropic fashion, can induce decaying 
turbulence to heat and ionize the plasma,
affect star formation etc. Thus galaxy formation in the
presence of strong primordial magnetic fields needs much
more work, perhaps using cosmological MHD simulations.
This will help to better pin down its effect on 
reionization and 21 cm signals.

Observations and experiments are the key to progress in any field.
In this respect it is heartening to note that the origin of
cosmic magnetism is one of the key projects of the SKA.
The determination of a large number of accurate rotation measures and its
statistical analysis, combined with numerical simulations of
structure formation including magnetic fields would be the way
to detect and constrain fields in the IGM which could be primordial.
Of course the detection of a single case, say in a cluster
of fields coherent on Mpc scales, would very much favor a primordial
origin. It was also an unexpected bonus that gamma-ray astronomy
started probing magnetic fields in voids at the level of
$10^{-16}$ G. Even though there could be other explanations
for the lack of GeV emission from some of the TeV blazars,
the positive detection of such a GeV halo around a TeV blazar, or the
time delayed GeV signal, could lead to positive 
detections of IGM fields; this is an exciting prospect.

Finally, it is appropriate to point out that the study of dynamos
and primordial fields are both important and complimentary.
Primordial magnetic fields would be interesting as a probe of the
physics of the early universe, even if they are not required
to seed the dynamo. And dynamos are required to maintain 
the field in collapsed objects, even if their initial origin were
primordial!

\ack

I thank John Barrow, Eric Blackman, 
Axel Brandenburg, T. R. Seshadri, Shiv Sethi and Anvar Shukurov 
for many discussions on these topics over the years. 
I thank my fellow student magnets, Pallavi Bhat, Luke Chamandy,
Sharanya Sur and Pranjal Trivedi and anti magnets, Charles Jose and
Saumyadip Samui, for forcing
me to learn more than I would have otherwise. The hospitality of 
Nordita, Sweden where the work began and the Institute for Advanced Study, 
Princeton where it ended are warmly acknowledged, along with my respective 
hosts Axel and Eric. Axel Brandenburg, Kanhaiya Lal Pandey and
Richard Shaw are thanked for providing figures from their work.
Two referees are thanked for very thoughtful reports which helped
improve the review.


\end{document}